\documentclass[aps,prb,twocolumn,floatfix,nofootinbib,superscriptaddress,longbibliography,nobibnotes]{revtex4-2}
\usepackage[utf8]{inputenc}
\usepackage{bm}
\usepackage{physics}
\usepackage{xcolor}
\usepackage[final,colorlinks,bookmarks=true,citecolor=red,linkcolor=red,urlcolor=blue]{hyperref}
\usepackage{graphicx}
\usepackage{mathbbol}

\usepackage{amsfonts}
\usepackage{amsmath}
\usepackage{listings}
\usepackage{enumerate}
\usepackage{latexsym}
\usepackage{float}
\usepackage{siunitx}
\usepackage{subfigure}
\usepackage{braket}
\usepackage{verbatim}
\usepackage{tikz}
\usepackage{mathtools}
\usepackage{ulem} % to strike things out
\usepackage{cleveref}

\newcommand{\bs}[1]{\boldsymbol{#1}}
\newcommand{\nis}{NiS$_2$}

\newcommand{\bk}{\ensuremath{\bs{k}}}

\newcommand\beginsupplementary{%
	\xdef\presupfigures{\arabic{figure}}% guarda el número de figura actual
	\renewcommand\thefigure{S\arabic{figure}}% redefine la numeración de figuras
}
\begin{document}
%\title{Origin of step-edge states in NiS$_2$ in terms obstructed Wannier charges}
\title{One-dimensional conduction channels in the correlated Mott NiS$_2$ arising from obstructed Wannier charges}

\author{Mikel Iraola$^\Vert$}
\email{m.iraola@ifw-dresden.de}
\affiliation{Donostia International Physics Center, 20018 Donostia-San Sebastian, Spain}
\affiliation{Institute for Theoretical Solid State Physics, IFW Dresden, Helmholtzstrasse 20, 01069 Dresden, Germany}
\def\thefootnote{$\Vert$}\footnotetext{These authors contributed equally to this work}
\author{Haojie Guo$^\Vert$}
\affiliation{Donostia International Physics Center, 20018 Donostia-San Sebastian, Spain}
\author{Fabio Orlandi}
\affiliation{ISIS Facility, STFC Rutherford Appleton Laboratory, Harwell Science and Innovation Campus, Oxfordshire OX11 0QX, United Kingdom}
\author{Sebastian Klemenz}
\affiliation{Fraunhofer Research Institution for Materials Recycling and Resource Strategies IWKS, Aschaffenburger Strasse 121, 64357 Hanau, Germany}
\author{Martina Soldini}
\affiliation{Department of Physics, University of Z\"urich, Winterthurerstrasse 190, CH-8057 Z\"urich, Switzerland}
\author{Sandra Sajan}
\affiliation{Donostia International Physics Center, 20018 Donostia-San Sebastian, Spain}
\author{Pascal Manuel}
\affiliation{ISIS Facility, STFC Rutherford Appleton Laboratory, Harwell Science and Innovation Campus, Oxfordshire OX11 0QX, United Kingdom}
\author{Jeroen van den Brink}
\affiliation{Institute for Theoretical Solid State Physics, IFW Dresden, Helmholtzstrasse 20, 01069 Dresden, Germany}
\author{Titus Neupert}
\affiliation{Department of Physics, University of Z\"urich, Winterthurerstrasse 190, CH-8057 Z\"urich, Switzerland}
\author{Miguel M. Ugeda}
\email{mmugeda@dipc.org}
\affiliation{Donostia International Physics Center, 20018 Donostia-San Sebastian, Spain}
\affiliation{KERBASQUE, Basque Foundation for Science, Maria Diaz de Haro 3, 48013 Bilbao, Spain}
\affiliation{Centro de Física de Materiales, Paseo Manuel de Lardizábal 5, 20018 San Sebastián, Spain}
\author{Leslie M. Schoop}
\email{lschoop@princeton.edu }
\affiliation{Department of Chemistry, Princeton University, New Jersey 08544, USA}
\author{Maia G. Vergniory}
\email{maia.vergniory@usherbrooke.ca}
\affiliation{Donostia International Physics Center, 20018 Donostia-San Sebastián, Spain}
\affiliation{D\'epartement de Physique et Institut Quantique,  
Universit\'e de Sherbrooke, Sherbrooke, J1K 2R1, Qu\'ebec, Canada.}
\affiliation{Regroupement Qu\'eb\'ecois sur les Mat\'eriaux de Pointe (RQMP), Quebec H3T 3J7, Canada}

\begin{abstract}

NiS$_2$, a compound characterized by its pyrite structure, uniquely bridges the realms of strong correlation physics and topology.  While bulk NiS$_2$ is known to be a Mott or charge-transfer insulator, its surface displays anomalous metallic behavior and finite conductivity. Using high-resolution neutron scattering data and symmetry analysis, we propose a refined description of NiS$_2$'s magnetic phases by introducing a novel model  for its ground state. Combined with high-resolution scanning tunneling microscopy and spectroscopy (STM/STS), we unveil the presence of edge states in both Ni- and S-terminated surfaces, which exhibit remarkable resilience to external magnetic fields. Although both types of edge states exhibit similar properties, only the edge states at the Ni termination populate the vicinity of the Fermi level and, therefore contribute to the surface conductivity. Utilizing \textit{ab initio} methods combined with a topological quantum chemistry analysis, we attribute these edge states to obstructed atomic charges originating from bulk topology. Overall, this work not only deepens our understanding of NiS$_2$ but also lays a robust experimental and theoretical foundation for further exploration of the interplay between one-dimensional step-edge states and the Wannier obstruction in correlated materials.
\end{abstract}

\maketitle

\section{Introduction}

The pyrite-type compound \nis{} originally garnered significant interest owing to the multiple magnetic phases it displays over different temperature ranges. The system undergoes a transition from a paramagnetic to an antiferromagnetic structure upon lowering the temperature. Upon further cooling, magnetic moments realign forming a weak-ferromagnetic (WFM) pattern~\cite{Hastings70, Nishihara75, Nagata76, Kikuchi78, Higo2015, Yano16, El-khatib2021}. In recent years, \nis{} has been under consideration as a platform to investigate metal-to-insulator transitions induced by chemical substitution~\cite{Yao1996, Kunes2010, Xu2014, Moon2015, Han2018, Jang2021}, pressure~\cite{Schuster2012, Park2024,  Friedemann2016} and gating~\cite{Day-Roberts2023}. 

Despite having been the subject of numerous studies, the electronic structure of \nis{} manifests intriguing aspects that require further analysis. 
For instance, while it is widely accepted that the material is a charge-transfer insulator exhibiting features of Mott physics in the bulk~\cite{Fujimori1996, Matuura1998, Iwaya2004, Kunes2010}, transport experiments have reported surface conduction~\cite{Thio1994, Sarma2003, Rao2011, Clak2016, El-khatib2021, El-khatib2023}. 
On the one hand, the contrast between the insulating bulk and conducting surface may suggest that the bulk of \nis{} hosts a topological Mott phase, which manifests as conducting states on the surface~\cite{Kivelson2010, Iraola2021, Soldini2023}. On the other hand, recent scanning tunneling microscopy/spectroscopy (STM/STS) measurements have corroborated the insulating nature of 2D surfaces of \nis{}, and observed metallic 1D step-edge states~\cite{Yasui2024}. While the existence of metallic 1D states could potentially explain the surface transport, their origin remains elusive.

In this work, first, we present a rigorous determination of the magnetic phases using high-quality neutron powder diffraction data, along with a new model for the weak ferromagnetic (WFM) ground state observed below $30$ K. Our model, developed from a detailed group-theory perspective, is a constant moment solution based on three irreducible representations of the parent space group, and provides a comprehensive framework to describe the magnetic ordering of the ground state.  Following the magnetic characterization, we investigated its surface structure and corresponding electronic properties by STM/STS.  High-resolution STM imaging of the cleaved surfaces perpendicular to the [001] direction in NiS$_2$ reveals two distinct atomic surface planes that unambiguously correspond to the S and Ni terminations. Tunneling spectroscopy measurements reveal the existence of in-gap states at  the step edges of both terminations with few nanometers in width. Moreover, we characterized the influence of out-of-plane magnetic fields on the edge states and demonstrate their robustness against fields of up to 10~T. 

\begin{figure}[htb!]
\centering
\includegraphics[width=1.0\linewidth]{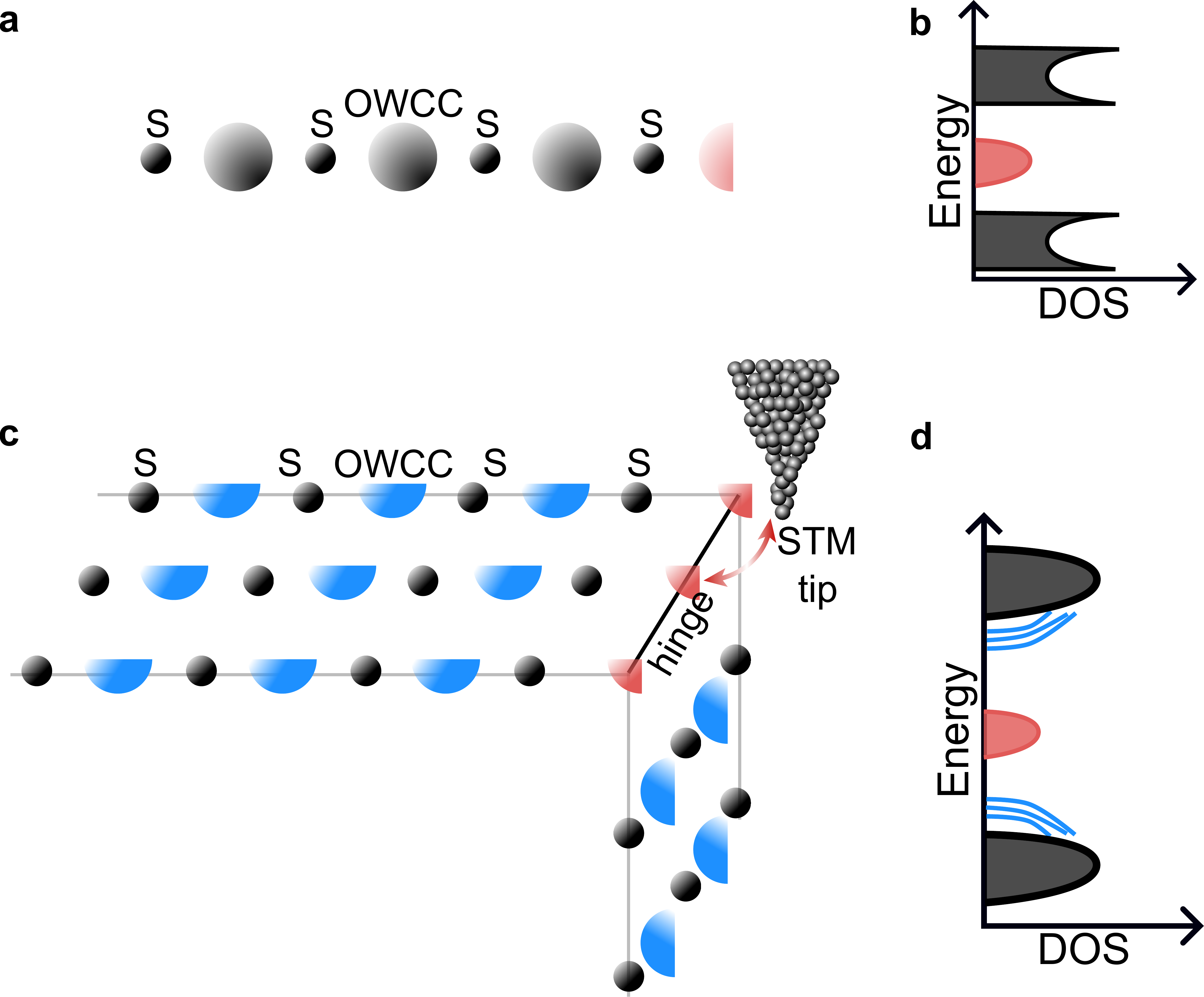}
\caption{\textbf{Illustration of the concept of OAI.} a) 1D chain where OWCC can be interpreted as bonding or antibonding states formed from neighboring S atoms. Grey spheres indicate the OWCCs. The half sphere in red denotes the OWCC on the boundary which leads to the filling anomaly. This situation in analogous to the one realized in the Su-Schrieffer-Heeger model~\cite{Su1979}. b) Density of states of the chain, where the contributions coming from bulk and boundary are indicated in black and red, respectively. c-d) Schematic illustration of the relation between OWCC and step-edge modes in \nis. The OWCC on the surfaces (blue) lead to a gapped DOS, while OWCC on the hinge (red) yield states are located within the gap.}
\label{Figure 1}
\end{figure}

In order to interpret the STM data, and to determine the origin of the observed 1D metallic states, we perform first-principle simulations and group theory. First, we identify an insulating state consistent with the symmetry of the antiferromagnetic phase, and further classify it within the framework of topological quantum chemistry (TQC)~\cite{Bradlyn2017, Cano2018, Bradlyn2018}. From the TQC analysis, we find that this state realizes an obstructed atomic insulator (OAI) phase. In an OAI, the valence bands are induced from Wannier functions localized on obstructed Wannier charge centers (OWCCs), i.\,e., sites that are not occupied by any ion, and these OWCC can manifest as in-gap states on certain surface cuts. In fact, our theoretical calculations show that the electronic structures of the Ni and S terminations and the step-edge configurations are consistent with our STS spectra, which suggests that the 1D states stem from the filling anomaly of obstructed charge centers on the edges of the crystal (see Fig.~\ref{Figure 1}c,d). Furthermore, we justify the robustness of step edge states against an external magnetic field observed in our STM/STS data, by arguing that symmetries prevent the field from displacing dangling charges out of the obstructed positions.

\section{Determination of magnetic orderings via neutron powder diffraction} \label{sec: neutron}

\begin{figure}[htb!]
\centering
\includegraphics[width=1.0\linewidth]{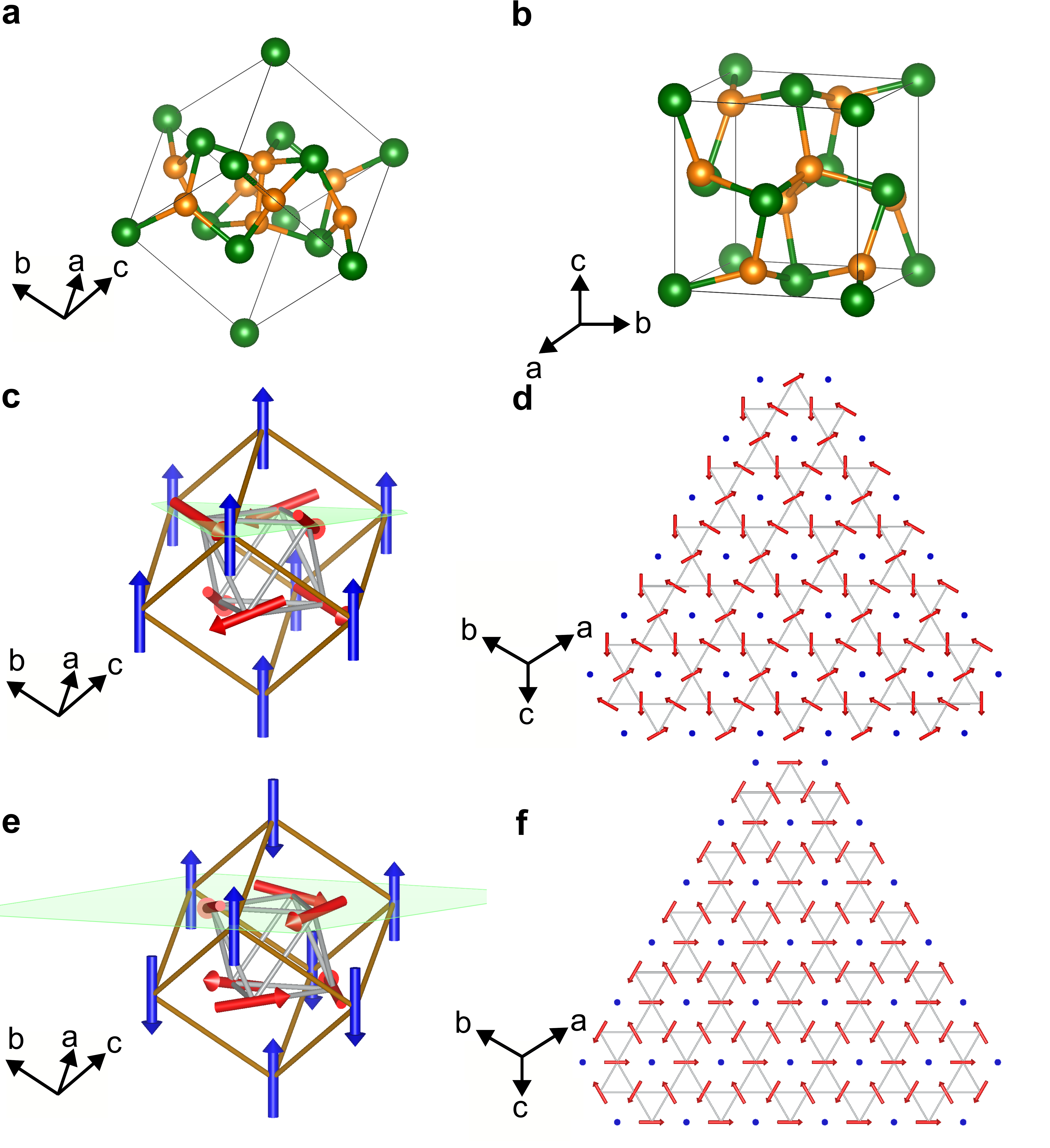}
\caption{
\textbf{Ionic and magnetic structures of the M1 and M2 phases.} a) and b) show two projections of the paramagnetic cubic cell (Ni atoms in green while the S atoms in orange). Panels c) and e) illustrate the arrangement of magnetic moments in the M1 and M2 phases showed in the paramagnetic cubic cell for simplicity. The magnetic structure of the M2 phase described in the rhombohedral cell is shown in Fig.~\ref{base_romb_cell}. The blue arrows represent moments located at the corner of the cubic cell, which correspond to the centers of hexagons when the crystal is viewed along the three-fold rotation axis, while red arrows depict moments sitting at the center of the cubic lattice faces, forming the Kagome pattern in the {111} planes (the green plane indicate the (111) plane). Panels d) and f) show views of the crystal in the (111) plane (green plane in panel c) and e)) along the three-fold rotation axis for the M1 and M2 phases, respectively.
}
\label{Exp-Figure-1}
\end{figure}

Before exploring the origin of the step-edge states, it is pivotal to determine and understand the lattice and magnetic symmetries of NiS$_2$ down to $1.5$ K. The arrangement of ions in NiS$_2$ forms a cubic structure, similar to the analog compound pyrite (FeS$_2$), which belongs to the crystallographic space group Pa$\bar{3}.1'$ (No. 205), as shown in Fig.~\ref{Exp-Figure-1}a,b. Ni atoms sit in the Wyckoff position (WP) 4a, creating a face-centered sublattice, whereas S atoms sit at the WP 8c, with $x=0.39364(10)$ at 100 K, forming S$_2$ dumbbells which in turn coordinate the Ni atoms forming NiS$_6$ octahedra. The system undergoes two phase transitions on cooling. 
At $T_{N1}\approx39$ K, it enters a non-collinear antiferromagnetic phase while, on further cooling below $T_{N2}\approx30$ K, it shifts into a M2 phase associated with a WFM moment~\cite{Thio1995} as observed in the magnetic susceptibility measurements  Fig.~\ref{Suscept}.

Despite several attempts to model the ground state of \nis{} ~\cite{Miyadai1975, Nishihara75, Kikuchi78, Yano16}, its spin arrangement and symmetry are still under debate (see Sec.~\ref{SI: hystory} in the SI for details). For this reason, we have re-investigated the nuclear and magnetic structure of NiS$_2$ using powder neutron diffraction at the time-of-flight diffractometer WISH (ISIS-UK)~\cite{wish}. The paramagnetic pattern collected at 100 K is well fitted with the $Pa \bar{3}1'$ space group (Fig.~\ref{Rietv}a,b)  and the well know pyrite structure (see Tab.~\ref{Tab_Para} for the structural parameters). On cooling below 39 K, we observe the rise of $\bs{k}_{\mathrm{M1}} = \bs{0}$ ($\Gamma$-point) magnetic reflections as shown in Fig.~\ref{Intensity_moment}, while at the second transition we see additional reflections with propagation vector $\bs{k}_{\mathrm{M2}} = (1/2, 1/2, 1/2)$ ($R$-point) as well as a sudden increase in intensity of the M1 reflections.

To solve the magnetic structures, we performed group theory calculation using ISODISTORT~\cite{Isodistort, isodistort2} and refined the diffraction pattern using Jana2006~\cite{Jana2006}. Below $T_{N1}$ we observe only the $\bs{k}_{\mathrm{M1}}$ propagation vector. 
The decomposition of the Ni site results in three irreducible representations that allow spin ordering. However, only $m\Gamma_{1}^{+}$ and $m\Gamma_{2}^{+}\Gamma_{3}^{+}$ are consistent with the absence of ferromagnetic moments in the M1 phase. 
The corresponding magnetic space groups (Fedorov groups) are $Pa \bar{3} $ and $Pbca$ respectively. The $Pa \bar{3} $ model gives a good fit of the data with moments constrained by symmetry to lie along the local $[111]$ direction (see Tab.~\ref{Tab_M1} for the structural parameters and Fig.~\ref{Exp-Figure-1}c,d). The magnetic space group of lower symmetry $Pbca$ allows the moments to deviate from the latter direction keeping the same relative orientation. Unfortunately, since the system is metrically cubic, it is not possible to distinguish between the two space groups from powder data alone. Nevertheless, the lack of orthorhombic distortion in the thermal expansion data~\cite{Nagata76} and synchrotron data~\cite{Thio1995} suggests the cubic magnetic space group as the best description of the system. The magnetic model agrees with previous reports and can be seen as the superposition in the [111] planes (Fig.~\ref{Exp-Figure-1}d) of a Kagome lattice with a 120 degree magnetic structure and a ferromagnetic triangular lattice of the Ni atoms at the center of the Kagome hexagons. The 120 degree structure is canted out of the plane, and this canting compensates exactly for the ferromagnetic pattern of the remaining Ni sites.

We now move to the solution of the magnetic ground state below $T_{N2}$. We assume that both M1 and M2 components form a multi-$k$ structure. This assumption is supported by the observation of an increase in the intensity of the M1 reflections at the second transition as well as the observation of $(1/2, 1/2, 1/2)$ nuclear superstructural peaks in synchrotron data~\cite{Feng2011} (see Sec.~\ref{SI: Symmetry M2} in the SI for details). We assume that the same $m\Gamma_1^+$ irrep of the M1 phase is still present in the ground state and we explore the isotropy subgroups derived by adding to this distortion another one with the propagation vector $\bs{k}_{\mathrm{M2}}$. In this case the decomposition of the Ni position with the latter propagation vector returns two irreps, $mR_1^+R_3^+$ and $mR_2^+R_2^+$. These two irreps and $m\Gamma_1^+$ constitute three possible isotropy subgroups. The best agreement with the data is achieved with the $R\bar{3}$ magnetic space group corresponding to $m\Gamma_1^+\oplus mR_1^+R_3^+$ (see Tab.~\ref{Tab_M2} for the structural parameters and cell transformation), as it can be seen in the Rietveld plot in Fig.~\ref{Rietv}e,f. The magnetic structure is shown in Fig.~\ref{Exp-Figure-1}e,f and in the rhombohedral cell in Fig.~\ref{base_romb_cell}. To obtain a constant moment solution, it is necessary to include a third $\Gamma$-point distortion that transforms as the $m\Gamma_4^+$ irrep. This mode is coupled to the previous two as detailed in Sec.~\ref{SI: Symmetry M2} in the SI, and does not change the magnetic symmetry. 

To understand the difference between the M1 and M2 phases it is useful to observe the structures projected along the [111] planes of the cubic paramagnetic cell (Fig.~\ref{Exp-Figure-1}d,f). In the M2 phase the Ni spins sitting on the Kagome lattice form again a 120-degree structure, with the spins close to being coplanar, whereas the Ni spin sitting in the triangular lattice point along the [111] parent cubic direction in a similar arrangement as the M1 phase. The difference arise when we look at the neighboring [111] layers. Indeed, in the M2 phase when moving to the next plane the Ni spins of the triangular lattice change sign, while the Ni spins on the Kagome layer change helicity. Contrary to the M1 phase, the magnetic moments of the Kagome and triangular sublattices do not cancel out, yielding a net magnetization along the out-of-plane direction in agreement with the macroscopic data which indicates the development of a WFM moment only below $T_{N2}$. The moment size at 1.5K is 1.205(7)$\mu_B$ and its evolution with temperature is shown in Fig.~\ref{Intensity_moment}.

\section{STM/STS experiments} \label{Sec: STM}

\begin{figure*}[htb!]
\centering
\includegraphics[width=17cm]{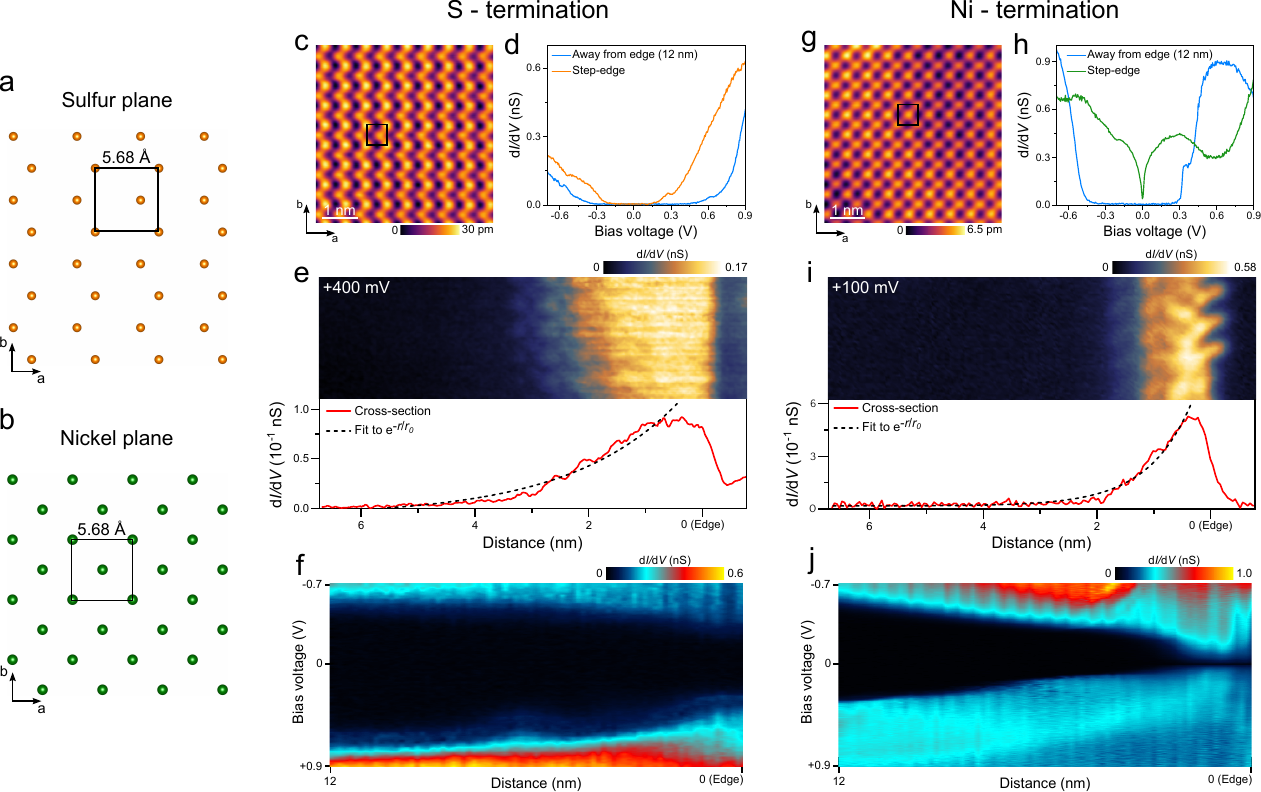}
\caption{
\textbf{Observation of step-edge state on the S and Ni surfaces termination.} a,b) Schematic representation of the cleaved \textit{ab} plane of the sulfur-terminated and nickel-terminated surfaces, respectively, with the black square indicating the surface unit cell. c) Atomically resolved STM image of the S-terminated surface. d) Representative d$I$/d$V$ spectra recorded at the step edge and 12 nm away from it on the S-terminated surface. e) Differential conductance map displaying the localized step-edge state on the S-termination (top) and the corresponding cross-sectional profile (bottom). The dashed black line represents a first-order exponential decay fit. f) Line spectroscopy illustrating the spatial evolution of the electronic structure as the step edge of the S-termination is approached. g) Typical atomically resolved STM image of the Ni-terminated surface. h) Differential conductance curves measured at the step edge and 12 nm away from it on the Ni-terminated surface. i) Spatially resolved d$I$/d$V$ map showing the localized step-edge state at the step edge of the Ni-termination (top) and the corresponding cross-sectional differential conductance profile (bottom). The dashed black line represents an exponential decay fit. j) Spatially resolved d$I$/d$V$ curves showing the population of the electronic gap by the step-edge as the step edge of the Ni-termination is approached. Acquisition parameters: c) \textit{V$_{\mathrm{set}}$} = 1 V, \textit{I$_{\mathrm{set}}$} = 0.3 nA. d-f) \textit{V$_{\mathrm{mod}}$} = 6 mV. g) \textit{V$_{\mathrm{set}}$} = 1.1 V, \textit{I$_{\mathrm{set}}$} = 0.2 nA. h-j) \textit{V$_{\mathrm{mod}}$} = 5 mV.
}
\label{Exp-Figure 3}
\end{figure*}

Although the existence of a Ni-termination requires breaking the strong S dimers~\cite{Le2024}, our high-resolution STM images reveal two inequivalent surfaces with atomic structure coincident with the  Ni and S atoms planes within the unit cell perpendicular to the [001] direction (Figs.~\ref{Exp-Figure 3}a,b and Sec.~\ref{SI-1} in the SI).
%The arrangement of ions in \nis{} described in Sec.~\ref{sec: neutron} results in inequivalent planes for S and Ni atoms along the \textit{c}-axis.
%Furthermore, the in-plane atomic arrangements of the S-terminated and Ni-terminated surfaces differ also from one another. Figures~\ref{Exp-Figure 2}a and~\ref{Exp-Figure 2}b schematically illustrate this, with each surface unit cell marked by a black square. The key difference lies in the central atom within the unit cell, which is shifted to one side in the S-terminated surface, creating a \textit{zigzag} bonding pattern among the S atoms. This shift disrupts the centers of inversion symmetry at S sites on the surface layer, unlike in the Ni termination. Although the Ni termination might seem unlikely because it requires breaking the strong S dimers~\cite{Le2024}, our high-resolution STM maps allow us to differentiate clearly between the two patterns depicted in Figs.~\ref{Exp-Figure 2}a,b, and to unequivocally identify them as Ni or S surfaces (see also Fig.~\ref{Exp-Figure S1} in the SI). Having access to both cleavage planes enables us to explore their potentially distinct electronic properties. Indeed, as we demonstrate below, both surface terminations display unique spectroscopic signatures.
We conducted a systematic experimental characterization of the structural and electronic properties of both surface terminations using STM/STS at 4.2 K. Below, we describe the findings for each surface termination individually, starting with the S atomic plane. 

Figure~\ref{Exp-Figure 3}c presents a typical atomically resolved STM image acquired on the S-termination, where the surface unit cell is outlined by the black square. The identification of this surface with the S-termination is based on the distinctive \textit{zigzag} atomic pattern observed, nearly matching the S atomic lattice depicted in Fig.~\ref{Exp-Figure 3}a (see also Fig.~\ref{Exp-Figure S1} in the SI). Our STM images reveal slightly different lattice constants along the \textit{a} and \textit{b} directions, with measured values of \textit{a} $\approx$ 5.70~{\AA} and \textit{b} $\approx$ 5.55~{\AA}. This indicates that the surface structure deviates from perfect square symmetry, instead exhibiting a tendency towards rectangular symmetry. This slight mismatch stems from a surface atomic relaxation minimizing the surface's free energy (see Sec.~\ref{S-12} in the SI), which has been predicted for the sister compound FeS$_2$~\cite{Hung2002}.

We experimentally determined the local electronic structure of the atomic terminations in the bulk and along the edges (Fig.~\ref{Exp-Figure S2} in the SI). Figure~\ref{Exp-Figure 3}d displays representative differential conductance (d$I$/d$V$) spectra recorded at the step edge and 12 nm (perpendicularly) away from it on the S-terminated surface (see also Fig.~\ref{Exp-Figure S3}a in the SI). At the bulk region, we observe a significant electronic gap of $\approx$ 0.8 eV. In contrast, the d$I$/d$V$ spectrum at the step edge reveals a reduced electronic gap of $\approx$ 0.4 eV, compatible with the presence of metallic states within the surface electronic gap at the step edge. This observation suggests the existence of a state localized at the step edge.

Spatially resolved d$I$/d$V$ maps further show that this step-edge state exhibits a robust, long-range spatial extension parallel to the step edge direction (see Fig.~\ref{Exp-Figure S4}a in the SI), while remaining confined in the direction perpendicular to the step edges (see top panel of Fig.~\ref{Exp-Figure 3}e). By fitting the cross-sectional d$I$/d$V$ profile (bottom panel of Fig.~\ref{Exp-Figure 3}e) to a first-order exponential decay equation, $y=y_{0}+Ae^{-r/r_{0}}$, we estimate a characteristic localization length, $r_{0}$, of $\approx$ 2 nm. Additionally, we observed that the bulk electronic gap shrinks gradually as the step edge is approached, as shown in color-coded d$I$/d$V(x,V)$ map, in Fig.~\ref{Exp-Figure 3}f.

%@Hao, do we observe the same behavior for other energies different from +400mV? if so, we should mention it

We now focus on the Ni-terminated surface, which exhibits a square atomic arrangement, as seen in the atomically resolved STM image in Fig.~\ref{Exp-Figure 3}g. This atomic configuration, distinguishable from that of the S-termination, aligns well with the schematic model of the Ni plane perpendicular to the cubic [001] direction depicted in Fig.~\ref{Exp-Figure 3}b (see also Fig.~\ref{Exp-Figure S1} in the SI). The lattice constants on the Ni-terminated surface also reveal a slight asymmetry, with  $a \approx$ 5.73~{\AA} and $b \approx$  5.55~{\AA}, indicating a deviation from a squared symmetry. This deviation results from surface atomic relaxation (see Sec.~\ref{S-12} in the SI).

Tunneling spectroscopy measurements were also conducted on the Ni-terminated surface, similarly at both the step edges and 12 nm away from it. Representative d$I$/d$V$ spectra are presented in Fig.~\ref{Exp-Figure 3}h (see also Fig.~\ref{Exp-Figure S3}b in the SI). They reveal an electronic gap of $\approx$ 0.7 eV in the bulk region, a value similar to that observed on the S-termination. However, in contrast to the S-termination, the d$I$/d$V$ curve measured at the step edge on the Ni-terminated surface disclose a gapless electronic structure with, however, a zero density of states (DOS) at the Fermi level at 4.2 K. Higher resolution d$I$/d$V$ curves acquired within of ± 90 mV confirm this result (Fig.~\ref{Exp-Figure S5}). Now, the electronic spectrum at the step edges develops a V-shaped dip around the Fermi level. We attribute the depletion of the DOS at E$_F$ to a STM tunneling-junction-related effect, as similar dip features have been observed on localized 1D conductive channels ~\cite{Pauly2015, Reis2017, Tang2017, Song2018, Ugeda2018} caused either by the Efros-Shklovskii mechanism ~\cite{Efros1975} or due to a 1D Luttinger-liquid behavior ~\cite{Voit1995}. Overall, our observation is compatible with the presence of a metallic state localized at the edges in the Ni-terminated surface. We argue that the network of 1D conductive channels from the Ni-plane step edges are likely the origin of the surface conductivity previously reported in NiS$_2$.

We conducted spatially resolved spectroscopy imaging at the edges of the Ni-planes. The top panel of Fig.~\ref{Exp-Figure 3}i shows a d$I$/d$V$ map acquired in a step edge, revealing that, as in the S-termination, the edge state is predominantly confined along the step edges, and propagates parallel to them (see Fig.~\ref{Exp-Figure S4}b in the SI). Fitting the cross-sectional d$I$/d$V$ profile to a first-order exponential decay (bottom panel of Fig.~\ref{Exp-Figure 3}i) yields a localization length $r_{0}$ $\approx$ 1 nm for the edge state on the Ni-terminated surface. Moreover, the shrink of the bulk electronic gap as the step edge is approached is roughly linear, as shown in Fig.~\ref{Exp-Figure 3}j. This gradual filling is further confirmed by d$I$/d$V$ maps recorded at other various energies: at high tunneling bias voltages, the local density of states (LDOS) is distributed widely across the surface, while at energies near the Fermi level, it remains localized at the step edges (Fig.~\ref{Exp-Figure S6} in the SI).

\begin{figure}[tb!]
\centering
\includegraphics[width=8.5cm]{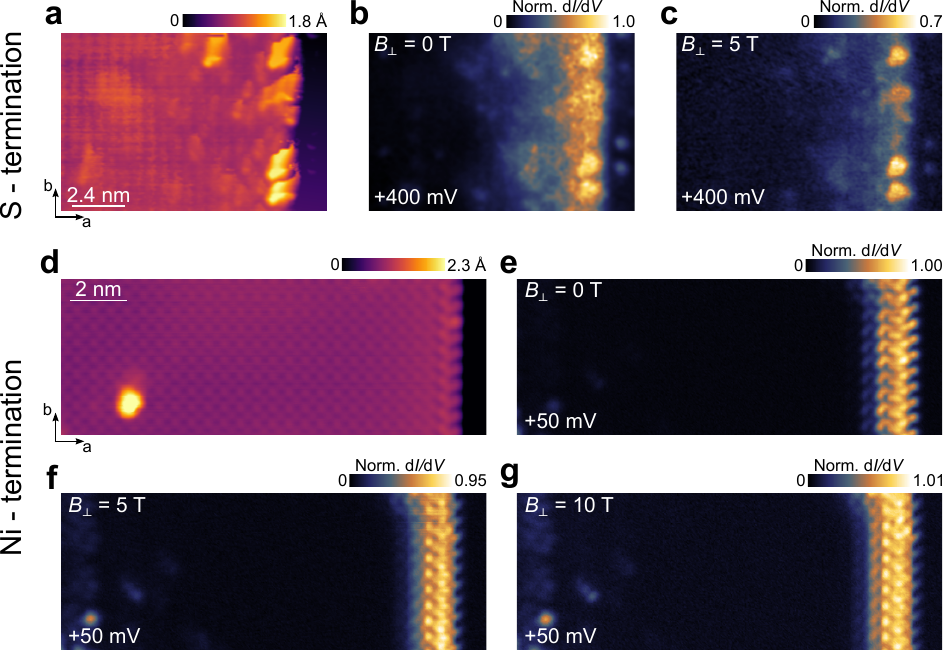}
\caption{
\textbf{Behavior of step-edge state against out-of-plane magnetic field.} a) STM image of the S-terminated surface near a step edge. b,c) Differential conductance maps measured at magnetic field strengths of 0 T and 5 T, respectively, within the same region shown in (a). The d$I$/d$V$ signals are normalized against the zero-field conductance map. d) Atomically resolved STM image showing a step edge of the Ni-terminated surfaces. e) Spatially resolved d$I$/d$V$ map of the edge state recorded at 0 T in the same region as the topography shown in (d). f,g) Same as in (e) but under out-of-plane magnetic fields of 5 T and 10 T, respectively. The d$I$/d$V$ signals are normalized to that measured at 0 T. Acquisition parameters: a) $V_{\mathrm{set}} = 1.1$ V, 
$I_{\mathrm{set}} = 0.15$ nA. b,c) 
$V_{\mathrm{mod}} = 6$ mV. d) 
$V_{\mathrm{set}} = 1.1$ V, 
$I_{\mathrm{set}} = 70$ pA. e-g) 
$V_{\mathrm{mod}} = 6$ mV.
}
\label{Exp-Figure 4}
\end{figure}

To gain insights into the nature and of the step-edge state, we conducted additional STS measurements under varying out-of-plane magnetic fields. Figure~\ref{Exp-Figure 4}a shows a STM topography image of a step edge region on the S-termination. On this area, we mapped the localization and intensity of the step-edge state at 0 T and 5 T through d$I$/d$V$ maps acquired at the same bias voltage, as shown in Fig.~\ref{Exp-Figure 4}b,c. Here, we normalized the d$I$/d$V$ scale ranges with respect the conductance map recorded at 0 T, allowing a straightforward qualitative comparison of the intensity. A direct comparison between both maps reveals that the presence of a magnetic field weakens the intensity of step-edge state, thus affecting its spatial extension, although it is not fully suppressed. This indicates that the edge states of the S termination remain robust against an external magnetic field applied perpendicular to the surface.

%Figure 4 is not referenced. FIX

%presence of an out-of-plane magnetic field affects the step-edge state of the S-termination mildly. 

We turn our attention now into the magnetic field dependence data obtained on the Ni atomic plane. Figure~\ref{Exp-Figure 4}d shows the topography of a Ni-terminated step edge. Within this same area, we characterized the intensity of the edge state at zero field and under the presence of an out-of-plane magnetic field of 5 T and 10 T. Similarly, to allow a direct qualitative comparison, we normalized each conductance scale range against the d$I$/d$V$ map recorded at 0 T. In contrast to the S-termination, neither the intensity nor the spatial location of the edge state are unaffected by the magnetic field. 
We have also mapped the LDOS at different bias voltage under the presence of magnetic fields (Fig.~\ref{Exp-Figure S7} in the SI). At all measured bias voltage, we observe a relative high resilience towards magnetic field perturbation, specially at energies close to the Fermi level.

%The following argument removed from the previous paragraph is not solid: This contrast can be explained by assuming that the magnetic field shifts the edge states to higher energies. Such a shift would weaken the intensity of edge states at certain values of the bias potential in the S-termination, while having a less pronounced effect on the Ni-termination, where the in-gap states are more evenly distributed throughout the gap.

\section{Theoretical analysis of the topology and 1D step-edge states} \label{sec: theory}

In this section, we provide a theoretical interpretation of the edge states observed in STM measurements.
Since this data has been taken on surfaces perpendicular to the crystallographic directions of the cubic lattice, we focus our theoretical analysis on the M1 phase first. We extend the conclusions of this analysis to the M2 phase in Sec.~\ref{SI: low-T phase}, proving that the low-temperature phase also hosts similar dangling surface states.

\begin{figure}[htb!]
\centering
\includegraphics[width=1.0\linewidth]{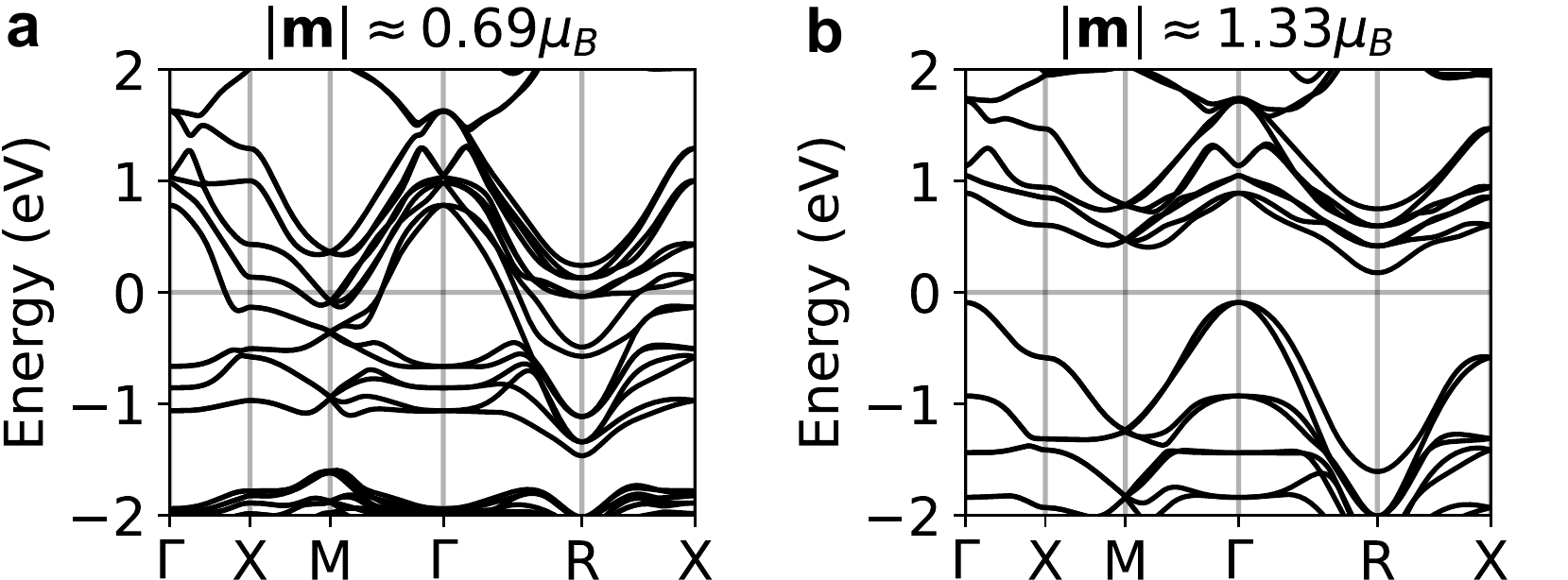}
\caption{\textbf{Band structure of \nis{} in the M1 phase, computed with different values for the parameter CMBJ within the mBJ approximation.} a) CMBJ=0.9, with the Ni moments matching the values observed experimentally (see Tab.~\ref{Tab_M1}). b) Insulating phase with Ni moments larger than the experimental $(\abs{\bs{m}}=1.33 \mu_B)$, achieved with CMBJ=1.15.}
\label{Theo-Figure-1}
\end{figure}

We have performed first-principle calculations in the antiferromagnetic phase via the Vienna Ab Initio Simulation Package (VASP) (see Sec.~\ref{Sec: Methods} for details). Since the DFT+U approach fails to reproduce an insulating band structure for the experimental magnetic moments, we apply the modified Becke-Johnson (mBJ) approximation for the exchange-correlation functional~\cite{Becke2006, Tran2009}. Ground states displaying magnetic moments of identical symmetry and different magnitude were achieved by tuning the parameter that controls the degree of mixing in the exchange potential within the modified Becke-Johnson (mBJ) approximation. Figure~\ref{Theo-Figure-1}c shows the band structure evaluated with magnetic moments consistent with experimental data (Sec.~\ref{sec: neutron}). The obtained spectrum corresponds to a metallic phase, in good agreement with the study performed within the generalized gradient approximation with Hubbard corrections (GGA+U) in Ref.~\cite{Reiss2022}. Upon increasing the magnitude of magnetic moments in Ni ions, bands around the Fermi level pull apart, and the system transitions to an insulating phase. The band structure of the insulator with magnetic moments $\abs{\bs{m}_{\mathrm{Ni}}} \approx 1.33 \mu_B$ is shown in Fig.~\ref{Theo-Figure-1}b. The irreducible representations at maximal points of the Brillouin zone (BZ) were computed with the software Mvasp2trace~\cite{Xu2020, Elcoro2021} (see Tab.~\ref{Theo-Table-S1} in SI). We verified that the set of irreducible representations of valence bands does not change after the transition, thus the valence bands topology is not affected by the increase of magnetic moments. 

\subsection{Topological classification of the insulating M1 phase}

The classification of the valence bands is based on the irreducible representations of their wave functions at maximal $\bs{k}$ points in the BZ, following the TQC approach.
Let $\rho_{\mathrm{val}}$ be the representation of the space group as which states in valence bands transform. To determine if a material is topological, we have to decompose $\rho_{\mathrm{val}}$ as a linear combination of the space group's elementary band representations (EBRs):
\begin{equation}\label{eq: tqc}
    \rho_{\mathrm{val}} = \bigoplus_{\alpha,i} C_{\alpha,i} (\rho_i @ \alpha),
\end{equation}
where $(\rho_i @ \alpha)$ denotes the EBR induced from the irreducible representation $\rho_i$ of the site-symmetry group of sites in maximal WP $\alpha$. If this equation does not have any solution involving only non-negative integer coefficients $C_{i,\alpha}$, the material is topological within the formalism of TQC~\cite{Bradlyn2017, Cano2018, Bradlyn2018}. Equation~\eqref{eq: tqc} has multiple solutions for \nis{}, due to the fact that the EBRs of the magnetic space group $Pa\bar 3$ form an overcomplete basis of atomic limits. It turns out that all solutions have in common two relevant properties: (i) all coefficients $C_{\alpha,i}$ are non-negative integer numbers, thus $\rho_{\mathrm{val}}$ matches an atomic limit within the scope of TQC, and (ii) the possible decompositions only involve EBRs induced from WPs $4a$ and $4b$. Combining (i) and (ii), Eq.~\eqref{eq: tqc}  can be rewritten for the valence bands of \nis{} as follows
\begin{equation} \label{eq: tqc for NiS2}
\rho_{\mathrm{val}} = \bigoplus_{i} C_{i,4a} (\rho_i @ 4a) \bigoplus_{i} C_{i,4b} (\rho_i @ 4b),
\end{equation}
where all the $C_{i, \alpha}$'s are non-negative integers.
From Eq.~\eqref{eq: tqc for NiS2}, we deduced that all possible decompositions of $\rho_{\mathrm{val}}$ are compatible with atomic limits induced from Wannier functions at WP $4a$ (occupied by Ni ions) and $4b$. The WP $4b$ coincide with the centers of the S dimers, and do not host any ion. Hence, the atomic limit of the valence bands involves an electronic charge located at empty spots of the crystal, i.\,e., the valence bands host an OAI.

To corroborate this characterization of \nis{} as an OAI, we verified that there is no combination of band representations induced from Ni and S sites -- located at WP $4a$ and $8c$, respectively -- compatible with $\rho_{\mathrm{val}}$. This analysis can be found in Sec.~\ref{SI-10} of the SI.

\subsection{Absence of filling anomaly on Ni and S terminations}
In this section, we show that the OAI phase is compatible with the absence of metallic states on the Ni and S surfaces explored via STS in Sec.~\ref{Sec: STM}.

When an OAI is cleaved along a plane that cuts through the OWCCs, it might be impossible to have an average charge per cell compatible with an insulator while simultaneously preserving the bulk's translation symmetry. This phenomenon is dubbed \textit{filling anomaly}, and manifests as dangling charges on the cleavage plane, which translate to metallic states in the slab's electronic structure~\cite{Song2017, Rhim2017, Benalcazar2019, Schindler2019}. A surface consistent with a filling anomaly will show half-filled states if the rest of the atoms are away from the cleavage plane, and the surface cuts through all the sites of the obstructed WP~\cite{Xu2021,Lu2024}.

\begin{figure*}
    \centering
    \includegraphics[width=0.85\linewidth]{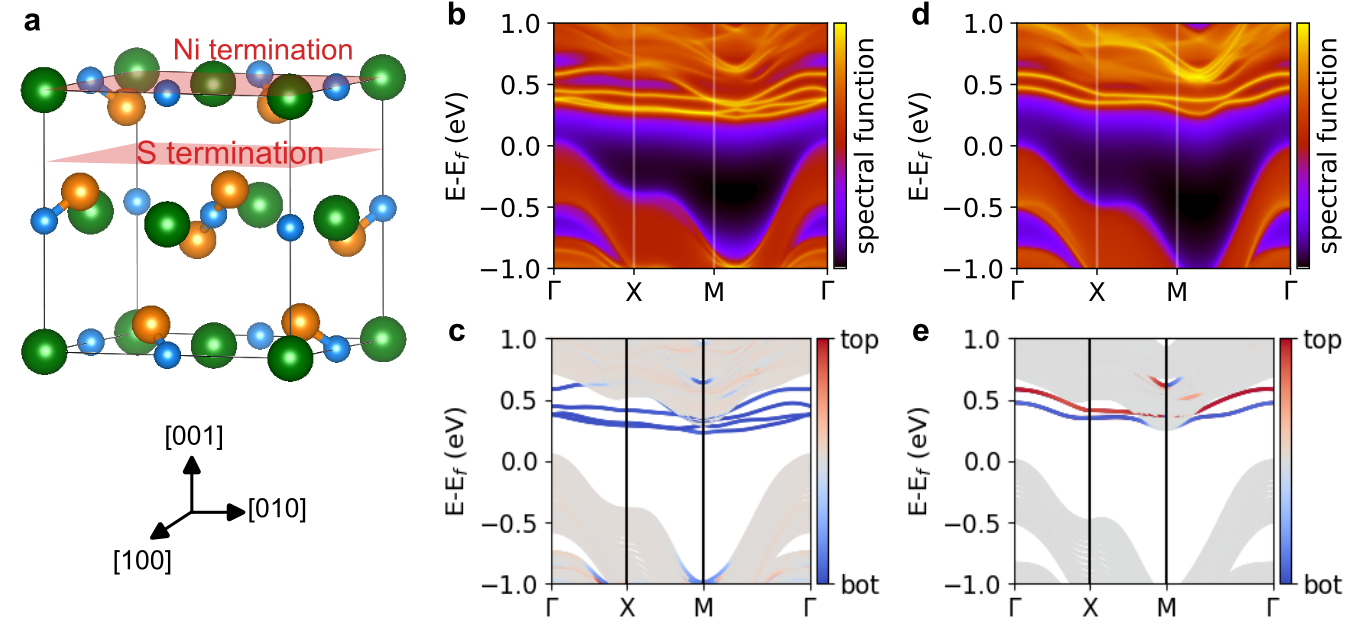}
    \caption{a) Cleavage planes perpendicular to the [001] direction considered to simulate the surface states. Green, orange and blue spheres denote Ni, S and OWCC states, respectively. b) Spectral function of a semi-infinite slab with a Ni termination. c) Band structure of the slab with a termination of Ni atoms. d) Spectral function of a semi-infinite slab with an S termination. e) Band structure of the slab with a termination of S atoms. The Fermi level is set to 0 in all of the subfigures.}
    \label{Theo-Figure-3}
\end{figure*}

Figure~\ref{Theo-Figure-3}a displays the cleavage planes yielding the Ni and S surfaces characterized by STS in section~\ref{Sec: STM}.
Both planes are oriented perpendicular to a principal axis of the cubic ionic lattice, specifically aligned along the [001] direction.
Figure~\ref{Theo-Figure-3}b shows the simulation of the spectral density for a semi-infinite slab with a Ni surface. Although this cleavage plane cuts through the OWCC, it does not exhibit surface metallic states related to a filling anomaly. This is due to the presence of Ni atoms on the surface, which are responsible for creating a crystal field that pushes the surface states to the bottom of the conduction band. The absence of metallic surface states is consistent with the gap observed in the STS experiments.

The spectral density of a semi-infinite slab with a termination exposing S atoms is shown in Fig.~\ref{Theo-Figure-3}d. This is most likely the S-termination observed in the STM experiments, as the alternative S-termination would consist of a freestanding monolayer of unpaired S atoms, which is energetically unfavorable and inconsistent with the observed stability in our measurements. Although the surface exhibits in-gap states, the counting of occupied bands in the slab's band structure (Fig.~\ref{Theo-Figure-3}e) confirms that these states are located above the Fermi level. The surface electronic structure is hence gapped, which is in good agreement with the STS data. This is consistent with the fact that the cleavage plane does not cut through any OWCC, so that no filling anomaly is expected for this case.

\subsection{Numerical simulation of step-edge states}

To study if the OAI state reproduces the step-edge states observed in our STS experiments, we computed a Wannier tight-binding model out of \textit{ab initio} wave functions via the Wannier90 package~\cite{Pizzi2020}. The $d$ orbitals of Ni atoms and the $p$ orbitals of S atoms were used as initial projections for the Wannier functions (see Sec.~\ref{Sec: Methods}). The Wannier model was computed for the band structure in Fig.~\ref{Theo-Figure-1}d, motivated by the fact that (i) it shows a gap consistent with the insulating behavior reported by previous works -- which also facilitates the identification of edge states within the gap -- and (ii) its valence bands have the same irreducible representations at maximal \bk{} points as the valence bands obtained with values of the magnetic moments compatible with the experiment. 

\begin{figure*}
    \centering
    \includegraphics[width=1.0\linewidth]{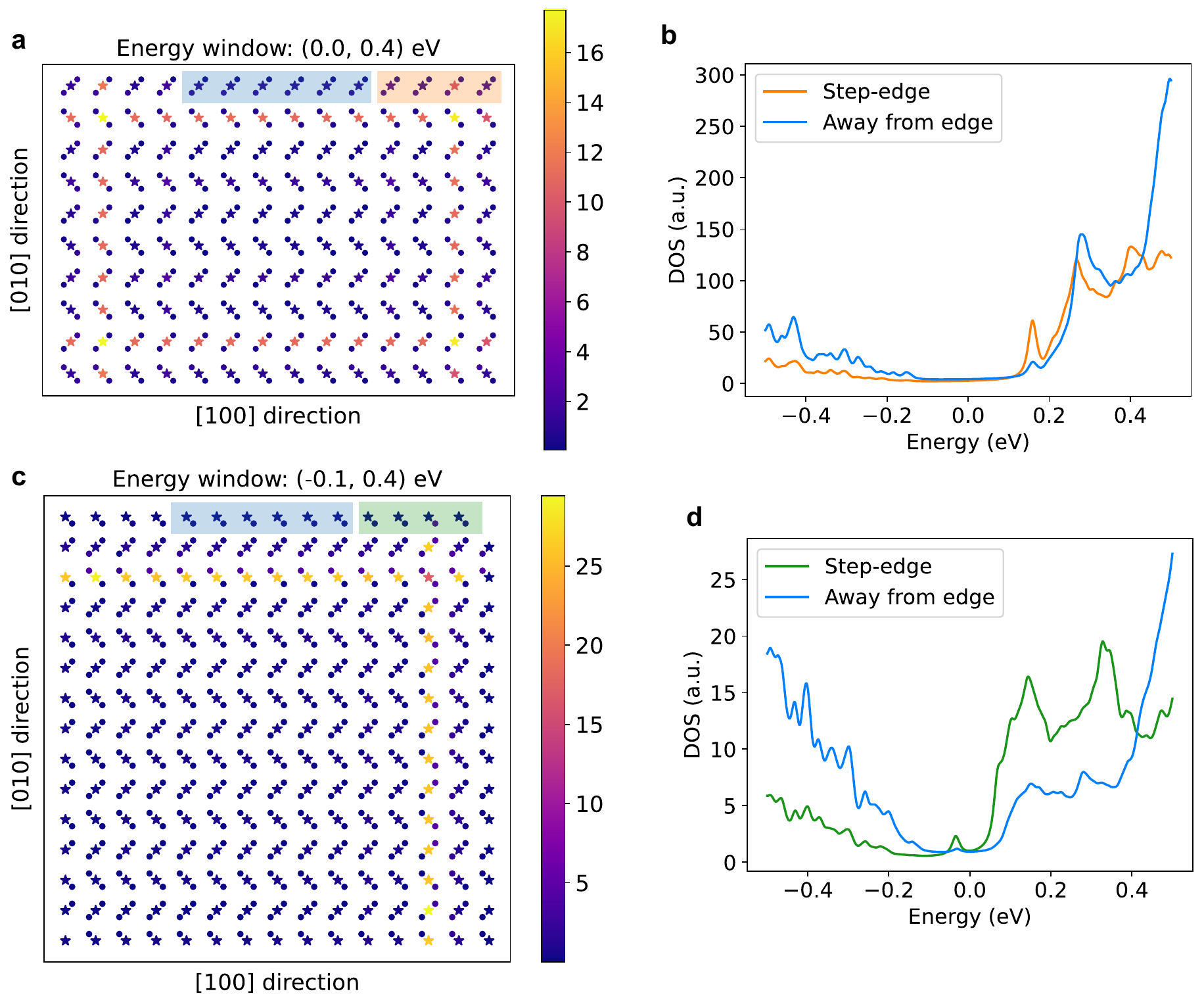}
    \caption{\textbf{Numerical simulations of step edges based on rods periodic along the [001] direction.} a) LDOS integrated over energies ranging between 0 and 0.4 eV in a rod terminated with S surfaces. The Ni and S positions are marked by stars and circles, respectively. The orange and blue shaded rectangles indicate regions near the step edge and away from it, respectively. b) LDOS of the rod with S terminations at the step edge and away from the edge, obtained by integrating the contributions of the atoms highlighted in the rectangles of panel a). c) LDOS integrated over energies ranging between $-0.1$ and 0.4 eV in a rod terminated with Ni surfaces, with Ni and S positions indicated by stars and circles, respectively. The green and blue rectangles denote regions near the step edge and away from it, respectively. d) LDOS of the rod with Ni terminations, analogous to panel b). The energy origin is set to the Fermi level of the 3D system for all plots, and the linear scale is employed to represent the LDOS.}
    \label{Theo-Figure-rods-linearDOS}
\end{figure*}

\begin{figure}[htb!]
    \centering
    \includegraphics[width=8.5cm]{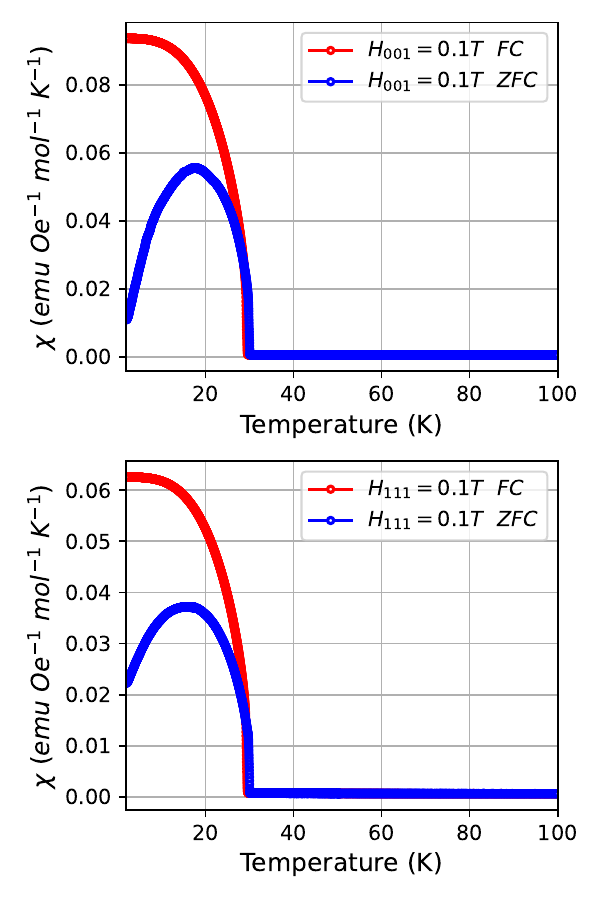}
    \caption{Magnetic susceptibility measurements performed on a NiS$_2$ single crystal with field applied along the [001] direction (top) and [111] direction (bottom). The applied magnetic field strength was 0.1T and the measurements were performed following both a field cooled (FC, red points) and zero field cooled (ZFC, blue points) protocol.}
    \label{Suscept}
\end{figure}

\begin{figure*}
    \centering
    \includegraphics[width=15cm]{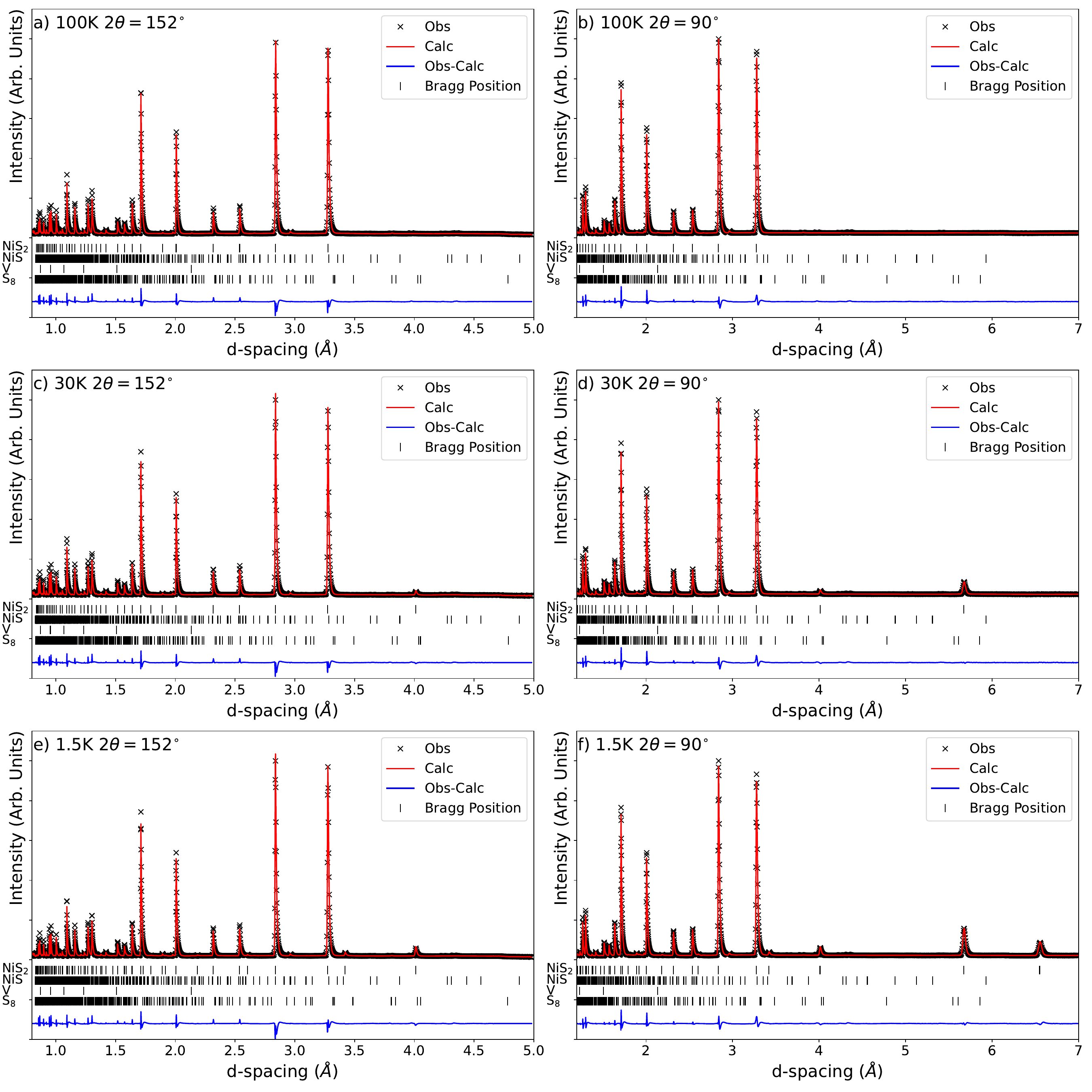}
    \caption{Rietveld plots of the neutron diffraction data collected at the WISH diffractometer at different temperatures and in two detectors banks with average $2\theta = 152^{\circ}$ and $90^{\circ}$. The x symbols represent the observed data whereas the red and blue lines indicate the calculated and difference pattern respectively. The tick marks represent the position of the Bragg reflection for the different phases present in the sample. The $NiS$ and $S_8$ impurities are $\approx 0.8\%$ and $1.13\%$ in volume respectively. The Vanadium peaks are due to the sample container. The reliability factors for each data set are a)$R_P=5.25\%$ and $R_{wP}=6.79\%$ b) $R_P=4.17\%$ and $R_{wP}=5.74\%$ c) $R_P=5.44\%$ and $R_{wP}=6.87\%$ d) $R_P=4.38\%$ and $R_{wP}=5.82\%$ e)$R_P=5.51\%$ and $R_{wP}=6.96\%$ f)$R_P=4.53\%$ and $R_{wP}=5.87\%$.}
    \label{Rietv}
\end{figure*}

\begin{figure}
    \centering
    \includegraphics[width=8.5cm]{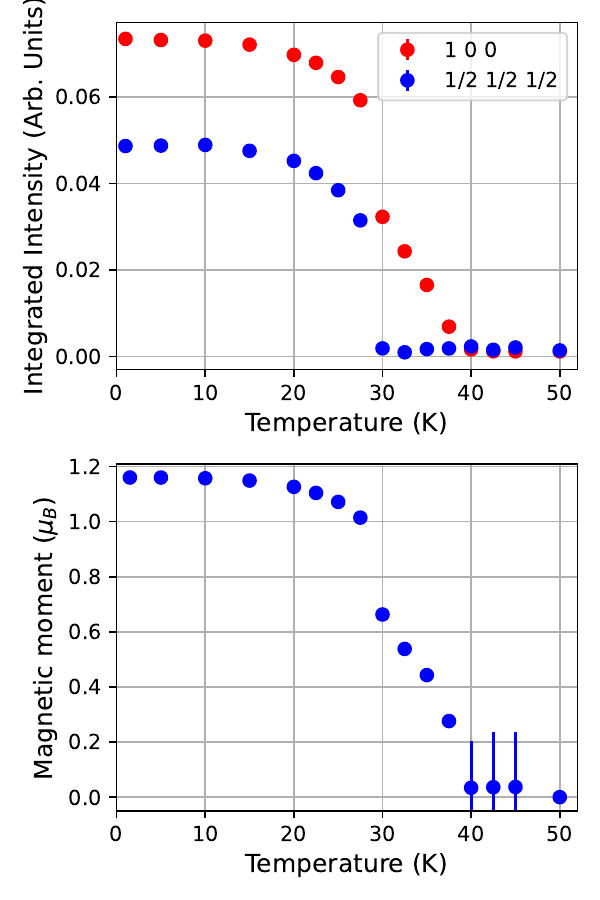}
    \caption{Top panel) Integrated intensities of the 1 0 0 and 1/2 1/2 1/2 Bragg peaks as function of temperature showing the two magnetic transition at $T_{N1}$ and $T_{N2}$. Bottom panel) Ni magnetic moment obtained from the fitting of the neutron data as function of temperature.}
    \label{Intensity_moment}
\end{figure}

\begin{figure}
    \centering
    \includegraphics[width=8.5cm]{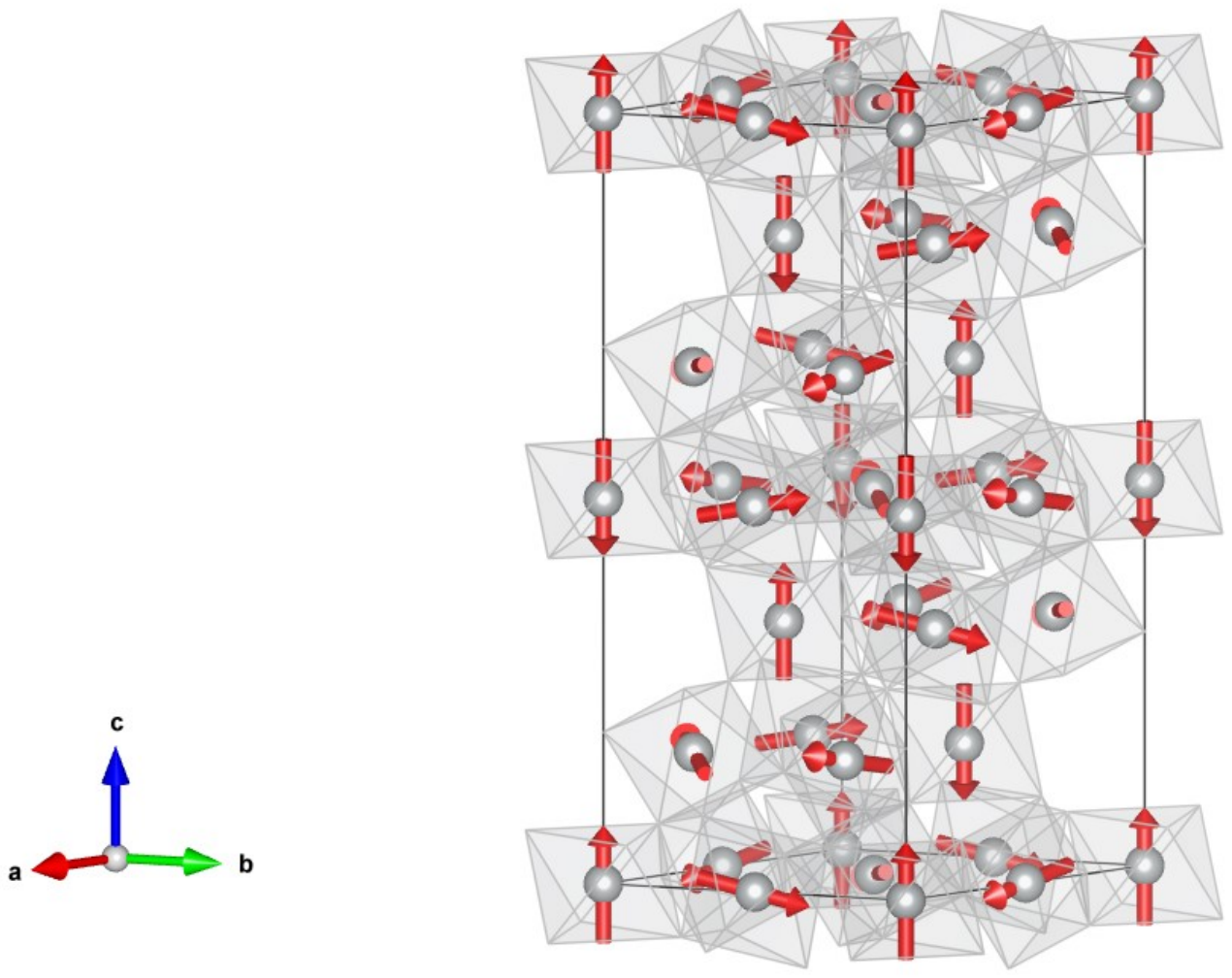}
    \caption{Drawing of the magnetic ground state of NiS$_2$ obtained from the refinement of the powder neutron diffraction data at 1.5K, for clarity only the Ni atoms are shown and the NiS$_6$ octahedra are drawn to help visualise the structure. The magnetic structure is described in the $R\bar{3}$ magnetic space group ($m\Gamma_1^+ \oplus m\Gamma_4^+ \oplus mR_1^+R_3^+$ irrep) defined in a unit cell obtained using the following transformation matrix $\{(-1,1,0),(1,0,-1),(-2,-2,-2)\}$, origin (0,0,0), on the parent cubic cell. The crystallographic data are reported in Tab.~\ref{Tab_M2}}
    \label{base_romb_cell}
\end{figure}

The Wannier model is then used to construct rods representing the step-edges observed in STM maps for both terminations. The rods were designed to be finite along the crystallographic [100] and [010] directions, and periodic along [001]. Consistent with the penetration lengths estimated in STM, we consider regions penetrating up to approximately $2$ nm into the surface as step edges. Figures~\ref{Theo-Figure-rods-linearDOS}a,b show the local density of states (LDOS) calculated for the rod with S terminations. Given that a surface exposing unpaired S atoms would require breaking the strong S dimers, we focus on a termination consisting of unbroken dimers. Figures~\ref{Theo-Figure-rods-linearDOS}b exhibits a dominant contribution coming from the step edge to the states between 0 and 0.2 eV, and a significant contribution between 0.2 and 0.4 eV. This result shows good qualitative agreement with the LDOS in Fig.~\ref{Exp-Figure 3}d, and suggests that the STM tip is sensitive principally to the signal coming from the outermost layer formed by S dimers and Ni atoms.
Figures~\ref{Theo-Figure-rods-linearDOS}c,d display the LDOS of a rod terminated with Ni atom layers. This LDOS shows a dominant contribution from the step edge to the states between 0 and 0.4 eV. In contrast to the S-terminated rod spectrum, the contribution of the step edge also prevails right below zero energy. This result is qualitatively consistent with the observation that the gap on the Ni surface is populated by step-edge states at negative and positive values of the bias potential (see Fig.~\ref{Exp-Figure 3}h), and it might be explained by a restoration of the filling anomaly due to the change in the coordination of S atoms at the edge.

\subsection{Robustness of OWCCs against magnetic field} \label{SI: robustness magnetic field}

The STM experiments described in Sec.~\ref{Sec: STM} demonstrate that step-edge states show resilience upon the application of an out-of-plane magnetic field. In this section, we focus on analyzing this robustness from a theoretical perspective.

Let us consider the crystal in the M1 phase to be subjected to a uniform magnetic field along the [001] direction. This external magnetic field breaks all crystalline symmetries except the identity, inversion, the two-fold screw rotations along the direction of the field, and the glide reflection with respect to the perpendicular plane. Therefore, the original space group is lowered to the monoclinic magnetic space group $P2_1/c$ (No. 14.75). This group can be decomposed in terms of cosets as follows
\begin{equation}
\begin{split}
    P2_1/c = T &\cup \{I | \boldsymbol{0} \} T
    \cup \{2_{001} | 1/2,0,1/2 \} T \\ &\cup \{m_{001} | 1/2,0,1/2 \} T,
\end{split}
\end{equation}
where $T$ denotes the translation subgroup and we used the Seitz notation to denote symmetry operations. Note that here the monoclinic axis is parallel to the [001] direction. The transformation between the cell vectors $\boldsymbol{a}_1$, $\boldsymbol{a}_2$ and $\boldsymbol{a}_3$ chosen here and the vectors $\boldsymbol{a}'_1$, $\boldsymbol{a}'_2$ and $\boldsymbol{a}'_3$ of the conventional \textit{b}-setting is given by:

\begin{equation}
(\boldsymbol{a}'_1 \ \boldsymbol{a}'_2 \ \boldsymbol{a}'_3) =
(\boldsymbol{a}_1 \ \boldsymbol{a}_2 \ \boldsymbol{a}_3)
\begin{pmatrix}
    0 & 1 & 0 \\
    0 & 0 & 1 \\
    1 & 0 & 0 
\end{pmatrix}.
\end{equation}

Under this reduction of the symmetry, the OWCCs in WP 4b of the M1 phase split into two independent pairs of WPs of the space group $P2_1/c$. The sites with direct coordinates $(0,0,0)$ and $(1/2,0,1/2)$ belong to WP 2a, while $(1/2,1/2,0)$ and $(0,1/2,1/2)$ sit in WP 2d. Therefore, the OWCCs remain pinned to the same sites upon lowering the symmetry due to the presence of the magnetic field. In fact, the WPs 2a and 2d are not connected to any other WP, thus the OWCC localized at these WPs cannot be moved away without breaking a symmetry or closing the bulk gap. Nevertheless, if the chemical potential is not pinned to the in-gap surface states by a filling anomaly, a strong magnetic field could displace the dangling states out of the gap, which would lead to a decrease of the signal at the edge. This robustness of the OWCCs is consistent with the resilience observed in STS experiments.

\section{Discussion} \label{sec: discussion}

In this work, we have proposed a new model for the M2 phase based on high-resolution neutron data, and a rigorous group theory analysis of the magnetic symmetry. Our model differs from the previously reported ones mainly because it involves three irreducible representations of the parent space group, which are needed to obtain a constant moment solution and a WFM moment matching the expected values. This model provides a detailed description of the magnetic structure in the WFM phase, shedding light on the long-standing debate regarding its magnetic moment arrangement.

Previous transport measurements, based on sample averaged data, have reported that the surface of \nis{} remains conductive at low temperatures, while the bulk of the material hosts a Mott-Hubbard insulating gap driven by strong electron-electron interactions~\cite{Sarma2003,El-khatib2021,Thio1994,El-khatib2023}. Contrary to the expectation, our STM/STS data provide clear evidence that the 2D surface of \nis{} also exhibits insulating behavior. Our data suggests that the observed discrepancy can be reconciled by attributing the conduction channels to be localized along the step-edges. This hypothesis is substantiated by our STS measurements, which provide atomic-scale confirmation of the channel localization. 
%Furthermore, based on a combined experimental and theoretical approach, we propose that the origin of metallic 1D states is the existence obstructed charges sitting at the centers of S dimers.

%This hypothesis is substantiated by our STS measurements, which provide atomic-scale confirmation of the channel localization.  Furthermore, the consistency between our STS measurements and numerical simulations of surface and step-edge spectra suggest a deep relation between the origin of the 1D conduction channels and the obstructed charges sitting at the center of S dimers in an OAI phase. This relation is further reinforced by the incapacity of an external magnetic field to displace the dangling charges from the obstructed positions, which is compatible with the robustness of step-edge states against the magnetic field observed in our STS analysis.

The observation of step-edges in this work is in full agreement with a recent STM/STS study, which similarly observed metallicity along the step edges of the [001] plane of \nis{}~\cite{Yasui2024}. Beyond this, our high-resolution STM maps enabled us to further (i) differentiate between S and Ni terminations, (ii) verify the existence of metallic states on both terminations' step edges, and (iii) analyze their spatial confinement on each of the cleavage planes. Additionally, we have gained insight into the nature of metallic step-edge states, reporting their resilience against magnetic fields up to 10 T.

To interpret the STM findings and elucidate the origin of edge states, we have performed first-principles calculations, and analyzed the results in terms of group theory.
The M1 phase of \nis{} with experimentally observed magnetic moments is a metal as for our DFT analysis based on the mBJ functional approximation. The irreducible representations of the valence bands of this metallic state at the maximal $\bs{k}$ points of the BZ correspond to an OAI phase. Indeed, our analysis reveals that the system transitions into an OAI insulator upon increasing the magnetic moments of the Ni atoms. A similar metal-to-insulator transition was observed in Ref.~\citep{Reiss2022} based on first-principle calculations performed within the GGA approximation, and the obstructed-atomic nature is compatible with the significant interstitial contributions reported in that reference. Although this transition cannot reproduce the evolution of the Fermi surface observed experimentally~\cite{Friedemann2016, Reiss2022}, it preserves the set of irreducible representations of the valence bands at maximal $\bs{k}$ points, and serves as a useful state for exploring the potential connection between edge states and the underlying OAI charges.

The Ni and S terminated surfaces exhibit gapped electronic spectra in our STS data and theoretical calculations. Additionally, our numerical simulations of step-edge states in the insulating OAI phase display good qualitative agreement with STS measurements taken on both terminations. 
%This consistency suggests a profound connection between OWCC and step-edge states, shedding light on the nature of the latter. 
Based on this consistency, we propose that the emergence of metallic edge states stems from the restoration of the filling anomaly at the step edges of these terminations.

Such a relation is consistent with the argument given by Yasui \textit{et al}. in Ref.~\cite{Yasui2024} to exclude the bandwidth-controlled Mott insulator-to-metal transition as the origin of 1D step edge states. As Yasui \textit{et al}. emphasize, such a transition would be feasible if the electron kinetic energy, quantified by the hopping term \textit{W}, would dominate over the on-site Coulomb repulsion \textit{U}. However, at the step edges, \textit{U} is expected to increase and \textit{W} to decrease due to reduced screening and diminished hopping, making a bandwidth-controlled Mott transition an unlikely explanation for the observed metallicity at the step edges.

%Determining the exact relation is complicated by the subtle chemistry of the system, and might require models that incorporate the OWCCs explicitly rather than S $p$ orbitals, which are beyond the numerical focus of this study's theoretical approach. Such models could reveal an interplay between the OWCC and Ni orbitals, and help to unveil the mechanism driving the transfer of charge to the later.
Determining the exact relation between OWCC and the step-edge states is complicated by the subtle chemistry and electronic structure of the system. The method developed -- and applied to NbSe$_2$ -- in Ref.~\cite{Calugaru2025} constitutes a promising approach to go beyond the relation established in our work. This method consists in first formulating a correlator that serves as an order parameter for the OAI phase, and then evaluating it by combining STM and \textit{ab initio} data. However, the application of this approach to NiS$_2$ might be more challenging than in NbSe$_2$. The complicated chemistry of NiS$_2$ hinders the derivation of a simple microscopic tight-binding model to assist the theoretical identification of an appropriate correlator. 
In fact, determining the exact number of OWCCs from the decomposition of the representation of valence bands in terms of EBRs is impossible, as multiple decompositions with different OWCC counts are equally valid.
In addition, the precise identification of the valence bands carrying the obstructed atomic character is difficult in NiS$_2$ due to its subtle electronic structure, which makes the evaluation of the correlator in terms of STM data challenging.

The role played by magnetic exchange interactions on the restoration of the filling anomaly at the step edges is another intriguing aspect that goes beyond the scope of this work, and hence requires further analysis. S atoms exhibit a different coordination number on each cleavage plane perpendicular to the cubic [001] direction. Furthermore, these coordination numbers differ from the value in the bulk. The difference in the number of bonds could  trigger the onset of net magnetic moments on S atoms, which could influence the filling of OWCC. Such magnetic moments would remain elusive to our STM measurements.

Electron-electron interactions have a significant impact on the electronic nature of the material. Indeed, they are responsible for the gap in the electronic structure~\cite{Fujimori1996, Matuura1998, Iwaya2004, Kunes2010, Schuster2012, Reiss2022}. On the other hand, OAIs have been theoretically shown to be robust against such interactions, meaning that obstruction properties and boundary modes can indeed survive the Mott transition~\cite{Jiang2023}. In light of this theoretical prediction, \nis{} might constitute, to the best of our knowledge, the first established material realization of an OAI surviving electron interactions. This resilience, as well as the agreement between our step-edge's simulations and STS measurements, suggest that the relation between OAI charges and edge modes is present despite strong electron interactions.

As \nis{} may embody the first realization of an OAI in a material with important electronic interactions, it represents an interesting testbed to further study the influence of interactions on obstructed charge centers. This is particularly true, given that \nis{} offers the possibility to control the metal-to-insulator transition in terms of several external factors, including pressure and chemical substitution. Such an analysis could indeed lead to the identification of new interacting topological phases, and push the boundaries of the classification of topology in interacting matter.

The resistance of step-edge modes against an out-of-plane magnetic field is consistent with the nature of the OWCCs. The lowering of the crystalline symmetries as a consequence of the magnetic field does not influence the OWCCs, so that they remain pinned to the same positions within the unit cell. This statement relies on a symmetry-based analysis, which is naturally unable to quantify the influence of the magnetic field on the energy levels of obstructed states, as these depend on the microscopic features of the sample. Consequently, we cannot rule out the possibility that strong magnetic fields, beyond the sensitivity of our probes, might shift the OWCC away from the low-energy region of the spectrum.
In fact, the displacement of surface state to higher energies can explain the observed behaviour of the edge states under a magnetic field.
Moreover, strong magnetic fields could potentially result in a closing of the bulk gap, driving the system into a different phase. 

Regarding the stability of the OAI nature in the transition from the M1 to the M2 phase, we show in Sec.~\ref{SI: low-T phase} that -- on a first-principle level -- the transition involves closing and reopening of the bulk gap. The fact that the termination perpendicular to the trigonal direction exhibits dangling surface states (see Fig.~\ref{Theo Fig S10}) suggests that the M2 phase is also an OAI with OWCC. This result is compatible with the observation of step-edge states below $T_{N2}$.

\section{Conclusions} \label{sec:conclusions}

In this study, we extend earlier analyses of NiS$_2$'s magnetic structure by introducing a novel model that clarifies both the intricate magnetic configuration and the properties of the low temperature M2 WFM phase. Our findings provide fresh insight into a debate spanning 50 years concerning the arrangement of magnetic moments in this phase.

Based on high-resolution STM/STS data, we have demonstrated the existence of step-edge states at low bias voltages on both Ni and S-terminated surfaces, and characterized quantitatively their spatial localization. Furthermore, we have unveiled the relation between the filling anomaly of OWCC and the origin of the step-edge states -- a direct consequence of the bulk's topology. To the best of our knowledge, this finding establishes NiS$_2$ as the first material realization of an OAI phase surviving strong electron interactions. This positions NiS$_2$ as an interesting testbed to study the interplay between topology and interactions, given that it offers the possibility
to control the metal-to-insulator transition in terms of several external factors.

Besides advancing the understanding of NiS$_2$, our findings motivate further explorations of the relation between 1D edge states and OAI phases in interacting materials. These new efforts could lead to the discovery of novel topological phases in interacting materials.

\section{Methods} \label{Sec: Methods}
\textbf{\textit{NiS$_2$ single crystal growth}}

Crystals were grown via flux synthesis using Te as flux.  1 g of Ni and S in a  1:3 or 1:4 stoichiometic ratio were mixed with 10 g of Te in a quartz tube and covered with a quartz wool plug. The tube was sealed under dynamic vacuum and placed upright into a furnace. The sample was heated to 900 °C within 10 h, held there for 48 h and subsequently cooled slowly to 550 °C over 100 h. The Te flux was removed at that temperature via centrifugation. Crystals were collected, re-sealed in a fresh quartz glass tube to evaporate remaining Te flux at 450 °C overnight.

\textbf{\textit{Neutron powder diffraction}}

The neutron powder diffraction data has been collected on the cold neutron diffractometer WISH situated on the second target station at the ISIS neutron and muon facility (UK) \cite{wish}. The diffraction data were collected on a powder sample in the temperature range 100-1.5 K with the use of a Oxford Instrument cryostat. The diffraction data refinement has been conducted with the help of the JANA2006 software \cite{Jana2006}. The diffraction data indicated the presence of minute impurities of NiS (0.8\% in volume) and $S_8$(1.13\% in volume). The Rietveld plot of the refinement at 100K, 30K and 1.5K are reported in the supporting information and mcif file of the obtained structure are provided as supporting information. Group theory calculations and symmetry analysis has been performed with the help of ISODISTORT and the ISOTROPY suite~\cite{Isodistort}. Raw file of the neutron diffraction data can be obtained at https://doi.org/10.5286/ISIS.E.RB1820132

\textbf{\textit{Scanning tunneling microscopy and spectroscopy}}

Experiments were performed in an ultra-high vacuum (UHV) chamber that houses a commercial STM (USM1300) capable to operate at 340 mK and under magnetic field strengths up to 11 T applied perpendicular to the sample surface. All the STM/STS data shown in this work have been measured at 4.2 K. Tunneling spectroscopy data were acquired using standard lock-in amplifier techniques with a lock-in frequency of 833 Hz and a modulation voltage (\textit{V$_{\mathrm{mod}}$}) indicated on each figures caption. Before each experimental run, the Pt/Ir STM-tip was treated on Cu(111) single crystal and calibrated against the Shockley surface state. WSxM software have been utilized to analyze and render all the acquired STM/STS data~\cite{Horcas2007}.

Single crystals of NiS$_2$ were glued to our sample holders using thermally and electrically conductive epoxy (EPO-TEK H20E) and cured for 1 hour at 130°C. The samples were then introduced into the UHV chamber without any additional treatment. To obtain a clean and flat surface, the NiS$_2$ crystals were mechanically cleaved at a temperature of 95 K and in a differential background pressure of 1.2×10$^{-10}$ Torr. Subsequently, the samples were transferred directly to the STM stage, maintained already at 4.2 K. In all measured samples, the cleavage plane was parallel to the \textit{ab} plane of the cubic crystal structure of NiS$_2$ (see Fig.~\ref{Exp-Figure-1}a).

\textbf{\textit{Theoretical methods}}

First principle calculations have been performed via the Vienna Ab Initio Simulation Package (VASP), incorporationg relativistic spin-orbit corrections. The BZ was sampled with a grid of $7 \times 7 \times 7$, and a plane-wave cutoff of 500 eV was applied. The atomic positions and magnetic moments determined by neutron scattering experiments were considered for the crystal structure and initial Ni moments, respectively. The Methfessel-Paxton method of order 1, with a width of 0.2 eV, was implemented for the smearing~\cite{Methfessel-Paxton}. 

The modified Becke-Johnson~\cite{Becke2006, Tran2009} scheme was used to approximate the exchange-correlation potential, as it a more sophisticated approach than the local density and general gradient approximations, and it is able to yield a finite magnetic ordering without implementing Hubbard interactions at mean-field level. It is also able to yield band gaps with an accuracy similar to hybrid functionals, but is computationally less expensive. The exchange part of the mBJ potential $v_{x}^{\mathrm{mBJ}}$ consists of two terms:

\begin{equation}
    v_{x}^{mBJ}(\bs{r}) = c v_{x}^{BR}(\bs{r}) + (3c-2)\frac{1}{\pi}\sqrt{\frac{5 \tau(\bs{r})}{6 n(\bs{r})}},
\end{equation}
where $v_x^{\mathrm{BR}}$ is the Becke-Roussel potential~\cite{Becke1989}, $n(\bs{r})$ is the electronic density and $\tau(\bs{r})$ is a term related to the kinetic energy density. The parameter $c$, which governs the relative weight of both terms, can be tuned in VASP via the input parameter CMBJ.

The magnetic ground states were obtained in two steps. First, we run a noncollinear calculation starting from a configuration with vanishing magnetic moments, which lead to a paramagnetic ground state. Second, the charge density of this paramagnetic ground state was used as the input to run a self-consistent calculation with the symmetry lowered to the magnetic structure determined via neutron scattering. 

The software Wannier90 and VASP's interface to it were employed to construct the Wannier models for the insulating OAI phase. Ni atoms' $d$ and S atoms' $p$ orbitals were used as initial projections. Identical disentanglement and frozen windows were used, whose lower and upper boundaries were set to 0 and 11.5 eV, respectively. Both the disentanglement and minimization of the non-gauge dependent spread functional run until the default convergence criteria were satisfied.

\section{Acknowledgments}
M.G.V. thanks support to PID2022-142008NB-I00 funded by  MICIU/AEI/10.13039/501100011033 and FEDER, UE, the Canada Excellence Research Chairs
Program for Topological Quantum Matter and to Diputación Foral de Gipuzkoa Programa Mujeres y Ciencia.  M.M.U. acknowledges support by the European Union ERC Starting grant LINKSPM (Grant \#758558) and and by the grant PID2023-153277NB-I00 funded by the Spanish Ministry of Science, Innovation and Universities. F.O. and P.M. acknowledge Dmitry Khalyavin for the very fruitful discussion regarding the magnetic structure solution. The authors acknowledge the Science and Technology Facility Council (STFC-UKRI) for the provision of beam time on the WISH diffractometer under the proposal RB1820132. The raw data of the neutron experiment can be found at https://doi.org/10.5286/ISIS.E.RB1820132. This work was supported by the Deutsche Forschungsgemeinschaft (DFG) through QUAST-FOR5249 and the Würzburg-Dresden Cluster of Excellence on Complexity and Topology in Quantum Matter, ct.qmat (EXC 2147, Project ID 390858490). M.I.I. and H.G. acknowledge funding from the EU NextGenerationEU/PRTR-C17.I1, as well as by the IKUR Strategy under the collaboration agreement between Ikerbasque Foundation and DIPC on behalf of the Department of Education of the Basque Government. S.S. acknowledges enrollment in the doctorate program “Physics of Nanostructures and Advanced Materials” from the “Advanced polymers and materials, physics, chemistry and technology” department of the Universidad del País Vasco (UPV/EHU). Work at Princeton was supported by NSF through the Princeton Center for Complex Materials, a Materials Research Science and Engineering Center DMR-2011750, the Gordon and Betty Moore Foundation (EPiQS Synthesis Award) through grant GBMF9064, and the David and Lucile Packard Foundation.

%\beginsupplement

\section{Supplementary Information}
\beginsupplementary

\subsection{ Previous magnetic models for \nis{} } \label{SI: hystory}

%The magnetic ground state of \nis{} is still under debate. Neutron powder diffraction experiments first reported the presence of two transitions – dubbed M1 and M2 – with emerging magnetic reflections consistent with $\bs{k}_{\mathrm{M1}} = \bs{0}$ ($\Gamma$-point) and $\bs{k}_{\mathrm{M2}} = (1/2, 1/2, 1/2)$ ($R$-point) propagation vectors at 40K and 30K, respectively~\cite{Hastings70}. After several attempts to model the ground state of \nis{} based on these observations~\cite{Miyadai1975, Nishihara75, Kikuchi78, Yano16}, the M1 structure has been confirmed to be noncollinear, while several open questions remain. For instance, current models do not explain the WFM moment observed below $T_{N2}$, as well as the evolution of the M1 reflection intensities with temperature.

Various contradicting reports regarding the magnetic structure of NiS$_2$ are present in the literature, leaving its magnetic ground state still under debate. Hastings and Corliss~\cite{Hastings70} first reported the presence of two transitions -- dubbed M1 and M2 -- in neutron powder diffraction, and identified their propagation vectors. Below 40 K they observed magnetic reflections consistent with a $\Gamma$-point distortion ($\bs{k}_{\mathrm{M1}}=\bs{0}$), whereas below 30 K they identified new reflections consistent with the propagation vector $\bs{k}_{\mathrm{M2}}=(1/2,1/2,1/2)$. Since then, various groups proposed different models for the M1 and M2 magnetic ground states. Miyadai et al.~\cite{Miyadai1975} proposed a model in which the M1 component is collinear along $[001]$ and the M2 component lies in the (110) plane. The two components are assumed to coexist in the same phase due to the observation of a anomaly in the M1 intensities at $T_{N2}$. Whereas the model has a constant moment as expected in a Mott insulator, the structure is not consistent with NMR data which indicate a non collinear structure for the M1 phase~\cite{Nishihara75}, and the proposed model also does not explain the base temperature NMR data and the anomaly in the M1 intensities at $T_{N2}$. Kikuchi et al.~\cite{Kikuchi78} improved Miyadai's model using single-crystal data in zero and applied field. They constructed a constant moment solution in which both the M1 and M2 components are noncollinear and the M1 structure is similar to the MnTe ground state described in the $Pa \bar{3} $ magnetic space group. Finally, a recent work from Yano et al.~\cite{Yano16}, confirmed the M1 noncollinear structure, and suggested a different model for the ground state assuming that the M2 component is perpendicular to M1 and the latter does not change at the second transition. In this case the moment is not constant on the Ni sites and the model does not explain the WFM moment observed below $T_{N2}$ as well as the evolution of the M1 reflections intensities with temperature. Moreover, to the best of our knowledge none of the previous work provide a detailed symmetry analysis and the magnetic space group for the material ground state.

\subsection{ Symmetry analysis and order parameter coupling in the M2 phase} \label{SI: Symmetry M2}

\begin{figure}
    \centering
    \includegraphics[width=8.5cm]{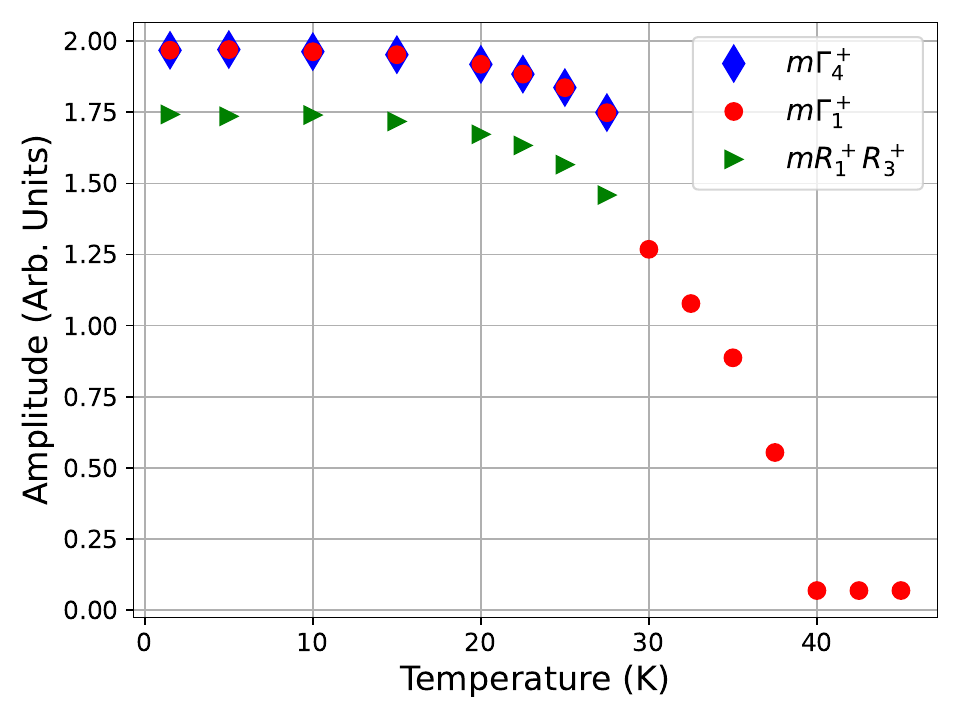}
    \caption{Amplitudes of the magnetic symmetry adapted modes as function of temperature obtained from the mode decomposition of the refined magnetic structures.}
    \label{Modes}
\end{figure}

Whereas in our analysis the isotropy subgroup $R\bar 3$ is obtained solely in terms of the irreps  $m\Gamma_1^+$ and $mR_1^{+}R_3^{+}$, a detailed symmetry analysis indicates that the refined magnetic ground state is actually the results of three irreducible representations of the parent space group. Indeed, 
a mode decomposition analysis performed on the refined constant moment model indicates the presence of distortion that transform as the $m\Gamma_4^+$ irreps of the parent space group with the same amplitude as the $m\Gamma_1^+$ one (Fig.~\ref{Modes}). In our analysis the order parameters associated to $m\Gamma_1^+$ and $mR_1^{+}R_3^{+}$ are, respectively, $(\mu)$ and $(\eta+\delta,(1-\sqrt{3})\delta,\eta-\delta,(1-\sqrt{3})\eta)$. These two order parameters can couple the  $m\Gamma_4^+$ $(\chi,\chi,\chi)$ distortion through a free energy invariant $A(3\mu\chi\eta^2 - \mu \chi \delta^2)+B(\mu \chi \eta \delta)+C(-\mu\chi\eta^2 + 3\mu \chi \delta^2)$ , where A, B and C are phenomenological coefficients. The $m\Gamma_4^+$ distortion is needed to obtain a constant moment solution and the amplitude of its symmetry adapted mode is equal to the $m\Gamma_1^+$, as shown in Fig.~\ref{Modes} justifying the sharp increase in intensity of the M1 reflections at $T_{N2}$. The $m\Gamma_4^+$ irreps also allows for a ferromagnetic moment along the [111] direction of the parent cubic cell explaining the susceptibility data showing a weak ferromagnetic moment below $T_{N2}$. Our proposed model also agrees with various observations reported in literature. Feng et al.~\cite{Feng2011} observed the presence of a nuclear distortion with the same propagation vector as the M2 reflections. This nuclear distortion is indeed expected as a secondary order parameter due to the coupling between the $m\Gamma_1^+$ and the $mR_1^+R_3^+$ distortions and it will transform as the $R_1^+R_3^+$ irreducible representation of the parent space group. This trigonal displacive distortion has also been observed by Nagata et al.~\cite{Nagata76} in thermal expansion measurements and by Thio et al.~\cite{Thio1995} in X-ray measurements. 

%\subsection{Supplemental plots for the neutron diffraction data} \label{S-13}

\begin{table}
\centering
\caption{Atomic positions, harmonic thermal parameters of the paramagnetic phase of NiS$_2$ described in the gray space group $Pa\bar31'$ obtained from the neutron diffraction data collected at 100 K. The cell parameters are $a=5.678335(13)$ ~{\AA}.}
\vspace{12 pt}
\label{Tab_Para}
\begin{tabular}{ccccc}
  Atom &$x$&$y$&$z$& U$_{\text{Iso}}$~({\AA}$^2$) \\ 
 \hline 
Ni1& 0& 0& 0& 0.00671(14)\\
S1 & 0.39366(9)& 0.39366(9)& 0.39366(9)&0.0060(3)\\
 \end{tabular}
\end{table}

\begin{table*}
\centering
\caption{Atomic positions, harmonic thermal parameters and magnetic moment components of the M1 phase of NiS$_2$ described in the magnetic space group $Pa\bar3$ obtained from the neutron diffraction data collected at 30 K. The cell parameters are $a=5.676656(14)$~{\AA}.}
\vspace{12 pt}
\label{Tab_M1}
\begin{tabular}{ccccccccc}
\hline \hline
  Atom & $x$ & $y$ & $z$ & U$_{\text{Iso}}$~({\AA}$^2$) & $M_x~(\mu_B)$ & $M_y~(\mu_B)$ & $M_z~(\mu_B)$ & $M_{tot} ~(\mu_B)$  \\ 
 \hline 
Ni1& 0& 0& 0& 0.00645(14) &0.389(4)&0.389(4)&0.389(4)&0.674(7)\\
S1 & 0.39361(10)& 0.39361(10)& 0.39361(10)&0.0063(3)& & &\\
\hline \hline
 \end{tabular}
\end{table*}

\begin{table*}
\centering
\caption{Atomic positions, harmonic thermal parameters and magnetic moment components of the M2 phase of NiS$_2$ described in the magnetic space group $R\bar3$ obtained from the neutron diffraction data collected at 1.5 K. The atomic position are reported without errorbars since they derive from the cubic structure and no rhombohedral distortions in the atomic position or cell parameters were observed within the resolution of the instrument. The cell parameters are $a=8.02740(6)$~{\AA},  $c=19.661(3)$~{\AA} and the latter cell is obtained from the cubic parent unit cell using the following transformation matrix $\{(-1,1,0),(1,0,-1),(-2,-2,-2)\}$ with origin shift (0,0,0)}
\vspace{12 pt}
\label{Tab_M2}
\begin{tabular}{ccccccccc}
\hline \hline
  Atom &$x$&$y$&$z$& U$_{\text{Iso}}$~({\AA}$^2$) & $M_x~(\mu_B)$ & $M_y~(\mu_B)$ & $M_z~(\mu_B)$ & $M_{\mathrm{tot}} ~(\mu_B)$  \\ 
 \hline 
Ni1-1 & 0      &  0      & 0      & 0.00732  & 0 & 0 & 1.195(11) & 1.195(11)\\
Ni1-2 & 0      &  0      & 0.5    & 0.00732  & 0 & 0 & -1.195(11) & 1.195(11)\\
Ni1-3 & 0.5    &  0      & 0.5    & 0.00732  & 1.180(5) & 0 & 0.254(18)&1.208(19) \\
Ni1-4 & 0.5    &  0      & 0      & 0.00732  & 1.180(5) & 0 & -0.254(18)&1.208(19)\\
S1-1  & 0      &  0      & 0.8032 & 0.007016 &  & & &\\
S1-2  & 0      &  0      & 0.3032 & 0.007016 & & & &\\
S1-3  & 0.5709 & -0.0709 & 0.2323 & 0.007016 & & & &\\
S1-4  & 0.2376 & 0.2624  & 0.0656 & 0.007016 & & & &\\
\hline \hline
 \end{tabular}
\end{table*}

\subsection{Identification of the cleavage plane} \label{SI-1}

\begin{figure}[htb!]
\centering
\includegraphics[width=8.5cm]{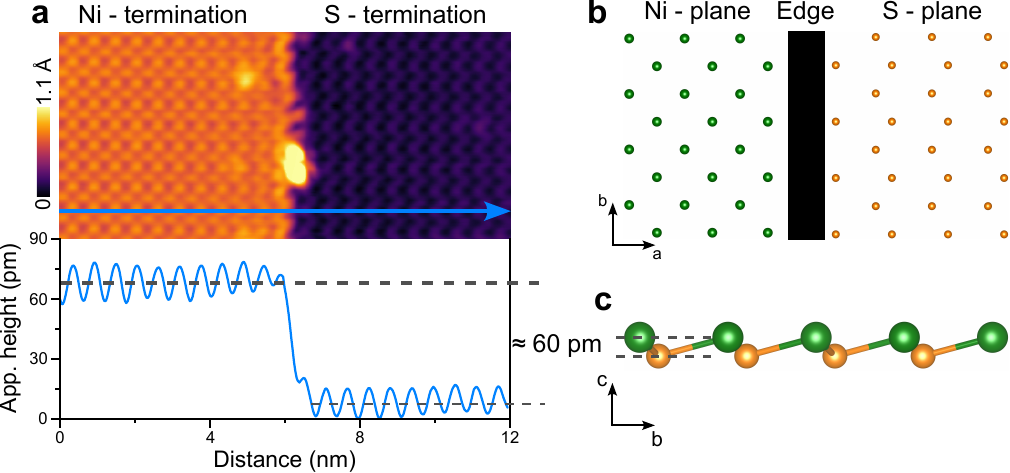}
\caption{
\textbf{Identification of the cleavage plane.} a) An atomically resolved STM image (top panel) reveals a region where two distinct atomic arrangements are simultaneously visible (left and right). The corresponding cross-sectional $Z$-profile (bottom panel) shows an apparent height difference of $\approx$ 60 pm between the left and right areas of the STM image. b) Tentative sketch representing the atomic arrangements of Ni and S atoms at both sides of the STM topography image shown in (a). c) Vertical shift along \textit{c}-axis direction of $\approx$ 60 pm between the Ni and S atomic plane, as deduced from the crystal structure model of NiS$_2$. Acquisition parameters: a) \textit{V$_{\mathrm{set}}$} = 1.1 V, \textit{I$_{\mathrm{set}}$} = 70 pA.
}
\label{Exp-Figure S1}
\end{figure}

    In the main text we stated on the existence of two type of surfaces that show distinctive atomic arrangements. We assigned both surfaces to be either the sulfur or nickel atomic termination based on the similarity between the atomic structure resolved from the STM images and the one extracted from the crystal structure model. Here, we provide further evidence that reinforce the existence of S-terminated and Ni-terminated surfaces on the as-cleaved samples of NiS$_2$. In the top panel of Fig.~\ref{Exp-Figure S1}a, we show an atomically resolved STM image that show simultaneously two types of atomic structures: i.e. a centrosymmetric (right) \textit{versus} a non-centrosymmetric (left) square structure. Both experimentally obtained atomic structures are in good agreement with the atomic arrangement of the Ni and S plane derived from the crystal model of NiS$_2$, as observed in Fig.~\ref{Exp-Figure S1}b. Additionally, from the cross-sectional profile extracted from the STM image (bottom panel of Fig.~\ref{Exp-Figure S1}a), we obtained a $Z$ apparent height differences of $\approx$ 60 pm, which closely matches the \textit{c}-axis separation between the Ni and S atomic planes.
    
\subsection{Numerical analysis of the surface reconstruction} \label{S-12}

The S and Ni surfaces explored via STM exhibit an orthorhombic distortion. This distortion is represented by a ratio $b/a \neq 1$ on the surface's pattern.

To understand the reconstruction of Ni and S surfaces, we have constructed slabs with various terminations and examined how the slab's energy depends on the ratio $b/a$. The slabs -- each $14.19$~{\AA} in height -- consist of three primitive orthorhombic cells stacked along the [001] direction, with a surrounding vacuum gap of $15.81$~{\AA}. The ration $b/a$ was varied by varying $b$ while the values for $a$ were fixed at $a=5.70$~{\AA} and $a=5.73$~{\AA} for the S and Ni surfaces, respectively -- i.e. the values for $a$ coincide with those observed in STM probes. The energy was calculated self consistently with a plane-wave cuttoff of 350 eV and a grid of $10 \times 10 \times 1$ for the BZ. The electronic minimization was allowed to run until the ground state's energy between two successive iterations displayed a difference smaller than $10^{-4}$ eV.

\begin{figure}
    \centering
    \includegraphics[width=1\linewidth]{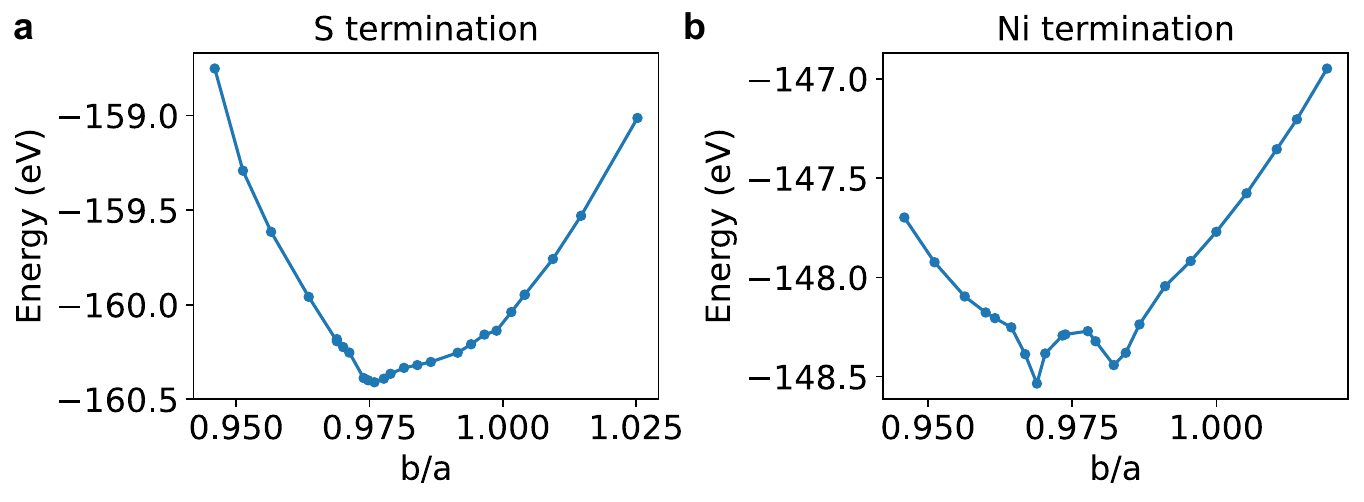}
    \caption{Energy of slabs formed by 3 unit cells in term of the orthorhombic distortion $b/a$. a) The upper surface of the slab consists of S dimers. b) The upper surface consists of Ni ions.}
    \label{Theo-Figure-S9}
\end{figure}

Figure~\ref{Theo-Figure-S9}a shows the data for the slab terminated with S dimers. The slab's energy shows a minimum around $b/a \approx 0.967$, which corresponds to a value $b \approx 5.56$~{\AA}. This result is consistent with the value of $b=5.55$~{\AA} determined by the STM experiment.

The data for the slab with a Ni surface is displayed in Fig.~\ref{Theo-Figure-S9}b. Two minima of the energy are identified. The absolute minimum is located around $b/a \approx 0.95$, which has in correspondence $b \approx 5.55$~{\AA}, in excellent agreement with the STM analysis.

\subsection{Overall surface morphology of \nis{} crystals after cleavage} \label{SI-2}

\begin{figure}[htb!]
\centering
\includegraphics[width=8.5cm]{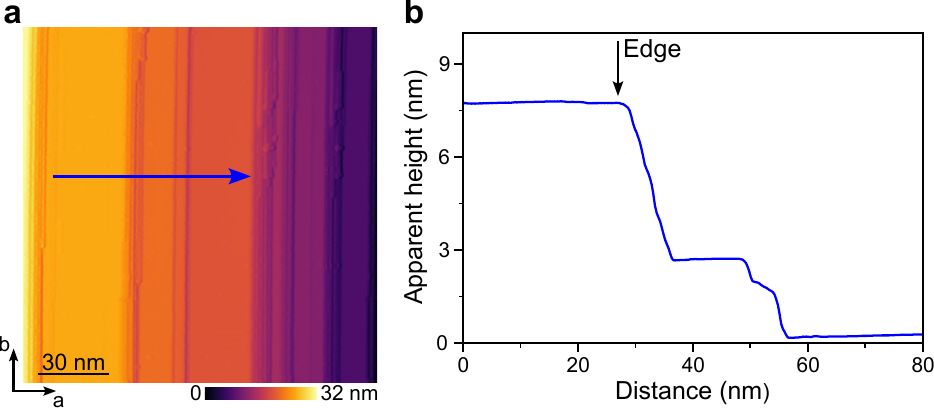}
\caption{
\textbf{Overall surface morphology of NiS$_2$ crystals after cleavage.} a) A typical STM topography image reveals atomic step bunches on the surface. A derivative filter has been applied to the raw constant-current image to enhance the visibility of features along the steps. b) Apparent height profile measured along the blue arrow indicated in b). Acquisition parameters: a) \textit{V$_{\mathrm{set}}$} = 1.1 V, \textit{I$_{\mathrm{set}}$} = 70 pA.
}
\label{Exp-Figure S2}
\end{figure}

    The mechanical cleavage of NiS$_2$ produces surfaces characterized by narrow atomic terraces (on average $\approx$ 30 nm) with multiple continuous atomic steps, as shown in Fig.~\ref{Exp-Figure S2}a. Rather than single atomic steps, most are step bunches formed by several contiguous atomic layers, as inferred from the height profile in Fig.~\ref{Exp-Figure S2}b. Additionally, the step edges do not show a sharp transition, but rather a continous trend. We note here that our experimental STM/STS data were collected from atomic step edge regions similar to those depicted in Fig.~\ref{Exp-Figure S2}.

\subsection{Logarithm representation of the d$I$/d$V$ spectra} \label{SI-3}
   
\begin{figure}[htb!]
\centering
\includegraphics[width=8.5cm]{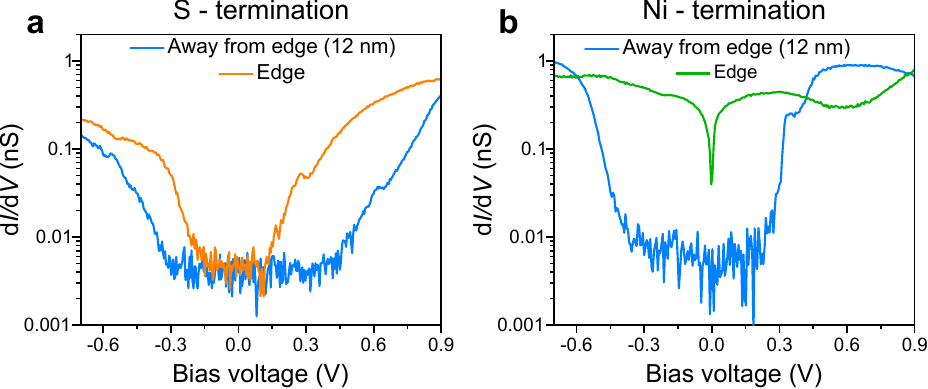}
\caption{
\textbf{Logarithm representation of the d\textit{I}/d\textit{V} spectra.} a,b) Differential conductance curves measured on the S- and Ni-terminated surfaces plotted on a Log$_{10}$ scale to better visualize the electronic gap. Acquisition parameters: a) \textit{V$_{\mathrm{mod}}$} = 6 mV. b) \textit{V$_{\mathrm{mod}}$} = 6 mV.
}
\label{Exp-Figure S3}
\end{figure}

    In Fig.~\ref{Exp-Figure 3}d and Fig.~\ref{Exp-Figure 3}h of the main text we have showed representative d$I$/d$V$ curves that capture the electronic structure of the 2D surface and step edge on the S- and Ni-termination, respectively. In Fig.~\ref{Exp-Figure S3}, we provide the same selection of d$I$/d$V$ curves but plotting the ordinate axis in a Log$_{10}$ scale. In this way, the development of a full electronic gap (vanishing conductance) can be corroborated. Additionally, the gap size can be estimated more accurately.

\subsection{Long-range extension of the step-edge state along the step edge direction} \label{SI-4}

\begin{figure}[htb!]
\centering
\includegraphics[width=8.5cm]{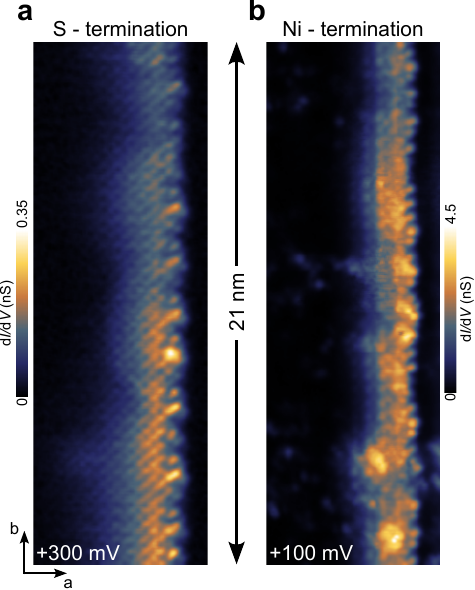}
\caption{
\textbf{Long-range extension of the step-edge state along the step edge direction.} a,b) Spatially resolved differential conductance maps highlighting the step-edge state on the S-terminated and Ni-terminated surfaces, respectively. Acquisition parameters: a) \textit{V$_{\mathrm{mod}}$} = 6 mV. b) \textit{V$_{\mathrm{mod}}$} = 5 mV
}
\label{Exp-Figure S4}
\end{figure}

    We observed that the step-edge state on both S- and Ni-terminated surfaces are predominantly localized along the step edges, exhibiting a significant long-range spatial extension. This is illustrated in Fig.~\ref{Exp-Figure S4}, where the dispersion perpendicular to the step edges is confined to just a few nanometers.

\subsection{Electronic structure of step edges around the Fermi level} \label{SI-5}

\begin{figure}[htb!]
\centering
\includegraphics[width=8.5cm]{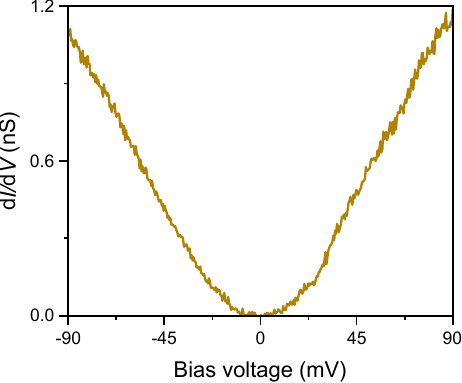}
\caption{
\textbf{Electronic structure near the Fermi level.} High-resolution differential conductance spectrum measured at the step edge of Ni-termination. Acquisition parameters: \textit{V$_{\mathrm{mod}}$} = 0.5 mV
}
\label{Exp-Figure S5}
\end{figure}

    The large-window d$I$/d$V$ spectra measured at the step edges of Ni-termination (Fig.~\ref{Exp-Figure 3}h in the main text) indicated a metallic character, featuring a V-shape dip structure near Fermi level. In Fig.~\ref{Exp-Figure S5}, we corroborate these observations by providing a high-resolution differential conductance curve measured in the window of $\pm$ 90 mV. Moreover, we rule out any possible additional electronic features close to Fermi level at the step edges. 

\subsection{Energy dependent mapping of the LDOS on the Ni-terminated surface} \label{SI-6}

\begin{figure*}[htb!]
\centering
\includegraphics[width=17cm]{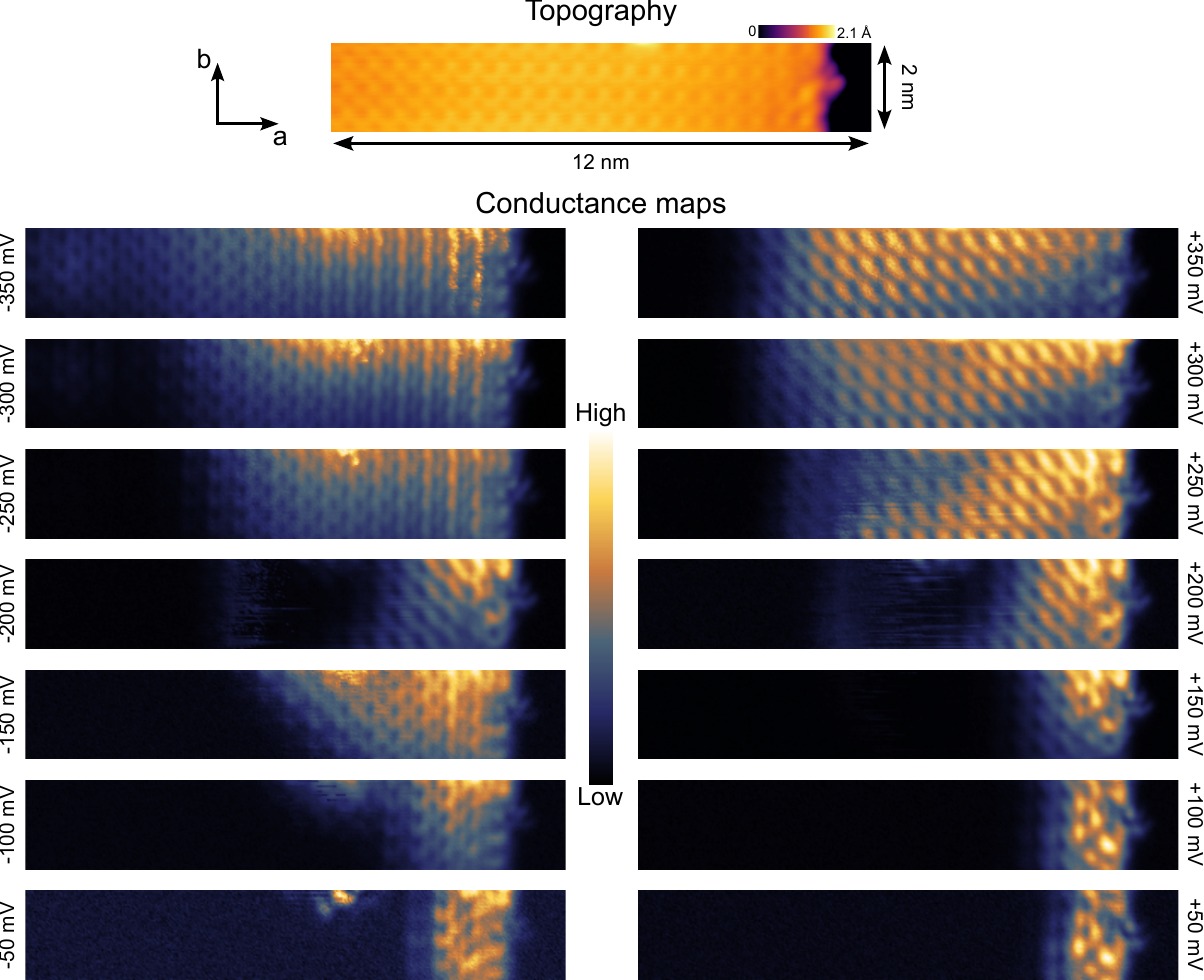}
\caption{
\textbf{Energy dependent mapping of the LDOS on the Ni-terminated surface.} Set of spatially resolved d$I$/d$V$ maps recorded at the specified bias voltages and measured within the same region as delimited by the STM topography image. Acquisition parameters: Topography image:  \textit{V$_{\mathrm{set}}$} = 1.1 V, \textit{I$_{\mathrm{set}}$} = 0.1 nA.  Conductance maps:  \textit{V$_{\mathrm{mod}}$} = 5 mV.
}
\label{Exp-Figure S6}
\end{figure*}

    In Fig.~\ref{Exp-Figure S6} we present a series of d$I$/d$V$ maps recorded at the step edge of the Ni-terminated surface accounting for the energy dependent evolution of the local density of states (LDOS). We observe that at higher voltages, the LDOS are widely distributed across the surface, while at lower voltages, the LDOS only remains confined to the step edge, suggesting the existence of a metallic states in this 1D region.

\subsection{Classification of the M2 phase} \label{SI: low-T phase}

Upon lowering the temperature, the system undergoes a transition into a weak-ferromagnetic state with the rhombohedral magnetic space group $R\bar3$ (No. 148.17). The transition is accompanied by a change in the direction of magnetic moments in Ni sites, whereas ions' positions remain unchanged within the precision of neutron scattering experiments (Sec.~\ref{sec: neutron}). Consequently, the crystal structure of the low-T phase can be represented with a unit cell whose basis vectors $(\boldsymbol{a}_1, \boldsymbol{a}_2, \boldsymbol{a}_3)_{M2}$ are given as:

\begin{equation}
    (\boldsymbol{a}_1 \ \boldsymbol{a}_2 \ \boldsymbol{a}_3)_{M2} =
    (\boldsymbol{a}_1 \ \boldsymbol{a}_2 \ \boldsymbol{a}_3)_{M1}
    \begin{pmatrix}
        1 & 0 & 2 \\
        0 & 1 & 2 \\
        -1 & -1 & 2
    \end{pmatrix}
\end{equation}
where $(\boldsymbol{a}_1, \boldsymbol{a}_2, \boldsymbol{a}_3)_{M1}$ are the primitive vectors of the antiferromagnetic phase.

\begin{figure}
    \centering
    \includegraphics[width=1.0\linewidth]{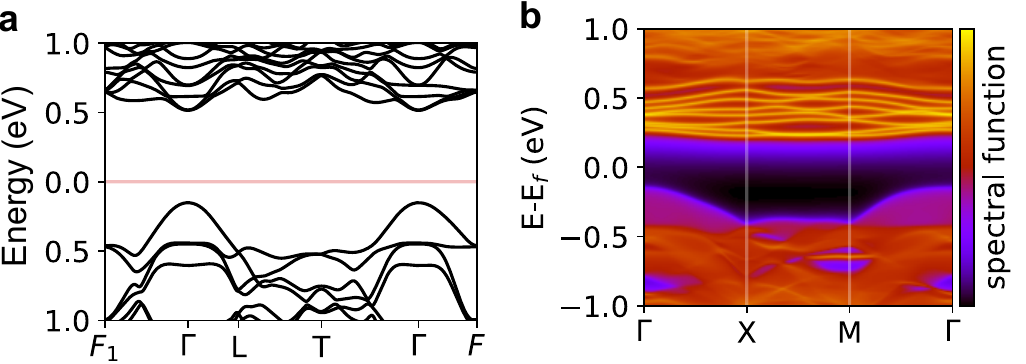}
    \caption{(a) Band structure in the M2 phase. (b) Spectral function of a semi-infinite slab with a termination perpendicular to the trigonal axis, exhibiting in-gap surface states.}
    \label{Theo Fig S10}
\end{figure}

Let us study if the transition into the low-T phase involves closing and reopening the bulk gap. We choose as starting point the last decomposition in Tab.~\ref{Theo-Table-S2}:

\begin{equation} \label{eq: decomposition OAI valence bands}
\begin{split}
    \rho_{M1} = &5(\bar E_g @ 4a) \oplus 4({}^{1}\bar{E}_{g}@4a) \oplus 5({}^{2}\bar{E}_{g}@4a) \oplus 2(\bar{E}_{u}@4a) \\
    &\oplus 2({}^{1}\bar{E}_{u}@4a) \oplus 2({}^{2}\bar{E}_{u}@4a) \oplus ({}^{1}\bar{E}_{g}@4b) \\ 
    & \oplus ({}^{2}\bar{E}_{g}@4b).
\end{split}
\end{equation}

This decomposition involves EBRs induced from WPs 4a and 4b. The conventional -- obverse triple hexagonal -- cell of the low-T rhombohedral phase contains 24 representative sites of the WP 4a. They distribute in four sets of independent orbits given by WPs 3a, 3b, 9d and 9e:

\begin{equation}
    4a \rightarrow 3a \cup 3b \cup 9d \cup 9e.
\end{equation}

The sites in WPs 3a and 3b have in correspondence site-symmetry groups isomorphic to $\bar 3$, whereas the sites in WPs 9d and 9e are left invariant by the inversion group $\bar 1$ (see Tab.~\ref{Theo-table-S3}). In the AFM phase, every set of Wannier functions in WP 4a transform as certain irreducible representations of the site-symmetry group $\bar 3$. Let us denote this irrep as $\rho$. Upon lowering the symmetry to the low-T phase, these orbitals transform as the subduced representations of their position's site-symmetry group $G_{i}$ -- with $i$ denoting the WP in the low-T conventional cell -- i.\,e., $\rho \downarrow G_i$. This subduced representation can be decomposed in terms of the irreps of $G_i$. The subducion/decomposition problem for every representation of WP 4a taking part in Eq.~\eqref{eq: decomposition OAI valence bands} is written in Tab.~\ref{Theo-table-S4}.

\begin{table}
    \centering
    \begin{tabular}{c|cc|cccccc}
    \hline \hline
      \rule{0pt}{2.2ex} Space group  & \multicolumn{2}{c|}{$Pa\bar 3$} & \multicolumn{6}{c}{$R\bar 3$}  \\
       \hline
        Wyckoff position & 4a & 4b & 3a & 3b & 6c & 9d & 9e & 18f \\
        \hline
        \rule{0pt}{2.2ex} Site-symmetry group & $\bar 3$ & $\bar 3$ & $\bar 3$ & $\bar 3$ & $3$ & $\bar 1$ & $\bar 1$ & 1 \\
        \hline \hline
    \end{tabular}
    \caption{Redistribution of sites in WP 4a and 4b of the AFM structure upon lowering the symmetry to the weak-ferromagnetic phase. The multiplicities of WPs belong to the conventional cells of the corresponding phase. Hermann-Mauguin notation is employed to denote site-symmetry groups.}
    \label{Theo-table-S3}
\end{table}

\begin{table}
    \centering
    \begin{tabular}{ccccccc}
    \hline \hline
      \rule{0pt}{2.2ex}4a  & $\bar E_g$ & ${}^{1}\bar E_g$ & ${}^{2}\bar E_g$ & $\bar E_u$ & ${}^{1}\bar E_u$ & ${}^{2}\bar E_u$ \\
       \hline
       \rule{0pt}{2.5ex}3a & $\bar E_g$ & $^{1}\bar E_g$ & $^{2}\bar E_g$ & $\bar E_u$ & $^{1}\bar E_u$ & $^{2}\bar E_u$ \\
        3b & $\bar E_g$ & $^{1}\bar E_g$ & $^{2}\bar E_g$ & $\bar E_u$ & $^{1}\bar E_u$ & $^{2}\bar E_u$ \\
        9d & $\bar A_g$ & $\bar A_g$ & $\bar A_g$ & $\bar A_u$ & $\bar A_u$ & $\bar A_u$\\
        9e & $\bar A_g$ & $\bar A_g$ & $\bar A_g$ & $\bar A_u$ & $\bar A_u$ & $\bar A_u$ \\
    \hline \hline
    \end{tabular}
    \caption{The first row contains the irreps of the site-symmetry group of WP 4a in the AFM phase that take part in the decomposition in Eq.~\eqref{eq: decomposition OAI valence bands}. The rest of rows contain the subduced representations of the site-symmetry groups in the WP of the WFM phase's conventional cell.}
    \label{Theo-table-S4}
\end{table}

Then, the EBRs induced from WP 4a in Eq.~\eqref{eq: decomposition OAI valence bands} subduce non-elementary band representations of the low-T group $R\bar 3$. These band representations can be decomposed in terms of the EBRs of the group:

\begin{equation} \label{eq: subduction contributions 4a}
    \begin{split}
        & \bar E_g @ 4a \rightarrow \bar E_g @ 3a \oplus \bar E_g @ 3b \oplus \bar A_g @ 9d \oplus \bar A_g @ 9e, \\
        & {}^{1} \bar E_g @ 4a \rightarrow {}^{1} \bar E_g @ 3a \oplus {}^{1} \bar E_g @ 3b \oplus \bar A_g @ 9d \oplus \bar A_g @ 9e, \\
        & {}^{2} \bar E_g @ 4a \rightarrow {}^{2} \bar E_g @ 3a \oplus {}^{2} \bar E_g @ 3b \oplus \bar A_g @ 9d \oplus \bar A_g @ 9e, \\
        & \bar E_u @ 4a \rightarrow \bar E_u @ 3a \oplus \bar E_u @ 3b \oplus \bar A_u @ 9d \oplus \bar A_u @ 9e, \\
        & {}^{1} \bar E_u @ 4a \rightarrow {}^{1} \bar E_u @ 3a \oplus {}^{1} \bar E_u @ 3b \oplus \bar A_u @ 9d \oplus \bar A_u @ 9e, \\
        & {}^{2} \bar E_u @ 4a \rightarrow {}^{2} \bar E_u @ 3a \oplus {}^{2} \bar E_u @ 3b \oplus \bar A_u @ 9d \oplus \bar A_u @ 9e. \\
    \end{split}
\end{equation}

Similarly, the conventional rhombohedral cell contains 24 representatives of sites in WP 4b of the AFM structure, which split into two sets of WP 6c and 18f:

\begin{equation}
    4b \rightarrow 6c \cup 18f.
\end{equation}

Whereas the sites in WP 4b of the AFM phase are left invariant by groups isomorphic to $\bar 3$, the site-symmetry groups of WP 6c and 18f in the low-T structure are isomorphic to the smaller groups $3$ and $1$, respectively (see Tab.~\ref{Theo-table-S4}). Consequently, the Wannier functions in WPs 6c and 18f transform as representations of their corresponding site-symmetry groups subduced by the irreps of the site-symmetry group of WP 4b in the AFM phase. This subduction is written in Tab.~\ref{Theo-Table-S5} for the EBRs induced from WP 4b taking part in Eq.~\eqref{eq: decomposition OAI valence bands}.

\begin{table}
    \centering
    \begin{tabular}{ccc}
    \hline \hline
        \rule{0pt}{2.5ex}4b & ${}^{1} \bar E_g$ & ${}^{2} \bar E_g $\\
        \hline
        \rule{0pt}{2.5ex}6c & ${}^{1} \bar E$ & ${}^{2} \bar E$ \\
        \rule{0pt}{2.5ex}18f & $\bar A$ & $\bar A$ \\
    \hline \hline
    \end{tabular}
    \caption{The first row contains the irreps of the site-symmetry group of WP 4b in the AFM phase that take part in the decomposition in Eq.~\eqref{eq: decomposition OAI valence bands}. The rest of rows contain the subduced representations of the site-symmetry groups in the WP of the WFM phase's conventional cell.}
    \label{Theo-Table-S5}
\end{table}

The EBRs induced from WP 4b in Eq.~\eqref{eq: decomposition OAI valence bands} subduce non-elementary band representations of the low-T group $R\bar 3$. Therefore, they can be decomposed in terms of the EBRs of the group $R\bar 3$:

\begin{equation} \label{eq: subduction contributions from 4b}
    \begin{split}
        & {}^{1} \bar E_g @ 4b \rightarrow {}^{1} \bar E @ 6c \oplus \bar A @ 18f, \\
        & {}^{2} \bar E_g @ 4b \rightarrow {}^{2} \bar E @ 6c \oplus \bar A @ 18f.
    \end{split}
\end{equation}

Applying the decompositions in Eq.~\eqref{eq: subduction contributions from 4b} together with the multiplicities in Eq.~\eqref{eq: decomposition OAI valence bands}, we conclude that the valence states of the AFM phase transform as the following subduced representation of the low-T phase upon lowering of symmetry:

\begin{align}
\rho_{M1}&\downarrow R\bar 3 = \nonumber\\
&5(\bar E_g @ 3a) \oplus 4({}^{1}\bar E_g @ 3a) \oplus 5({}^{2}\bar E_g @ 3a) \nonumber\\ 
&\oplus 2(\bar E_u @ 3a) \oplus 2({}^{1}\bar E_u @ 3a) \oplus 2({}^{2}\bar E_u @ 3a) \nonumber\\
&\oplus 5(\bar E_g @ 3b) \oplus 4({}^{1}\bar E_g @ 3b) \oplus 5({}^{2}\bar E_g @ 3b) \nonumber\\ 
&\oplus 2(\bar E_u @ 3b) \oplus 2({}^{1}\bar E_u @ 3b) \oplus 2({}^{2}\bar E_u @ 3b) \nonumber\\
&\oplus 14(\bar A_g @ 9d) \oplus 6(\bar A_u @ 9d) \nonumber\\ &\oplus 14(\bar A_g @ 9d) \oplus 6(\bar A_u @ 9d) \nonumber\\
&\oplus 14(\bar A_g @ 9e) \oplus 6(\bar A_u @ 9e) \nonumber\\ &\oplus 14(\bar A_g @ 9e) \oplus 6(\bar A_u @ 9e) \nonumber\\
& \oplus ({}^{1}\bar E @ 6c) \oplus ({}^{2}\bar E @ 6c) \oplus 2 (\bar A @ 18f).
\end{align}

We can then determine the irreducible representations of maximal $\bk$ points' little groups corresponding to this band representation (see Tab.~\ref{Theo-Table-S6}). If the transition into the low-T phase does not involve closing the gap between occupied and unoccupied states, valence bands should have the irreducible representations predicted here.

To check if this is the case, we have calculated within DFT the valence wave functions and irreducible representations at maximal $\bk$ points in the low-T magnetic phase. For that, we started from an initial configuration with the same symmetry as the low-T rhombohedral structure. The exchange-correlation term was implemented within the mBJ approximation with the VASP parameter CMBJ=1.15. The obtained irreducible representations are written in Tab.~\ref{Theo-Table-S6}. In fact, they do not match  those predicted above, which indicates that the temperature-driven transition from the cubic to the rhombohedral symmetry involves band inversions between occupied and unoccupied states. Therefore, the transition does not consist merely on the lowering of symmetry, but it requires closing and reopening the bulk gap between valence and conduction states.

\begin{table*}
    \centering
    \begin{tabular}{c|cccccc|cc|cc|cccccc}
    \hline \hline
       \rule{0pt}{2.5ex}Irreducible representation  & $\bar \Gamma_4$ & $\bar \Gamma_5$ & $\bar \Gamma_6$ & $\bar \Gamma_7$ & $\bar \Gamma_8$ & $\bar \Gamma_9$ & $\bar F_2$ & $\bar F_3$ & $\bar L_2$ & $\bar L_3$ & $\bar T_4$ & $\bar T_5$ & $\bar T_6$ & $\bar T_7$ & $\bar T_8$ & $\bar T_9$ \\ \hline
       Predicted multiplicities & 40 & 39 & 41 & 18 & 19 & 19 & 88 & 88 & 88 & 88 & 29 & 29 & 30 & 29 & 29 & 30 \\
       DFT multiplicities & 40 & 40 & 40 & 18 & 19 & 19 & 88 & 88 & 88 & 88 & 29 & 29 & 30 & 29 & 30 & 29 \\
       \hline \hline
    \end{tabular}
    \caption{Multiplicities of irreducible representation at maximal $\bk$ points of the low-T phase's BZ. The second row contains the multiplicities that valence bands would have if the transition did not band inversion between valence and conduction states. The third row contains the multiplicities identified via first-principle calculation.}
    \label{Theo-Table-S6}
\end{table*}

%First, mention that getting the WFM phase requires running the DFT calculation by breaking down the from the beginning to this setting.

It turns out that the OAI positions 9c and 18f are connected to the WPs 3a, 3b, 9d and 9e occupied by Ni atoms. This means that the Wannier functions located in OWCCs can be moved away to Ni sites without breaking any symmetry. Therefore, the decomposition of the representation of valence bands obtained via DFT in terms of the EBRs of $R\bar 3$ involves only Ni sites. Based on symmetry considerations, we can not differentiate between having charges on OWCCs or having them only on Ni sites. 

In order to go beyond the prediction based on symmetries, we have simulated the spectral function for a semi-infinite structure with a termination perpendicular to the trigonal direction in the M2 phase (see Fig.~\ref{Theo Fig S10}). The presence of in-gap states localized on the surface suggests that \nis{} hosts an OAI phase below $T_{N2}$.

\subsection{Topological classification of the OAI phase} \label{SI-10}

The BZ of the antiferromagnetic phase's magnetic space group $Pa\bar3$ has four maximal $\bs{k}$ points, denoted $\Gamma$, X, M and R (see Fig.~\ref{Theo-Figure-1}b). The irreducible representations of valence bands at these maximal points are given in Tab.~\ref{Theo-Table-S1}. This set of representations can be written as many linear combinations of EBRs according to Eq.~\eqref{eq: tqc for NiS2}. Table~\ref{Theo-Table-S2} contains the coefficients of all these decompositions. They match with atomic limits as the coefficients of all EBRs are non-negative integer numbers. 

Furthermore, we verified the representation of valence bands $\rho_{\mathrm{val}}$ does not admit any decomposition in terms of band representations induced solely from WPs 4a and 8c, corresponding to sites occupied by Ni and S atoms, respectively. This verification was performed in two steps. First, we calculated the irreducible representations at maximal points of the BZ of all band representations of dimension identical to $\rho_{\mathrm{val}}$ that can be constructed as the following  combination:
\begin{equation} \label{eq: tqc in terms of 8c}
    \rho = \bigoplus_{i} \tilde C_{i,4a} (\rho_i @ 4a) \bigoplus_{i} \tilde C_{i,8c} (\rho_i @ 8c).
\end{equation}
where coefficients $\tilde C_{i,4a}$ and $\tilde C_{i,8c}$ take non-negative integer values. Then, we verified that the set of irreducible representations of $\rho_{\mathrm{val}}$ and$\rho$ differ at least in a maximal $\bs{k}$ point. This confirms that in none of the solutions of Eq.~\eqref{eq: tqc for NiS2} can the EBRs induced from WP 4b originate from the separation of a band representation of S states, which confirms that $\rho_{\mathrm{val}}$ hosts an OAI within the scope of TQC.

\begin{table}[h]
    \centering
    \begin{tabular}{cc}
    \hline \hline
        $\bs{k}$ point & valence bands irreps\\ \hline
        \rule{0pt}{10pt}
        $\Gamma$ & $11 \bar{\Gamma}_{5} \oplus 11 \bar{\Gamma}_{6} \oplus 10 \bar{\Gamma}_{7} \oplus 4 \bar{\Gamma}_{8} \oplus 4 \bar{\Gamma}_{9} \oplus 4 \bar{\Gamma}_{10}$ \\
        X & $22 \bar{X}_{3} \oplus 22 \bar{X}_{4}$ \\
        M & $22 \bar{M}_{3} \oplus 22 \bar{M}_{4}$ \\
        R & $5 \bar{R}_{4} \oplus 4 \bar{R}_{5} \oplus 5 \bar{R}_{6} \oplus 2 \bar{R}_{7} \oplus 3 \bar{R}_{8} \oplus 3 \bar{R}_{9} \oplus 14 \bar{R}_{10} \oplus 8 \bar{R}_{11}$ \\ \hline \hline
    \end{tabular}
    \caption{Irreducible representations of valence bands at maximal $\bs{k}$ points in the antiferromagnetic phase.}
    \label{Theo-Table-S1}
\end{table}

\begin{table*}
    \centering
    \begin{tabular}{ccccccccccccc}
    \hline \hline
    \rule{0pt}{10pt}
       EBR  & $\bar{E}_{g}@4a$ & ${}^{1}\bar{E}_{g}@4a$ & ${}^{2}\bar{E}_{g}@4a$ & $\bar{E}_{u}@4a$ & ${}^{1}\bar{E}_{u}@4a$ & ${}^{2}\bar{E}_{u}@4a$ & $\bar{E}_{g}@4b$ & ${}^{1}\bar{E}_{g}@4b$ & ${}^{2}\bar{E}_{g}@4b$ & $\bar{E}_{u}@4b$ & ${}^{1}\bar{E}_{u}@4b$ & ${}^{2}\bar{E}_{u}@4b$ \\ \hline
         & 3 & 2 & 3 & 0 & 0 & 0 & 2 & 3 & 3 & 2 & 2 & 2\\
         & 3 & 2 & 4 & 0 & 0 & 1 & 2 & 3 & 2 & 2 & 2 & 1\\
         & 3 & 2 & 5 & 0 & 0 & 2 & 2 & 3 & 1 & 2 & 2 & 0\\
         & 3 & 3 & 3 & 0 & 1 & 0 & 2 & 2 & 3 & 2 & 1 & 2\\
         & 3 & 3 & 4 & 0 & 1 & 1 & 2 & 2 & 2 & 2 & 1 & 1\\
         & 3 & 3 & 5 & 0 & 1 & 2 & 2 & 2 & 1 & 2 & 1 & 0\\
         & 3 & 4 & 3 & 0 & 2 & 0 & 2 & 1 & 3 & 2 & 0 & 2\\
         & 3 & 4 & 4 & 0 & 2 & 1 & 2 & 1 & 2 & 2 & 0 & 1\\
         & 3 & 4 & 5 & 0 & 2 & 2 & 2 & 1 & 1 & 2 & 0 & 0\\
         & 4 & 2 & 3 & 1 & 0 & 0 & 1 & 3 & 3 & 1 & 2 & 2\\
         & 4 & 2 & 4 & 1 & 0 & 1 & 1 & 3 & 2 & 1 & 2 & 1\\
         & 4 & 2 & 5 & 1 & 0 & 2 & 1 & 3 & 1 & 1 & 2 & 0\\
         & 4 & 3 & 3 & 1 & 1 & 0 & 1 & 2 & 3 & 1 & 1 & 2\\
         & 4 & 3 & 4 & 1 & 1 & 1 & 1 & 2 & 2 & 1 & 1 & 1\\
         & 4 & 3 & 5 & 1 & 1 & 2 & 1 & 2 & 1 & 1 & 1 & 0\\
         & 4 & 4 & 3 & 1 & 2 & 0 & 1 & 1 & 3 & 1 & 0 & 2\\
         & 4 & 4 & 4 & 1 & 2 & 1 & 1 & 1 & 2 & 1 & 0 & 1\\
         & 4 & 4 & 5 & 1 & 2 & 2 & 1 & 1 & 1 & 1 & 0 & 0\\
         & 5 & 2 & 3 & 2 & 0 & 0 & 0 & 3 & 3 & 0 & 2 & 2\\
         & 5 & 2 & 4 & 2 & 0 & 1 & 0 & 3 & 2 & 0 & 2 & 1\\
         & 5 & 2 & 5 & 2 & 0 & 2 & 0 & 3 & 1 & 0 & 2 & 0\\
         & 5 & 3 & 3 & 2 & 1 & 0 & 0 & 2 & 3 & 0 & 1 & 2\\
         & 5 & 3 & 4 & 2 & 1 & 1 & 0 & 2 & 2 & 0 & 1 & 1\\
         & 5 & 3 & 5 & 2 & 1 & 2 & 0 & 2 & 1 & 0 & 1 & 0\\
         & 5 & 4 & 3 & 2 & 2 & 0 & 0 & 1 & 3 & 0 & 0 & 2\\
         & 5 & 4 & 4 & 2 & 2 & 1 & 0 & 1 & 2 & 0 & 0 & 1\\
         & 5 & 4 & 5 & 2 & 2 & 2 & 0 & 1 & 1 & 0 & 0 & 0\\ \hline \hline
    \end{tabular}
    \caption{Coefficients of EBRs in the decompositions of the set of irreducible representation of the antiferromagnetic phase's valence bands at maximal $\bs{k}$ as linear combinations of the EBRs of the magnetic space group $Pa\bar 3$.}
    \label{Theo-Table-S2}
\end{table*}

\subsection{Resilience of the electronic structure against magnetic field on the Ni-termination} \label{SI-7}

\begin{figure*}[htb!]
\centering
\includegraphics[width=17cm]{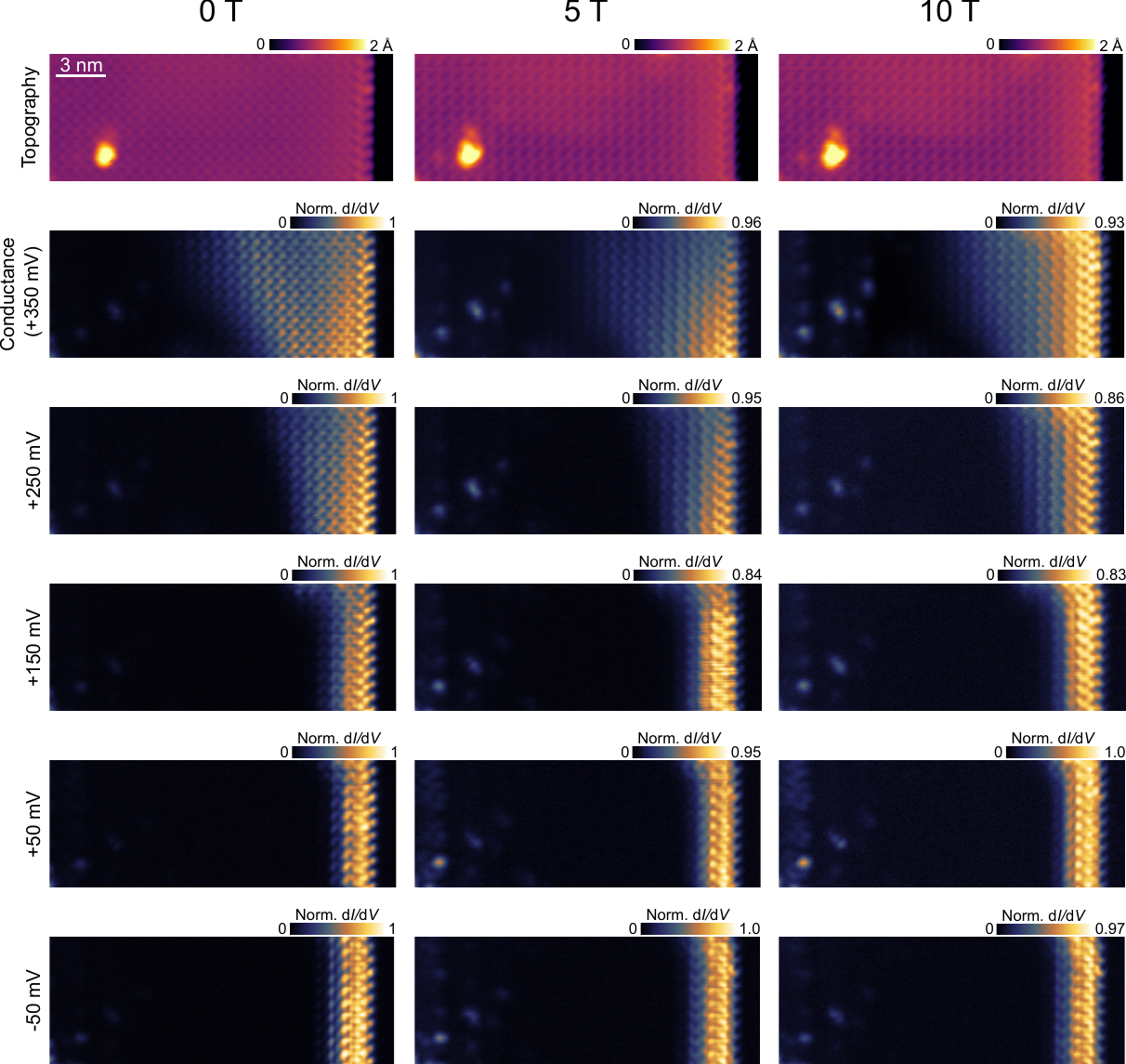}
\caption{
\textbf{Resilience of the electronic structure against magnetic field on the Ni-termination.} Set of magnetic field-dependent differential conductance maps measured at different bias voltage on the very same region as imaged by the STM topography image. The d$I$/d$V$ signal values from the differential conductance maps acquired at 5 T and 10 T were normalized against those obtained at 0 T at each bias voltage. Acquisition parameters: Topography images: \textit{V$_{\mathrm{set}}$} = 1.1 V, \textit{I$_{\mathrm{set}}$} = 70 pA. Conductance maps: \textit{V$_{\mathrm{mod}}$} = 6 mV.
}
\label{Exp-Figure S7}
\end{figure*}

    We have performed mapping of LDOS near the step edge on the Ni-termination as function of the bias voltage and magnetic field strength. The obtained results are shown in Fig.~\ref{Exp-Figure S7}, where we infer that the magnetic field do not affect the overall electronic structure of the Ni-terminated surfaces. Hence, it indicates a strong resilience of the step-edge state against magnetic fields.

\subsection{Evolution of magnetic moments and gap within the mBJ functional approximation} \label{SI-9}

The magnetic moments on Ni ions have been calculated for several values of the parameter CMBJ to study their evolution within the mBJ approximation. The comparison between the moments computed and the experimentally determined ones made in two steps: First, we verified the consistency of the moments' directions, i.\,e., that the alignment of moments is compatible with the symmetries of the magnetic space group $Pa\bar 3$. The orientation of Ni moments remains consistent with the experimental observation for all values of CMBJ. 

\begin{figure}
    \centering
    \includegraphics[width=0.9\linewidth]{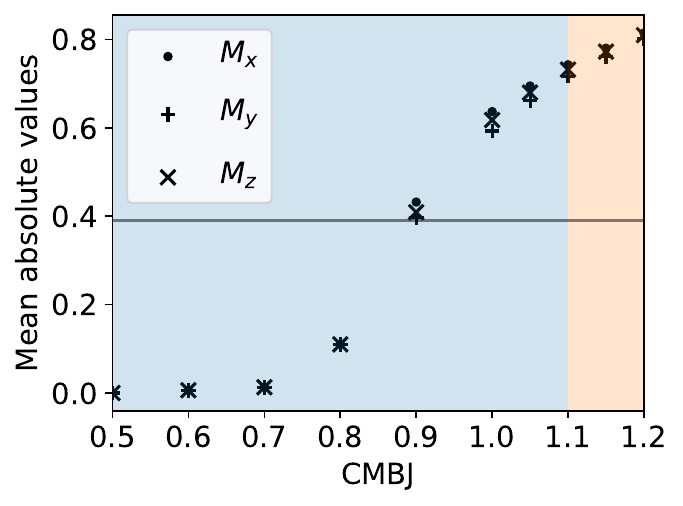}
    \caption{Evolution of the mean value in Eq.~\eqref{eq: average value moments} (in units of the Bohr's magneton) in terms of CMBJ, in the antiferromagnetic phase . The values $\abs{m_i} \sim 0.39$ determined in neutron scattering experiments are denoted by the black line. The blue and orange regions indicate the values of CMBJ for which metallic and insulating electronic structures are obtained, respectively.}
    \label{Theo-Fig-S8}
\end{figure}

Then, we analyzed the consistency of the magnitude of local the moments by computing the average of the moments' components over all Ni sites:

\begin{equation} \label{eq: average value moments}
    \ev{\abs{m_i}} = \frac{1}{4} \sum_{j=1}^{4} |m_i^{(j)}|,
\end{equation}
where $m_{i}^{(j)}$ is the component $i$ of the magnetic moment vector on the Ni site denoted $j$.
The evolution of $\ev{\abs{m_i}}$ is illustrated in Fig.~\ref{Theo-Fig-S8}. For $\mathrm{CMBJ} < 0.7$, the calculation converges into a metallic ground state with vanishing magnetic moments. Then, the magnitude of moments increases upon increasing CMBJ. A set of moments consistent with those determined in the antiferromagnetic phase via neutron scattering is obtained at $\mathrm{CMBJ} \approx 0.9$. The ground state remains metallic for $\mathrm{CMBJ} \lesssim 1.1$. Around this value, valence and conduction bands tear apart, opening a spectral gap and leading to an insulating phase. 

\clearpage{}

\bibliography{references}

\begin{thebibliography}{66}%
\makeatletter
\providecommand \@ifxundefined [1]{%
 \@ifx{#1\undefined}
}%
\providecommand \@ifnum [1]{%
 \ifnum #1\expandafter \@firstoftwo
 \else \expandafter \@secondoftwo
 \fi
}%
\providecommand \@ifx [1]{%
 \ifx #1\expandafter \@firstoftwo
 \else \expandafter \@secondoftwo
 \fi
}%
\providecommand \natexlab [1]{#1}%
\providecommand \enquote  [1]{``#1''}%
\providecommand \bibnamefont  [1]{#1}%
\providecommand \bibfnamefont [1]{#1}%
\providecommand \citenamefont [1]{#1}%
\providecommand \href@noop [0]{\@secondoftwo}%
\providecommand \href [0]{\begingroup \@sanitize@url \@href}%
\providecommand \@href[1]{\@@startlink{#1}\@@href}%
\providecommand \@@href[1]{\endgroup#1\@@endlink}%
\providecommand \@sanitize@url [0]{\catcode `\\12\catcode `\$12\catcode
  `\&12\catcode `\#12\catcode `\^12\catcode `\_12\catcode `\%12\relax}%
\providecommand \@@startlink[1]{}%
\providecommand \@@endlink[0]{}%
\providecommand \url  [0]{\begingroup\@sanitize@url \@url }%
\providecommand \@url [1]{\endgroup\@href {#1}{\urlprefix }}%
\providecommand \urlprefix  [0]{URL }%
\providecommand \Eprint [0]{\href }%
\providecommand \doibase [0]{https://doi.org/}%
\providecommand \selectlanguage [0]{\@gobble}%
\providecommand \bibinfo  [0]{\@secondoftwo}%
\providecommand \bibfield  [0]{\@secondoftwo}%
\providecommand \translation [1]{[#1]}%
\providecommand \BibitemOpen [0]{}%
\providecommand \bibitemStop [0]{}%
\providecommand \bibitemNoStop [0]{.\EOS\space}%
\providecommand \EOS [0]{\spacefactor3000\relax}%
\providecommand \BibitemShut  [1]{\csname bibitem#1\endcsname}%
\let\auto@bib@innerbib\@empty
%</preamble>
\bibitem [{\citenamefont {Hastings}\ and\ \citenamefont
  {Corliss}(1970)}]{Hastings70}%
  \BibitemOpen
  \bibfield  {author} {\bibinfo {author} {\bibfnamefont {J.~M.}\ \bibnamefont
  {Hastings}}\ and\ \bibinfo {author} {\bibfnamefont {L.~M.}\ \bibnamefont
  {Corliss}},\ }\bibfield  {title} {\bibinfo {title} {Ordered moment of nis2},\
  }\href {https://doi.org/10.1147/rd.143.0227} {\bibfield  {journal} {\bibinfo
  {journal} {IBM Journal of Research and Development}\ }\textbf {\bibinfo
  {volume} {14}},\ \bibinfo {pages} {227} (\bibinfo {year} {1970})}\BibitemShut
  {NoStop}%
\bibitem [{\citenamefont {Nishihara}\ \emph {et~al.}(1975)\citenamefont
  {Nishihara}, \citenamefont {Ogawa},\ and\ \citenamefont
  {Waki}}]{Nishihara75}%
  \BibitemOpen
  \bibfield  {author} {\bibinfo {author} {\bibfnamefont {Y.}~\bibnamefont
  {Nishihara}}, \bibinfo {author} {\bibfnamefont {S.}~\bibnamefont {Ogawa}},\
  and\ \bibinfo {author} {\bibfnamefont {S.}~\bibnamefont {Waki}},\ }\bibfield
  {title} {\bibinfo {title} {Mössbauer study of ni0.99557fe0.005s2–magnetic
  structure of nis2–},\ }\href {https://doi.org/10.1143/JPSJ.39.63}
  {\bibfield  {journal} {\bibinfo  {journal} {Journal of the Physical Society
  of Japan}\ }\textbf {\bibinfo {volume} {39}},\ \bibinfo {pages} {63}
  (\bibinfo {year} {1975})}\BibitemShut {NoStop}%
\bibitem [{\citenamefont {Nagata}\ \emph {et~al.}(1976)\citenamefont {Nagata},
  \citenamefont {Ito},\ and\ \citenamefont {Miyadai}}]{Nagata76}%
  \BibitemOpen
  \bibfield  {author} {\bibinfo {author} {\bibfnamefont {H.}~\bibnamefont
  {Nagata}}, \bibinfo {author} {\bibfnamefont {H.}~\bibnamefont {Ito}},\ and\
  \bibinfo {author} {\bibfnamefont {T.}~\bibnamefont {Miyadai}},\ }\bibfield
  {title} {\bibinfo {title} {Thermal expansion and crystal distortion of
  nis2},\ }\href {https://doi.org/10.1143/JPSJ.41.2133} {\bibfield  {journal}
  {\bibinfo  {journal} {Journal of the Physical Society of Japan}\ }\textbf
  {\bibinfo {volume} {41}},\ \bibinfo {pages} {2133} (\bibinfo {year}
  {1976})}\BibitemShut {NoStop}%
\bibitem [{\citenamefont {Kikuchi}\ \emph {et~al.}(1978)\citenamefont
  {Kikuchi}, \citenamefont {Miyadai}, \citenamefont {Fukui}, \citenamefont
  {It\^{o}},\ and\ \citenamefont {Takizawa}}]{Kikuchi78}%
  \BibitemOpen
  \bibfield  {author} {\bibinfo {author} {\bibfnamefont {K.}~\bibnamefont
  {Kikuchi}}, \bibinfo {author} {\bibfnamefont {T.}~\bibnamefont {Miyadai}},
  \bibinfo {author} {\bibfnamefont {T.}~\bibnamefont {Fukui}}, \bibinfo
  {author} {\bibfnamefont {H.}~\bibnamefont {It\^{o}}},\ and\ \bibinfo {author}
  {\bibfnamefont {K.}~\bibnamefont {Takizawa}},\ }\bibfield  {title} {\bibinfo
  {title} {Spin structure and magnetic properties of nis2},\ }\href
  {https://doi.org/10.1143/JPSJ.44.410} {\bibfield  {journal} {\bibinfo
  {journal} {Journal of the Physical Society of Japan}\ }\textbf {\bibinfo
  {volume} {44}},\ \bibinfo {pages} {410} (\bibinfo {year} {1978})}\BibitemShut
  {NoStop}%
\bibitem [{\citenamefont {Higo}\ and\ \citenamefont
  {Nakatsuji}(2015)}]{Higo2015}%
  \BibitemOpen
  \bibfield  {author} {\bibinfo {author} {\bibfnamefont {T.}~\bibnamefont
  {Higo}}\ and\ \bibinfo {author} {\bibfnamefont {S.}~\bibnamefont
  {Nakatsuji}},\ }\bibfield  {title} {\bibinfo {title} {{Magnetization Anomaly
  due to the Non-Coplanar Spin Structure in NiS2}},\ }\href
  {https://doi.org/10.7566/JPSJ.84.053702} {\bibfield  {journal} {\bibinfo
  {journal} {Journal of the Physical Society of Japan}\ }\textbf {\bibinfo
  {volume} {84}},\ \bibinfo {pages} {053702} (\bibinfo {year}
  {2015})}\BibitemShut {NoStop}%
\bibitem [{\citenamefont {Yano}\ \emph {et~al.}(2016)\citenamefont {Yano},
  \citenamefont {Louca}, \citenamefont {Yang}, \citenamefont {Chatterjee},
  \citenamefont {Bugaris}, \citenamefont {Chung}, \citenamefont {Peng},
  \citenamefont {Grayson},\ and\ \citenamefont {Kanatzidis}}]{Yano16}%
  \BibitemOpen
  \bibfield  {author} {\bibinfo {author} {\bibfnamefont {S.}~\bibnamefont
  {Yano}}, \bibinfo {author} {\bibfnamefont {D.}~\bibnamefont {Louca}},
  \bibinfo {author} {\bibfnamefont {J.}~\bibnamefont {Yang}}, \bibinfo {author}
  {\bibfnamefont {U.}~\bibnamefont {Chatterjee}}, \bibinfo {author}
  {\bibfnamefont {D.~E.}\ \bibnamefont {Bugaris}}, \bibinfo {author}
  {\bibfnamefont {D.~Y.}\ \bibnamefont {Chung}}, \bibinfo {author}
  {\bibfnamefont {L.}~\bibnamefont {Peng}}, \bibinfo {author} {\bibfnamefont
  {M.}~\bibnamefont {Grayson}},\ and\ \bibinfo {author} {\bibfnamefont {M.~G.}\
  \bibnamefont {Kanatzidis}},\ }\bibfield  {title} {\bibinfo {title} {Magnetic
  structure of ${\mathrm{nis}}_{2\ensuremath{-}x}{\mathrm{se}}_{x}$},\ }\href
  {https://doi.org/10.1103/PhysRevB.93.024409} {\bibfield  {journal} {\bibinfo
  {journal} {Phys. Rev. B}\ }\textbf {\bibinfo {volume} {93}},\ \bibinfo
  {pages} {024409} (\bibinfo {year} {2016})}\BibitemShut {NoStop}%
\bibitem [{\citenamefont {El-Khatib}\ \emph {et~al.}(2021)\citenamefont
  {El-Khatib}, \citenamefont {Voigt}, \citenamefont {Das}, \citenamefont
  {Stahl}, \citenamefont {Moore}, \citenamefont {Maiti},\ and\ \citenamefont
  {Leighton}}]{El-khatib2021}%
  \BibitemOpen
  \bibfield  {author} {\bibinfo {author} {\bibfnamefont {S.}~\bibnamefont
  {El-Khatib}}, \bibinfo {author} {\bibfnamefont {B.}~\bibnamefont {Voigt}},
  \bibinfo {author} {\bibfnamefont {B.}~\bibnamefont {Das}}, \bibinfo {author}
  {\bibfnamefont {A.}~\bibnamefont {Stahl}}, \bibinfo {author} {\bibfnamefont
  {W.}~\bibnamefont {Moore}}, \bibinfo {author} {\bibfnamefont
  {M.}~\bibnamefont {Maiti}},\ and\ \bibinfo {author} {\bibfnamefont
  {C.}~\bibnamefont {Leighton}},\ }\bibfield  {title} {\bibinfo {title}
  {Conduction via surface states in antiferromagnetic mott-insulating
  $\mathrm{Ni}{\mathrm{s}}_{2}$ single crystals},\ }\href
  {https://doi.org/10.1103/PhysRevMaterials.5.115003} {\bibfield  {journal}
  {\bibinfo  {journal} {Phys. Rev. Mater.}\ }\textbf {\bibinfo {volume} {5}},\
  \bibinfo {pages} {115003} (\bibinfo {year} {2021})}\BibitemShut {NoStop}%
\bibitem [{\citenamefont {Yao}\ \emph {et~al.}(1996)\citenamefont {Yao},
  \citenamefont {Honig}, \citenamefont {Hogan}, \citenamefont {Kannewurf},\
  and\ \citenamefont {Spa\l{}ek}}]{Yao1996}%
  \BibitemOpen
  \bibfield  {author} {\bibinfo {author} {\bibfnamefont {X.}~\bibnamefont
  {Yao}}, \bibinfo {author} {\bibfnamefont {J.~M.}\ \bibnamefont {Honig}},
  \bibinfo {author} {\bibfnamefont {T.}~\bibnamefont {Hogan}}, \bibinfo
  {author} {\bibfnamefont {C.}~\bibnamefont {Kannewurf}},\ and\ \bibinfo
  {author} {\bibfnamefont {J.}~\bibnamefont {Spa\l{}ek}},\ }\bibfield  {title}
  {\bibinfo {title} {Electrical properties of
  $\mathrm{Ni}{\mathrm{s}}_{2\ensuremath{-}x}{\mathrm{se}}_{x}$ single
  crystals: From mott insulator to paramagnetic metal},\ }\href
  {https://doi.org/10.1103/PhysRevB.54.17469} {\bibfield  {journal} {\bibinfo
  {journal} {Phys. Rev. B}\ }\textbf {\bibinfo {volume} {54}},\ \bibinfo
  {pages} {17469} (\bibinfo {year} {1996})}\BibitemShut {NoStop}%
\bibitem [{\citenamefont {Kune\ifmmode~\check{s}\else \v{s}\fi{}}\ \emph
  {et~al.}(2010)\citenamefont {Kune\ifmmode~\check{s}\else \v{s}\fi{}},
  \citenamefont {Baldassarre}, \citenamefont {Sch\"achner}, \citenamefont
  {Rabia}, \citenamefont {Kuntscher}, \citenamefont {Korotin}, \citenamefont
  {Anisimov}, \citenamefont {McLeod}, \citenamefont {Kurmaev},\ and\
  \citenamefont {Moewes}}]{Kunes2010}%
  \BibitemOpen
  \bibfield  {author} {\bibinfo {author} {\bibfnamefont {J.}~\bibnamefont
  {Kune\ifmmode~\check{s}\else \v{s}\fi{}}}, \bibinfo {author} {\bibfnamefont
  {L.}~\bibnamefont {Baldassarre}}, \bibinfo {author} {\bibfnamefont
  {B.}~\bibnamefont {Sch\"achner}}, \bibinfo {author} {\bibfnamefont
  {K.}~\bibnamefont {Rabia}}, \bibinfo {author} {\bibfnamefont {C.~A.}\
  \bibnamefont {Kuntscher}}, \bibinfo {author} {\bibfnamefont {D.~M.}\
  \bibnamefont {Korotin}}, \bibinfo {author} {\bibfnamefont {V.~I.}\
  \bibnamefont {Anisimov}}, \bibinfo {author} {\bibfnamefont {J.~A.}\
  \bibnamefont {McLeod}}, \bibinfo {author} {\bibfnamefont {E.~Z.}\
  \bibnamefont {Kurmaev}},\ and\ \bibinfo {author} {\bibfnamefont
  {A.}~\bibnamefont {Moewes}},\ }\bibfield  {title} {\bibinfo {title}
  {Metal-insulator transition in
  ${\text{nis}}_{2\ensuremath{-}x}{\text{se}}_{x}$},\ }\href
  {https://doi.org/10.1103/PhysRevB.81.035122} {\bibfield  {journal} {\bibinfo
  {journal} {Phys. Rev. B}\ }\textbf {\bibinfo {volume} {81}},\ \bibinfo
  {pages} {035122} (\bibinfo {year} {2010})}\BibitemShut {NoStop}%
\bibitem [{\citenamefont {Xu}\ \emph {et~al.}(2014)\citenamefont {Xu},
  \citenamefont {Zhang}, \citenamefont {Xu}, \citenamefont {Peng},
  \citenamefont {Shen}, \citenamefont {Strocov}, \citenamefont {Shi},
  \citenamefont {Kobayashi}, \citenamefont {Schmitt}, \citenamefont {Xie},\
  and\ \citenamefont {Feng}}]{Xu2014}%
  \BibitemOpen
  \bibfield  {author} {\bibinfo {author} {\bibfnamefont {H.~C.}\ \bibnamefont
  {Xu}}, \bibinfo {author} {\bibfnamefont {Y.}~\bibnamefont {Zhang}}, \bibinfo
  {author} {\bibfnamefont {M.}~\bibnamefont {Xu}}, \bibinfo {author}
  {\bibfnamefont {R.}~\bibnamefont {Peng}}, \bibinfo {author} {\bibfnamefont
  {X.~P.}\ \bibnamefont {Shen}}, \bibinfo {author} {\bibfnamefont {V.~N.}\
  \bibnamefont {Strocov}}, \bibinfo {author} {\bibfnamefont {M.}~\bibnamefont
  {Shi}}, \bibinfo {author} {\bibfnamefont {M.}~\bibnamefont {Kobayashi}},
  \bibinfo {author} {\bibfnamefont {T.}~\bibnamefont {Schmitt}}, \bibinfo
  {author} {\bibfnamefont {B.~P.}\ \bibnamefont {Xie}},\ and\ \bibinfo {author}
  {\bibfnamefont {D.~L.}\ \bibnamefont {Feng}},\ }\bibfield  {title} {\bibinfo
  {title} {Direct observation of the bandwidth control mott transition in the
  ${\mathrm{nis}}_{2\ensuremath{-}x}{\mathrm{se}}_{x}$ multiband system},\
  }\href {https://doi.org/10.1103/PhysRevLett.112.087603} {\bibfield  {journal}
  {\bibinfo  {journal} {Phys. Rev. Lett.}\ }\textbf {\bibinfo {volume} {112}},\
  \bibinfo {pages} {087603} (\bibinfo {year} {2014})}\BibitemShut {NoStop}%
\bibitem [{\citenamefont {Moon}\ \emph {et~al.}(2015)\citenamefont {Moon},
  \citenamefont {Kang}, \citenamefont {Jang},\ and\ \citenamefont
  {Shim}}]{Moon2015}%
  \BibitemOpen
  \bibfield  {author} {\bibinfo {author} {\bibfnamefont {C.-Y.}\ \bibnamefont
  {Moon}}, \bibinfo {author} {\bibfnamefont {H.}~\bibnamefont {Kang}}, \bibinfo
  {author} {\bibfnamefont {B.~G.}\ \bibnamefont {Jang}},\ and\ \bibinfo
  {author} {\bibfnamefont {J.~H.}\ \bibnamefont {Shim}},\ }\bibfield  {title}
  {\bibinfo {title} {{Composition and temperature dependent electronic
  structures of ${\mathrm{NiS}}_{2\ensuremath{-}x}{\mathrm{Se}}_{x}$ alloys:
  First-principles dynamical mean-field theory approach}},\ }\href
  {https://doi.org/10.1103/PhysRevB.92.235130} {\bibfield  {journal} {\bibinfo
  {journal} {Phys. Rev. B}\ }\textbf {\bibinfo {volume} {92}},\ \bibinfo
  {pages} {235130} (\bibinfo {year} {2015})}\BibitemShut {NoStop}%
\bibitem [{\citenamefont {Han}\ \emph {et~al.}(2018)\citenamefont {Han},
  \citenamefont {Choi}, \citenamefont {Cho}, \citenamefont {Sohn},
  \citenamefont {Park},\ and\ \citenamefont {Kim}}]{Han2018}%
  \BibitemOpen
  \bibfield  {author} {\bibinfo {author} {\bibfnamefont {G.}~\bibnamefont
  {Han}}, \bibinfo {author} {\bibfnamefont {S.}~\bibnamefont {Choi}}, \bibinfo
  {author} {\bibfnamefont {H.}~\bibnamefont {Cho}}, \bibinfo {author}
  {\bibfnamefont {B.}~\bibnamefont {Sohn}}, \bibinfo {author} {\bibfnamefont
  {J.-G.}\ \bibnamefont {Park}},\ and\ \bibinfo {author} {\bibfnamefont
  {C.}~\bibnamefont {Kim}},\ }\bibfield  {title} {\bibinfo {title} {Structural
  investigation of the insulator-metal transition in
  ${\mathrm{nis}}_{2\ensuremath{-}x}{\mathrm{se}}_{x}$ compounds},\ }\href
  {https://doi.org/10.1103/PhysRevB.98.125114} {\bibfield  {journal} {\bibinfo
  {journal} {Phys. Rev. B}\ }\textbf {\bibinfo {volume} {98}},\ \bibinfo
  {pages} {125114} (\bibinfo {year} {2018})}\BibitemShut {NoStop}%
\bibitem [{\citenamefont {Jang}\ \emph {et~al.}(2021)\citenamefont {Jang},
  \citenamefont {Han}, \citenamefont {Park}, \citenamefont {Kim}, \citenamefont
  {Koh}, \citenamefont {Kim}, \citenamefont {Kyung}, \citenamefont {Kim},
  \citenamefont {Cheng}, \citenamefont {Tsuei}, \citenamefont {Lee},
  \citenamefont {Hur}, \citenamefont {Shim}, \citenamefont {Kim},\ and\
  \citenamefont {Kotliar}}]{Jang2021}%
  \BibitemOpen
  \bibfield  {author} {\bibinfo {author} {\bibfnamefont {B.~G.}\ \bibnamefont
  {Jang}}, \bibinfo {author} {\bibfnamefont {G.}~\bibnamefont {Han}}, \bibinfo
  {author} {\bibfnamefont {I.}~\bibnamefont {Park}}, \bibinfo {author}
  {\bibfnamefont {D.}~\bibnamefont {Kim}}, \bibinfo {author} {\bibfnamefont
  {Y.~Y.}\ \bibnamefont {Koh}}, \bibinfo {author} {\bibfnamefont
  {Y.}~\bibnamefont {Kim}}, \bibinfo {author} {\bibfnamefont {W.}~\bibnamefont
  {Kyung}}, \bibinfo {author} {\bibfnamefont {H.-D.}\ \bibnamefont {Kim}},
  \bibinfo {author} {\bibfnamefont {C.-M.}\ \bibnamefont {Cheng}}, \bibinfo
  {author} {\bibfnamefont {K.-D.}\ \bibnamefont {Tsuei}}, \bibinfo {author}
  {\bibfnamefont {K.~D.}\ \bibnamefont {Lee}}, \bibinfo {author} {\bibfnamefont
  {N.}~\bibnamefont {Hur}}, \bibinfo {author} {\bibfnamefont {J.~H.}\
  \bibnamefont {Shim}}, \bibinfo {author} {\bibfnamefont {C.}~\bibnamefont
  {Kim}},\ and\ \bibinfo {author} {\bibfnamefont {G.}~\bibnamefont {Kotliar}},\
  }\bibfield  {title} {\bibinfo {title} {Direct observation of kink evolution
  due to hund's coupling on approach to metal-insulator transition in
  ${\mathrm{nis}}_{2\ensuremath{-}x}{\mathrm{se}}_{x}$},\ }\href
  {https://doi.org/10.1038/s41467-021-21460-5} {\bibfield  {journal} {\bibinfo
  {journal} {Nature Communications}\ }\textbf {\bibinfo {volume} {12}},\
  \bibinfo {pages} {1208} (\bibinfo {year} {2021})}\BibitemShut {NoStop}%
\bibitem [{\citenamefont {{Schuster}}\ \emph {et~al.}(2012)\citenamefont
  {{Schuster}}, \citenamefont {{Gatti}},\ and\ \citenamefont
  {{Rubio}}}]{Schuster2012}%
  \BibitemOpen
  \bibfield  {author} {\bibinfo {author} {\bibfnamefont {C.}~\bibnamefont
  {{Schuster}}}, \bibinfo {author} {\bibfnamefont {M.}~\bibnamefont
  {{Gatti}}},\ and\ \bibinfo {author} {\bibfnamefont {A.}~\bibnamefont
  {{Rubio}}},\ }\bibfield  {title} {\bibinfo {title} {{Electronic and magnetic
  properties of NiS$_{2}$, NiSSe and NiSe$_{2}$ by a combination of theoretical
  methods}},\ }\href {https://doi.org/10.1140/epjb/e2012-30384-7} {\bibfield
  {journal} {\bibinfo  {journal} {European Physical Journal B}\ }\textbf
  {\bibinfo {volume} {85}},\ \bibinfo {eid} {325} (\bibinfo {year}
  {2012})}\BibitemShut {NoStop}%
\bibitem [{\citenamefont {Park}\ \emph {et~al.}(2024)\citenamefont {Park},
  \citenamefont {Jang}, \citenamefont {Kim}, \citenamefont {Shim},\ and\
  \citenamefont {Kotliar}}]{Park2024}%
  \BibitemOpen
  \bibfield  {author} {\bibinfo {author} {\bibfnamefont {I.}~\bibnamefont
  {Park}}, \bibinfo {author} {\bibfnamefont {B.~G.}\ \bibnamefont {Jang}},
  \bibinfo {author} {\bibfnamefont {D.~W.}\ \bibnamefont {Kim}}, \bibinfo
  {author} {\bibfnamefont {J.~H.}\ \bibnamefont {Shim}},\ and\ \bibinfo
  {author} {\bibfnamefont {G.}~\bibnamefont {Kotliar}},\ }\bibfield  {title}
  {\bibinfo {title} {{Clean realization of Hund's physics near the Mott
  transition: ${\mathrm{NiS}}_{2}$ under pressure}},\ }\href
  {https://doi.org/10.1103/PhysRevB.109.045146} {\bibfield  {journal} {\bibinfo
   {journal} {Phys. Rev. B}\ }\textbf {\bibinfo {volume} {109}},\ \bibinfo
  {pages} {045146} (\bibinfo {year} {2024})}\BibitemShut {NoStop}%
\bibitem [{\citenamefont {Friedemann}\ \emph {et~al.}(2016)\citenamefont
  {Friedemann}, \citenamefont {Chang}, \citenamefont {Gam{\.{z}}a},
  \citenamefont {Reiss}, \citenamefont {Chen}, \citenamefont {Alireza},
  \citenamefont {Coniglio}, \citenamefont {Graf}, \citenamefont {Tozer},\ and\
  \citenamefont {Grosche}}]{Friedemann2016}%
  \BibitemOpen
  \bibfield  {author} {\bibinfo {author} {\bibfnamefont {S.}~\bibnamefont
  {Friedemann}}, \bibinfo {author} {\bibfnamefont {H.}~\bibnamefont {Chang}},
  \bibinfo {author} {\bibfnamefont {M.~B.}\ \bibnamefont {Gam{\.{z}}a}},
  \bibinfo {author} {\bibfnamefont {P.}~\bibnamefont {Reiss}}, \bibinfo
  {author} {\bibfnamefont {X.}~\bibnamefont {Chen}}, \bibinfo {author}
  {\bibfnamefont {P.}~\bibnamefont {Alireza}}, \bibinfo {author} {\bibfnamefont
  {W.~A.}\ \bibnamefont {Coniglio}}, \bibinfo {author} {\bibfnamefont
  {D.}~\bibnamefont {Graf}}, \bibinfo {author} {\bibfnamefont {S.}~\bibnamefont
  {Tozer}},\ and\ \bibinfo {author} {\bibfnamefont {F.~M.}\ \bibnamefont
  {Grosche}},\ }\bibfield  {title} {\bibinfo {title} {Large fermi surface of
  heavy electrons at the border of mott insulating state in nis2},\ }\href
  {https://doi.org/10.1038/srep25335} {\bibfield  {journal} {\bibinfo
  {journal} {Scientific Reports}\ }\textbf {\bibinfo {volume} {6}},\ \bibinfo
  {pages} {25335} (\bibinfo {year} {2016})}\BibitemShut {NoStop}%
\bibitem [{\citenamefont {Day-Roberts}\ \emph {et~al.}(2023)\citenamefont
  {Day-Roberts}, \citenamefont {Fernandes},\ and\ \citenamefont
  {Birol}}]{Day-Roberts2023}%
  \BibitemOpen
  \bibfield  {author} {\bibinfo {author} {\bibfnamefont {E.}~\bibnamefont
  {Day-Roberts}}, \bibinfo {author} {\bibfnamefont {R.~M.}\ \bibnamefont
  {Fernandes}},\ and\ \bibinfo {author} {\bibfnamefont {T.}~\bibnamefont
  {Birol}},\ }\bibfield  {title} {\bibinfo {title} {{Gating-induced Mott
  transition in ${\mathrm{NiS}}_{2}$}},\ }\href
  {https://doi.org/10.1103/PhysRevB.107.085150} {\bibfield  {journal} {\bibinfo
   {journal} {Phys. Rev. B}\ }\textbf {\bibinfo {volume} {107}},\ \bibinfo
  {pages} {085150} (\bibinfo {year} {2023})}\BibitemShut {NoStop}%
\bibitem [{\citenamefont {Fujimori}\ \emph {et~al.}(1996)\citenamefont
  {Fujimori}, \citenamefont {Mamiya}, \citenamefont {Mizokawa}, \citenamefont
  {Miyadai}, \citenamefont {Sekiguchi}, \citenamefont {Takahashi},
  \citenamefont {M\^ori},\ and\ \citenamefont {Suga}}]{Fujimori1996}%
  \BibitemOpen
  \bibfield  {author} {\bibinfo {author} {\bibfnamefont {A.}~\bibnamefont
  {Fujimori}}, \bibinfo {author} {\bibfnamefont {K.}~\bibnamefont {Mamiya}},
  \bibinfo {author} {\bibfnamefont {T.}~\bibnamefont {Mizokawa}}, \bibinfo
  {author} {\bibfnamefont {T.}~\bibnamefont {Miyadai}}, \bibinfo {author}
  {\bibfnamefont {T.}~\bibnamefont {Sekiguchi}}, \bibinfo {author}
  {\bibfnamefont {H.}~\bibnamefont {Takahashi}}, \bibinfo {author}
  {\bibfnamefont {N.}~\bibnamefont {M\^ori}},\ and\ \bibinfo {author}
  {\bibfnamefont {S.}~\bibnamefont {Suga}},\ }\bibfield  {title} {\bibinfo
  {title} {Resonant photoemission study of pyrite-type ${\mathrm{nis}}_{2}$,
  ${\mathrm{cos}}_{2}$ and ${\mathrm{fes}}_{2}$},\ }\href
  {https://doi.org/10.1103/PhysRevB.54.16329} {\bibfield  {journal} {\bibinfo
  {journal} {Phys. Rev. B}\ }\textbf {\bibinfo {volume} {54}},\ \bibinfo
  {pages} {16329} (\bibinfo {year} {1996})}\BibitemShut {NoStop}%
\bibitem [{\citenamefont {Matsuura}\ \emph {et~al.}(1998)\citenamefont
  {Matsuura}, \citenamefont {Watanabe}, \citenamefont {Kim}, \citenamefont
  {Doniach}, \citenamefont {Shen}, \citenamefont {Thio},\ and\ \citenamefont
  {Bennett}}]{Matuura1998}%
  \BibitemOpen
  \bibfield  {author} {\bibinfo {author} {\bibfnamefont {A.~Y.}\ \bibnamefont
  {Matsuura}}, \bibinfo {author} {\bibfnamefont {H.}~\bibnamefont {Watanabe}},
  \bibinfo {author} {\bibfnamefont {C.}~\bibnamefont {Kim}}, \bibinfo {author}
  {\bibfnamefont {S.}~\bibnamefont {Doniach}}, \bibinfo {author} {\bibfnamefont
  {Z.-X.}\ \bibnamefont {Shen}}, \bibinfo {author} {\bibfnamefont
  {T.}~\bibnamefont {Thio}},\ and\ \bibinfo {author} {\bibfnamefont {J.~W.}\
  \bibnamefont {Bennett}},\ }\bibfield  {title} {\bibinfo {title}
  {Metal-insulator transition in
  ${\mathrm{nis}}_{2\ensuremath{-}x}{\mathrm{se}}_{x}$ and the local impurity
  self-consistent approximation model},\ }\href
  {https://doi.org/10.1103/PhysRevB.58.3690} {\bibfield  {journal} {\bibinfo
  {journal} {Phys. Rev. B}\ }\textbf {\bibinfo {volume} {58}},\ \bibinfo
  {pages} {3690} (\bibinfo {year} {1998})}\BibitemShut {NoStop}%
\bibitem [{\citenamefont {Iwaya}\ \emph {et~al.}(2004)\citenamefont {Iwaya},
  \citenamefont {Kohsaka}, \citenamefont {Satow}, \citenamefont {Hanaguri},
  \citenamefont {Miyasaka},\ and\ \citenamefont {Takagi}}]{Iwaya2004}%
  \BibitemOpen
  \bibfield  {author} {\bibinfo {author} {\bibfnamefont {K.}~\bibnamefont
  {Iwaya}}, \bibinfo {author} {\bibfnamefont {Y.}~\bibnamefont {Kohsaka}},
  \bibinfo {author} {\bibfnamefont {S.}~\bibnamefont {Satow}}, \bibinfo
  {author} {\bibfnamefont {T.}~\bibnamefont {Hanaguri}}, \bibinfo {author}
  {\bibfnamefont {S.}~\bibnamefont {Miyasaka}},\ and\ \bibinfo {author}
  {\bibfnamefont {H.}~\bibnamefont {Takagi}},\ }\bibfield  {title} {\bibinfo
  {title} {Evolution of local electronic states from a metal to a correlated
  insulator in a $\mathrm{Ni}{\mathrm{s}}_{2\ensuremath{-}x}{\mathrm{se}}_{x}$
  solid solution},\ }\href {https://doi.org/10.1103/PhysRevB.70.161103}
  {\bibfield  {journal} {\bibinfo  {journal} {Phys. Rev. B}\ }\textbf {\bibinfo
  {volume} {70}},\ \bibinfo {pages} {161103} (\bibinfo {year}
  {2004})}\BibitemShut {NoStop}%
\bibitem [{\citenamefont {Thio}\ and\ \citenamefont
  {Bennett}(1994)}]{Thio1994}%
  \BibitemOpen
  \bibfield  {author} {\bibinfo {author} {\bibfnamefont {T.}~\bibnamefont
  {Thio}}\ and\ \bibinfo {author} {\bibfnamefont {J.~W.}\ \bibnamefont
  {Bennett}},\ }\bibfield  {title} {\bibinfo {title} {{Hall effect and
  conductivity in pyrite ${\mathrm{NiS}}_{2}$}},\ }\href
  {https://doi.org/10.1103/PhysRevB.50.10574} {\bibfield  {journal} {\bibinfo
  {journal} {Phys. Rev. B}\ }\textbf {\bibinfo {volume} {50}},\ \bibinfo
  {pages} {10574} (\bibinfo {year} {1994})}\BibitemShut {NoStop}%
\bibitem [{\citenamefont {Sarma}\ \emph {et~al.}(2003)\citenamefont {Sarma},
  \citenamefont {Krishnakumar}, \citenamefont {Weschke}, \citenamefont
  {Sch\"u\ss{}ler-Langeheine}, \citenamefont {Mazumdar}, \citenamefont
  {Kilian}, \citenamefont {Kaindl}, \citenamefont {Mamiya}, \citenamefont
  {Fujimori}, \citenamefont {Fujimori},\ and\ \citenamefont
  {Miyadai}}]{Sarma2003}%
  \BibitemOpen
  \bibfield  {author} {\bibinfo {author} {\bibfnamefont {D.~D.}\ \bibnamefont
  {Sarma}}, \bibinfo {author} {\bibfnamefont {S.~R.}\ \bibnamefont
  {Krishnakumar}}, \bibinfo {author} {\bibfnamefont {E.}~\bibnamefont
  {Weschke}}, \bibinfo {author} {\bibfnamefont {C.}~\bibnamefont
  {Sch\"u\ss{}ler-Langeheine}}, \bibinfo {author} {\bibfnamefont
  {C.}~\bibnamefont {Mazumdar}}, \bibinfo {author} {\bibfnamefont
  {L.}~\bibnamefont {Kilian}}, \bibinfo {author} {\bibfnamefont
  {G.}~\bibnamefont {Kaindl}}, \bibinfo {author} {\bibfnamefont
  {K.}~\bibnamefont {Mamiya}}, \bibinfo {author} {\bibfnamefont {S.-I.}\
  \bibnamefont {Fujimori}}, \bibinfo {author} {\bibfnamefont {A.}~\bibnamefont
  {Fujimori}},\ and\ \bibinfo {author} {\bibfnamefont {T.}~\bibnamefont
  {Miyadai}},\ }\bibfield  {title} {\bibinfo {title} {{Metal-insulator
  crossover behavior at the surface of ${\mathrm{NiS}}_{2}$}},\ }\href
  {https://doi.org/10.1103/PhysRevB.67.155112} {\bibfield  {journal} {\bibinfo
  {journal} {Phys. Rev. B}\ }\textbf {\bibinfo {volume} {67}},\ \bibinfo
  {pages} {155112} (\bibinfo {year} {2003})}\BibitemShut {NoStop}%
\bibitem [{\citenamefont {Rao}\ \emph {et~al.}(2011)\citenamefont {Rao},
  \citenamefont {Bhuvana}, \citenamefont {Radha}, \citenamefont {Kurra},
  \citenamefont {Vidhyadhiraja},\ and\ \citenamefont {Kulkarni}}]{Rao2011}%
  \BibitemOpen
  \bibfield  {author} {\bibinfo {author} {\bibfnamefont {K.~D.~M.}\
  \bibnamefont {Rao}}, \bibinfo {author} {\bibfnamefont {T.}~\bibnamefont
  {Bhuvana}}, \bibinfo {author} {\bibfnamefont {B.}~\bibnamefont {Radha}},
  \bibinfo {author} {\bibfnamefont {N.}~\bibnamefont {Kurra}}, \bibinfo
  {author} {\bibfnamefont {N.~S.}\ \bibnamefont {Vidhyadhiraja}},\ and\
  \bibinfo {author} {\bibfnamefont {G.~U.}\ \bibnamefont {Kulkarni}},\
  }\bibfield  {title} {\bibinfo {title} {Metallic conduction in nis2
  nanocrystalline structures},\ }\href {https://doi.org/10.1021/jp201740q}
  {\bibfield  {journal} {\bibinfo  {journal} {The Journal of Physical Chemistry
  C}\ }\textbf {\bibinfo {volume} {115}},\ \bibinfo {pages} {10462} (\bibinfo
  {year} {2011})}\BibitemShut {NoStop}%
\bibitem [{\citenamefont {Clark}\ and\ \citenamefont
  {Friedemann}(2016)}]{Clak2016}%
  \BibitemOpen
  \bibfield  {author} {\bibinfo {author} {\bibfnamefont {C.}~\bibnamefont
  {Clark}}\ and\ \bibinfo {author} {\bibfnamefont {S.}~\bibnamefont
  {Friedemann}},\ }\bibfield  {title} {\bibinfo {title} {Atomic diffusion in
  the surface state of mott insulator nis2},\ }\href
  {https://doi.org/https://doi.org/10.1016/j.jmmm.2015.08.012} {\bibfield
  {journal} {\bibinfo  {journal} {Journal of Magnetism and Magnetic Materials}\
  }\textbf {\bibinfo {volume} {400}},\ \bibinfo {pages} {56} (\bibinfo {year}
  {2016})},\ \bibinfo {note} {proceedings of the 20th International Conference
  on Magnetism (Barcelona) 5-10 July 2015}\BibitemShut {NoStop}%
\bibitem [{\citenamefont {El-Khatib}\ \emph {et~al.}(2023)\citenamefont
  {El-Khatib}, \citenamefont {Mustafa}, \citenamefont {Egilmez}, \citenamefont
  {Das}, \citenamefont {Tao}, \citenamefont {Maiti}, \citenamefont {Lee},\ and\
  \citenamefont {Leighton}}]{El-khatib2023}%
  \BibitemOpen
  \bibfield  {author} {\bibinfo {author} {\bibfnamefont {S.}~\bibnamefont
  {El-Khatib}}, \bibinfo {author} {\bibfnamefont {F.}~\bibnamefont {Mustafa}},
  \bibinfo {author} {\bibfnamefont {M.}~\bibnamefont {Egilmez}}, \bibinfo
  {author} {\bibfnamefont {B.}~\bibnamefont {Das}}, \bibinfo {author}
  {\bibfnamefont {Y.}~\bibnamefont {Tao}}, \bibinfo {author} {\bibfnamefont
  {M.}~\bibnamefont {Maiti}}, \bibinfo {author} {\bibfnamefont
  {Y.}~\bibnamefont {Lee}},\ and\ \bibinfo {author} {\bibfnamefont
  {C.}~\bibnamefont {Leighton}},\ }\bibfield  {title} {\bibinfo {title} {Exotic
  surface magnetotransport phenomena in the antiferromagnetic mott insulator
  $\mathrm{Ni}{\mathrm{s}}_{2}$},\ }\href
  {https://doi.org/10.1103/PhysRevMaterials.7.104401} {\bibfield  {journal}
  {\bibinfo  {journal} {Phys. Rev. Mater.}\ }\textbf {\bibinfo {volume} {7}},\
  \bibinfo {pages} {104401} (\bibinfo {year} {2023})}\BibitemShut {NoStop}%
\bibitem [{\citenamefont {Yao}\ and\ \citenamefont
  {Kivelson}(2010)}]{Kivelson2010}%
  \BibitemOpen
  \bibfield  {author} {\bibinfo {author} {\bibfnamefont {H.}~\bibnamefont
  {Yao}}\ and\ \bibinfo {author} {\bibfnamefont {S.~A.}\ \bibnamefont
  {Kivelson}},\ }\bibfield  {title} {\bibinfo {title} {Fragile mott
  insulators},\ }\href {https://doi.org/10.1103/PhysRevLett.105.166402}
  {\bibfield  {journal} {\bibinfo  {journal} {Phys. Rev. Lett.}\ }\textbf
  {\bibinfo {volume} {105}},\ \bibinfo {pages} {166402} (\bibinfo {year}
  {2010})}\BibitemShut {NoStop}%
\bibitem [{\citenamefont {Iraola}\ \emph {et~al.}(2021)\citenamefont {Iraola},
  \citenamefont {Heinsdorf}, \citenamefont {Tiwari}, \citenamefont {Lessnich},
  \citenamefont {Mertz}, \citenamefont {Ferrari}, \citenamefont {Fischer},
  \citenamefont {Winter}, \citenamefont {Pollmann}, \citenamefont {Neupert},
  \citenamefont {Valent\'{\i}},\ and\ \citenamefont {Vergniory}}]{Iraola2021}%
  \BibitemOpen
  \bibfield  {author} {\bibinfo {author} {\bibfnamefont {M.}~\bibnamefont
  {Iraola}}, \bibinfo {author} {\bibfnamefont {N.}~\bibnamefont {Heinsdorf}},
  \bibinfo {author} {\bibfnamefont {A.}~\bibnamefont {Tiwari}}, \bibinfo
  {author} {\bibfnamefont {D.}~\bibnamefont {Lessnich}}, \bibinfo {author}
  {\bibfnamefont {T.}~\bibnamefont {Mertz}}, \bibinfo {author} {\bibfnamefont
  {F.}~\bibnamefont {Ferrari}}, \bibinfo {author} {\bibfnamefont {M.~H.}\
  \bibnamefont {Fischer}}, \bibinfo {author} {\bibfnamefont {S.~M.}\
  \bibnamefont {Winter}}, \bibinfo {author} {\bibfnamefont {F.}~\bibnamefont
  {Pollmann}}, \bibinfo {author} {\bibfnamefont {T.}~\bibnamefont {Neupert}},
  \bibinfo {author} {\bibfnamefont {R.}~\bibnamefont {Valent\'{\i}}},\ and\
  \bibinfo {author} {\bibfnamefont {M.~G.}\ \bibnamefont {Vergniory}},\
  }\bibfield  {title} {\bibinfo {title} {Towards a topological quantum
  chemistry description of correlated systems: The case of the hubbard diamond
  chain},\ }\href {https://doi.org/10.1103/PhysRevB.104.195125} {\bibfield
  {journal} {\bibinfo  {journal} {Phys. Rev. B}\ }\textbf {\bibinfo {volume}
  {104}},\ \bibinfo {pages} {195125} (\bibinfo {year} {2021})}\BibitemShut
  {NoStop}%
\bibitem [{\citenamefont {Soldini}\ \emph {et~al.}(2023)\citenamefont
  {Soldini}, \citenamefont {Astrakhantsev}, \citenamefont {Iraola},
  \citenamefont {Tiwari}, \citenamefont {Fischer}, \citenamefont
  {Valent\'{\i}}, \citenamefont {Vergniory}, \citenamefont {Wagner},\ and\
  \citenamefont {Neupert}}]{Soldini2023}%
  \BibitemOpen
  \bibfield  {author} {\bibinfo {author} {\bibfnamefont {M.~O.}\ \bibnamefont
  {Soldini}}, \bibinfo {author} {\bibfnamefont {N.}~\bibnamefont
  {Astrakhantsev}}, \bibinfo {author} {\bibfnamefont {M.}~\bibnamefont
  {Iraola}}, \bibinfo {author} {\bibfnamefont {A.}~\bibnamefont {Tiwari}},
  \bibinfo {author} {\bibfnamefont {M.~H.}\ \bibnamefont {Fischer}}, \bibinfo
  {author} {\bibfnamefont {R.}~\bibnamefont {Valent\'{\i}}}, \bibinfo {author}
  {\bibfnamefont {M.~G.}\ \bibnamefont {Vergniory}}, \bibinfo {author}
  {\bibfnamefont {G.}~\bibnamefont {Wagner}},\ and\ \bibinfo {author}
  {\bibfnamefont {T.}~\bibnamefont {Neupert}},\ }\bibfield  {title} {\bibinfo
  {title} {Interacting topological quantum chemistry of mott atomic limits},\
  }\href {https://doi.org/10.1103/PhysRevB.107.245145} {\bibfield  {journal}
  {\bibinfo  {journal} {Phys. Rev. B}\ }\textbf {\bibinfo {volume} {107}},\
  \bibinfo {pages} {245145} (\bibinfo {year} {2023})}\BibitemShut {NoStop}%
\bibitem [{\citenamefont {Yasui}\ \emph {et~al.}(2024)\citenamefont {Yasui},
  \citenamefont {Iwata}, \citenamefont {Okazaki}, \citenamefont {Miyasaka},
  \citenamefont {Sugimoto}, \citenamefont {Hanaguri}, \citenamefont {Takagi},\
  and\ \citenamefont {Sasagawa}}]{Yasui2024}%
  \BibitemOpen
  \bibfield  {author} {\bibinfo {author} {\bibfnamefont {Y.}~\bibnamefont
  {Yasui}}, \bibinfo {author} {\bibfnamefont {K.}~\bibnamefont {Iwata}},
  \bibinfo {author} {\bibfnamefont {S.}~\bibnamefont {Okazaki}}, \bibinfo
  {author} {\bibfnamefont {S.}~\bibnamefont {Miyasaka}}, \bibinfo {author}
  {\bibfnamefont {Y.}~\bibnamefont {Sugimoto}}, \bibinfo {author}
  {\bibfnamefont {T.}~\bibnamefont {Hanaguri}}, \bibinfo {author}
  {\bibfnamefont {H.}~\bibnamefont {Takagi}},\ and\ \bibinfo {author}
  {\bibfnamefont {T.}~\bibnamefont {Sasagawa}},\ }\bibfield  {title} {\bibinfo
  {title} {Closing of the mott gap near step edges in ${\mathrm{nis}}_{2}$},\
  }\href {https://doi.org/10.1103/PhysRevB.110.045139} {\bibfield  {journal}
  {\bibinfo  {journal} {Phys. Rev. B}\ }\textbf {\bibinfo {volume} {110}},\
  \bibinfo {pages} {045139} (\bibinfo {year} {2024})}\BibitemShut {NoStop}%
\bibitem [{\citenamefont {Bradlyn}\ \emph {et~al.}(2017)\citenamefont
  {Bradlyn}, \citenamefont {Elcoro}, \citenamefont {Cano}, \citenamefont
  {Vergniory}, \citenamefont {Wang}, \citenamefont {Felser}, \citenamefont
  {Aroyo},\ and\ \citenamefont {Bernevig}}]{Bradlyn2017}%
  \BibitemOpen
  \bibfield  {author} {\bibinfo {author} {\bibfnamefont {B.}~\bibnamefont
  {Bradlyn}}, \bibinfo {author} {\bibfnamefont {L.}~\bibnamefont {Elcoro}},
  \bibinfo {author} {\bibfnamefont {J.}~\bibnamefont {Cano}}, \bibinfo {author}
  {\bibfnamefont {M.~G.}\ \bibnamefont {Vergniory}}, \bibinfo {author}
  {\bibfnamefont {Z.}~\bibnamefont {Wang}}, \bibinfo {author} {\bibfnamefont
  {C.}~\bibnamefont {Felser}}, \bibinfo {author} {\bibfnamefont {M.~I.}\
  \bibnamefont {Aroyo}},\ and\ \bibinfo {author} {\bibfnamefont {B.~A.}\
  \bibnamefont {Bernevig}},\ }\bibfield  {title} {\bibinfo {title} {Topological
  quantum chemistry},\ }\href {https://doi.org/10.1038/nature23268} {\bibfield
  {journal} {\bibinfo  {journal} {Nature}\ }\textbf {\bibinfo {volume} {547}},\
  \bibinfo {pages} {298} (\bibinfo {year} {2017})}\BibitemShut {NoStop}%
\bibitem [{\citenamefont {Cano}\ \emph {et~al.}(2018)\citenamefont {Cano},
  \citenamefont {Bradlyn}, \citenamefont {Wang}, \citenamefont {Elcoro},
  \citenamefont {Vergniory}, \citenamefont {Felser}, \citenamefont {Aroyo},\
  and\ \citenamefont {Bernevig}}]{Cano2018}%
  \BibitemOpen
  \bibfield  {author} {\bibinfo {author} {\bibfnamefont {J.}~\bibnamefont
  {Cano}}, \bibinfo {author} {\bibfnamefont {B.}~\bibnamefont {Bradlyn}},
  \bibinfo {author} {\bibfnamefont {Z.}~\bibnamefont {Wang}}, \bibinfo {author}
  {\bibfnamefont {L.}~\bibnamefont {Elcoro}}, \bibinfo {author} {\bibfnamefont
  {M.~G.}\ \bibnamefont {Vergniory}}, \bibinfo {author} {\bibfnamefont
  {C.}~\bibnamefont {Felser}}, \bibinfo {author} {\bibfnamefont {M.~I.}\
  \bibnamefont {Aroyo}},\ and\ \bibinfo {author} {\bibfnamefont {B.~A.}\
  \bibnamefont {Bernevig}},\ }\bibfield  {title} {\bibinfo {title} {Building
  blocks of topological quantum chemistry: Elementary band representations},\
  }\href {https://doi.org/10.1103/physrevb.97.035139} {\bibfield  {journal}
  {\bibinfo  {journal} {Physical Review B}\ }\textbf {\bibinfo {volume} {97}},\
  \bibinfo {pages} {035139} (\bibinfo {year} {2018})}\BibitemShut {NoStop}%
\bibitem [{\citenamefont {Bradlyn}\ \emph {et~al.}(2018)\citenamefont
  {Bradlyn}, \citenamefont {Elcoro}, \citenamefont {Vergniory}, \citenamefont
  {Cano}, \citenamefont {Wang}, \citenamefont {Felser}, \citenamefont {Aroyo},\
  and\ \citenamefont {Bernevig}}]{Bradlyn2018}%
  \BibitemOpen
  \bibfield  {author} {\bibinfo {author} {\bibfnamefont {B.}~\bibnamefont
  {Bradlyn}}, \bibinfo {author} {\bibfnamefont {L.}~\bibnamefont {Elcoro}},
  \bibinfo {author} {\bibfnamefont {M.~G.}\ \bibnamefont {Vergniory}}, \bibinfo
  {author} {\bibfnamefont {J.}~\bibnamefont {Cano}}, \bibinfo {author}
  {\bibfnamefont {Z.}~\bibnamefont {Wang}}, \bibinfo {author} {\bibfnamefont
  {C.}~\bibnamefont {Felser}}, \bibinfo {author} {\bibfnamefont {M.~I.}\
  \bibnamefont {Aroyo}},\ and\ \bibinfo {author} {\bibfnamefont {B.~A.}\
  \bibnamefont {Bernevig}},\ }\bibfield  {title} {\bibinfo {title} {Band
  connectivity for topological quantum chemistry: Band structures as a graph
  theory problem},\ }\href {https://doi.org/10.1103/physrevb.97.035138}
  {\bibfield  {journal} {\bibinfo  {journal} {Physical Review B}\ }\textbf
  {\bibinfo {volume} {97}},\ \bibinfo {pages} {035138} (\bibinfo {year}
  {2018})}\BibitemShut {NoStop}%
\bibitem [{\citenamefont {Su}\ \emph {et~al.}(1979)\citenamefont {Su},
  \citenamefont {Schrieffer},\ and\ \citenamefont {Heeger}}]{Su1979}%
  \BibitemOpen
  \bibfield  {author} {\bibinfo {author} {\bibfnamefont {W.~P.}\ \bibnamefont
  {Su}}, \bibinfo {author} {\bibfnamefont {J.~R.}\ \bibnamefont {Schrieffer}},\
  and\ \bibinfo {author} {\bibfnamefont {A.~J.}\ \bibnamefont {Heeger}},\
  }\bibfield  {title} {\bibinfo {title} {Solitons in polyacetylene},\ }\href
  {https://doi.org/10.1103/PhysRevLett.42.1698} {\bibfield  {journal} {\bibinfo
   {journal} {Phys. Rev. Lett.}\ }\textbf {\bibinfo {volume} {42}},\ \bibinfo
  {pages} {1698} (\bibinfo {year} {1979})}\BibitemShut {NoStop}%
\bibitem [{\citenamefont {Thio}\ \emph {et~al.}(1995)\citenamefont {Thio},
  \citenamefont {Bennett},\ and\ \citenamefont {Thurston}}]{Thio1995}%
  \BibitemOpen
  \bibfield  {author} {\bibinfo {author} {\bibfnamefont {T.}~\bibnamefont
  {Thio}}, \bibinfo {author} {\bibfnamefont {J.~W.}\ \bibnamefont {Bennett}},\
  and\ \bibinfo {author} {\bibfnamefont {T.~R.}\ \bibnamefont {Thurston}},\
  }\bibfield  {title} {\bibinfo {title} {Surface and bulk magnetic properties
  of pyrite ${\mathrm{nis}}_{2}$: Magnetization and neutron-scattering
  studies},\ }\href {https://doi.org/10.1103/PhysRevB.52.3555} {\bibfield
  {journal} {\bibinfo  {journal} {Phys. Rev. B}\ }\textbf {\bibinfo {volume}
  {52}},\ \bibinfo {pages} {3555} (\bibinfo {year} {1995})}\BibitemShut
  {NoStop}%
\bibitem [{\citenamefont {Miyadai}\ \emph {et~al.}(1975)\citenamefont
  {Miyadai}, \citenamefont {Takizawa}, \citenamefont {Nagata}, \citenamefont
  {Ito}, \citenamefont {Miyahara},\ and\ \citenamefont
  {Hirakawa}}]{Miyadai1975}%
  \BibitemOpen
  \bibfield  {author} {\bibinfo {author} {\bibfnamefont {T.}~\bibnamefont
  {Miyadai}}, \bibinfo {author} {\bibfnamefont {K.}~\bibnamefont {Takizawa}},
  \bibinfo {author} {\bibfnamefont {H.}~\bibnamefont {Nagata}}, \bibinfo
  {author} {\bibfnamefont {H.}~\bibnamefont {Ito}}, \bibinfo {author}
  {\bibfnamefont {S.}~\bibnamefont {Miyahara}},\ and\ \bibinfo {author}
  {\bibfnamefont {K.}~\bibnamefont {Hirakawa}},\ }\bibfield  {title} {\bibinfo
  {title} {Neutron diffraction study of nis2 with pyrite structure},\ }\href
  {https://doi.org/10.1143/JPSJ.38.115} {\bibfield  {journal} {\bibinfo
  {journal} {Journal of the Physical Society of Japan}\ }\textbf {\bibinfo
  {volume} {38}},\ \bibinfo {pages} {115} (\bibinfo {year} {1975})}\BibitemShut
  {NoStop}%
\bibitem [{\citenamefont {Chapon}\ \emph {et~al.}(2011)\citenamefont {Chapon},
  \citenamefont {Manuel}, \citenamefont {Radaelli}, \citenamefont {Benson},
  \citenamefont {Perrott}, \citenamefont {Ansell}, \citenamefont {Rhodes},
  \citenamefont {Raspino}, \citenamefont {Duxbury}, \citenamefont {Spill},\
  and\ \citenamefont {Norris}}]{wish}%
  \BibitemOpen
  \bibfield  {author} {\bibinfo {author} {\bibfnamefont {L.~C.}\ \bibnamefont
  {Chapon}}, \bibinfo {author} {\bibfnamefont {P.}~\bibnamefont {Manuel}},
  \bibinfo {author} {\bibfnamefont {P.~G.}\ \bibnamefont {Radaelli}}, \bibinfo
  {author} {\bibfnamefont {C.}~\bibnamefont {Benson}}, \bibinfo {author}
  {\bibfnamefont {L.}~\bibnamefont {Perrott}}, \bibinfo {author} {\bibfnamefont
  {S.}~\bibnamefont {Ansell}}, \bibinfo {author} {\bibfnamefont {N.~J.}\
  \bibnamefont {Rhodes}}, \bibinfo {author} {\bibfnamefont {D.}~\bibnamefont
  {Raspino}}, \bibinfo {author} {\bibfnamefont {D.}~\bibnamefont {Duxbury}},
  \bibinfo {author} {\bibfnamefont {E.}~\bibnamefont {Spill}},\ and\ \bibinfo
  {author} {\bibfnamefont {J.}~\bibnamefont {Norris}},\ }\bibfield  {title}
  {\bibinfo {title} {{WISH: The New Powder and Single Crystal Magnetic
  Diffractometer on the Second Target Station}},\ }\href
  {https://doi.org/10.1080/10448632.2011.569650} {\bibfield  {journal}
  {\bibinfo  {journal} {Neutron News}\ }\textbf {\bibinfo {volume} {22}},\
  \bibinfo {pages} {22} (\bibinfo {year} {2011})},\ \Eprint
  {https://arxiv.org/abs/https://doi.org/10.1080/10448632.2011.}
  {https://doi.org/10.1080/10448632.2011.} \BibitemShut {NoStop}%
\bibitem [{\citenamefont {Campbell}\ \emph {et~al.}(2006)\citenamefont
  {Campbell}, \citenamefont {Stokes}, \citenamefont {Tanner},\ and\
  \citenamefont {Hatch}}]{Isodistort}%
  \BibitemOpen
  \bibfield  {author} {\bibinfo {author} {\bibfnamefont {B.~J.}\ \bibnamefont
  {Campbell}}, \bibinfo {author} {\bibfnamefont {H.~T.}\ \bibnamefont
  {Stokes}}, \bibinfo {author} {\bibfnamefont {D.~E.}\ \bibnamefont {Tanner}},\
  and\ \bibinfo {author} {\bibfnamefont {D.~M.}\ \bibnamefont {Hatch}},\
  }\bibfield  {title} {\bibinfo {title} {{{\it ISODISPLACE}: a web-based tool
  for exploring structural distortions}},\ }\href
  {https://doi.org/10.1107/S0021889806014075} {\bibfield  {journal} {\bibinfo
  {journal} {Journal of Applied Crystallography}\ }\textbf {\bibinfo {volume}
  {39}},\ \bibinfo {pages} {607} (\bibinfo {year} {2006})}\BibitemShut
  {NoStop}%
\bibitem [{\citenamefont {Stokes}\ \emph {et~al.}()\citenamefont {Stokes},
  \citenamefont {Hatch},\ and\ \citenamefont {Campbell}}]{isodistort2}%
  \BibitemOpen
  \bibfield  {author} {\bibinfo {author} {\bibfnamefont {H.~T.}\ \bibnamefont
  {Stokes}}, \bibinfo {author} {\bibfnamefont {D.~M.}\ \bibnamefont {Hatch}},\
  and\ \bibinfo {author} {\bibfnamefont {B.~J.}\ \bibnamefont {Campbell}},\
  }\href@noop {} {\bibinfo {title} {Isodistort, isotropy software suite}},\
  \bibinfo {howpublished} {\url{https://iso.byu.edu}}\BibitemShut {NoStop}%
\bibitem [{\citenamefont {Pet{\v r}{\'i}{\v c}ek}\ \emph
  {et~al.}(2023)\citenamefont {Pet{\v r}{\'i}{\v c}ek}, \citenamefont
  {Palatinus}, \citenamefont {Pl{\'a}{\v s}il},\ and\ \citenamefont {Du{\v
  s}ek}}]{Jana2006}%
  \BibitemOpen
  \bibfield  {author} {\bibinfo {author} {\bibfnamefont {V.}~\bibnamefont
  {Pet{\v r}{\'i}{\v c}ek}}, \bibinfo {author} {\bibfnamefont {L.}~\bibnamefont
  {Palatinus}}, \bibinfo {author} {\bibfnamefont {J.}~\bibnamefont {Pl{\'a}{\v
  s}il}},\ and\ \bibinfo {author} {\bibfnamefont {M.}~\bibnamefont {Du{\v
  s}ek}},\ }\bibfield  {title} {\bibinfo {title} {Jana2020 -- a new version of
  the crystallographic computing system jana},\ }\href
  {https://doi.org/10.1515/zkri-2023-0005} {\bibfield  {journal} {\bibinfo
  {journal} {Zeitschrift f{\"u}r Kristallographie - Crystalline Materials}\
  }\textbf {\bibinfo {volume} {238}},\ \bibinfo {pages} {271} (\bibinfo {year}
  {2023})}\BibitemShut {NoStop}%
\bibitem [{\citenamefont {Feng}\ \emph {et~al.}(2011)\citenamefont {Feng},
  \citenamefont {Jaramillo}, \citenamefont {Banerjee}, \citenamefont {Honig},\
  and\ \citenamefont {Rosenbaum}}]{Feng2011}%
  \BibitemOpen
  \bibfield  {author} {\bibinfo {author} {\bibfnamefont {Y.}~\bibnamefont
  {Feng}}, \bibinfo {author} {\bibfnamefont {R.}~\bibnamefont {Jaramillo}},
  \bibinfo {author} {\bibfnamefont {A.}~\bibnamefont {Banerjee}}, \bibinfo
  {author} {\bibfnamefont {J.~M.}\ \bibnamefont {Honig}},\ and\ \bibinfo
  {author} {\bibfnamefont {T.~F.}\ \bibnamefont {Rosenbaum}},\ }\bibfield
  {title} {\bibinfo {title} {Magnetism, structure, and charge correlation at a
  pressure-induced mott-hubbard insulator-metal transition},\ }\href
  {https://doi.org/10.1103/PhysRevB.83.035106} {\bibfield  {journal} {\bibinfo
  {journal} {Phys. Rev. B}\ }\textbf {\bibinfo {volume} {83}},\ \bibinfo
  {pages} {035106} (\bibinfo {year} {2011})}\BibitemShut {NoStop}%
\bibitem [{\citenamefont {{Le}}\ \emph {et~al.}(2024)\citenamefont {{Le}},
  \citenamefont {{Zhang}}, \citenamefont {{Li}}, \citenamefont {{Jiang}},
  \citenamefont {{Sheng}}, \citenamefont {{Tu}}, \citenamefont {{Cao}},
  \citenamefont {{Lyu}}, \citenamefont {{Shen}}, \citenamefont {{Liu}},
  \citenamefont {{Liu}}, \citenamefont {{Wang}}, \citenamefont {{Lu}},\ and\
  \citenamefont {{Qu}}}]{Le2024}%
  \BibitemOpen
  \bibfield  {author} {\bibinfo {author} {\bibfnamefont {T.}~\bibnamefont
  {{Le}}}, \bibinfo {author} {\bibfnamefont {R.}~\bibnamefont {{Zhang}}},
  \bibinfo {author} {\bibfnamefont {C.}~\bibnamefont {{Li}}}, \bibinfo {author}
  {\bibfnamefont {R.}~\bibnamefont {{Jiang}}}, \bibinfo {author} {\bibfnamefont
  {H.}~\bibnamefont {{Sheng}}}, \bibinfo {author} {\bibfnamefont
  {L.}~\bibnamefont {{Tu}}}, \bibinfo {author} {\bibfnamefont {X.}~\bibnamefont
  {{Cao}}}, \bibinfo {author} {\bibfnamefont {Z.}~\bibnamefont {{Lyu}}},
  \bibinfo {author} {\bibfnamefont {J.}~\bibnamefont {{Shen}}}, \bibinfo
  {author} {\bibfnamefont {G.}~\bibnamefont {{Liu}}}, \bibinfo {author}
  {\bibfnamefont {F.}~\bibnamefont {{Liu}}}, \bibinfo {author} {\bibfnamefont
  {Z.}~\bibnamefont {{Wang}}}, \bibinfo {author} {\bibfnamefont
  {L.}~\bibnamefont {{Lu}}},\ and\ \bibinfo {author} {\bibfnamefont
  {F.}~\bibnamefont {{Qu}}},\ }\bibfield  {title} {\bibinfo {title} {{Magnetic
  field filtering of the boundary supercurrent in unconventional metal
  NiTe$_{2}$-based Josephson junctions}},\ }\href
  {https://doi.org/10.1038/s41467-024-47103-z} {\bibfield  {journal} {\bibinfo
  {journal} {Nature Communications}\ }\textbf {\bibinfo {volume} {15}},\
  \bibinfo {eid} {2785} (\bibinfo {year} {2024})},\ \Eprint
  {https://arxiv.org/abs/2303.05041} {arXiv:2303.05041} \BibitemShut {NoStop}%
\bibitem [{\citenamefont {Hung}\ \emph {et~al.}(2002)\citenamefont {Hung},
  \citenamefont {Muscat}, \citenamefont {Yarovsky},\ and\ \citenamefont
  {Russo}}]{Hung2002}%
  \BibitemOpen
  \bibfield  {author} {\bibinfo {author} {\bibfnamefont {A.}~\bibnamefont
  {Hung}}, \bibinfo {author} {\bibfnamefont {J.}~\bibnamefont {Muscat}},
  \bibinfo {author} {\bibfnamefont {I.}~\bibnamefont {Yarovsky}},\ and\
  \bibinfo {author} {\bibfnamefont {S.~P.}\ \bibnamefont {Russo}},\ }\bibfield
  {title} {\bibinfo {title} {Density-functional theory studies of pyrite
  fes2(100) and (110) surfaces},\ }\href
  {https://doi.org/https://doi.org/10.1016/S0039-6028(02)01849-6} {\bibfield
  {journal} {\bibinfo  {journal} {Surface Science}\ }\textbf {\bibinfo {volume}
  {513}},\ \bibinfo {pages} {511} (\bibinfo {year} {2002})}\BibitemShut
  {NoStop}%
\bibitem [{\citenamefont {Pauly}\ \emph {et~al.}(2015)\citenamefont {Pauly},
  \citenamefont {Rasche}, \citenamefont {Koepernik}, \citenamefont {Liebmann},
  \citenamefont {Pratzer}, \citenamefont {Richter}, \citenamefont {Kellner},
  \citenamefont {Eschbach}, \citenamefont {Kaufmann}, \citenamefont
  {Plucinski}, \citenamefont {Schneider}, \citenamefont {Ruck}, \citenamefont
  {van~den Brink},\ and\ \citenamefont {Morgenstern}}]{Pauly2015}%
  \BibitemOpen
  \bibfield  {author} {\bibinfo {author} {\bibfnamefont {C.}~\bibnamefont
  {Pauly}}, \bibinfo {author} {\bibfnamefont {B.}~\bibnamefont {Rasche}},
  \bibinfo {author} {\bibfnamefont {K.}~\bibnamefont {Koepernik}}, \bibinfo
  {author} {\bibfnamefont {M.}~\bibnamefont {Liebmann}}, \bibinfo {author}
  {\bibfnamefont {M.}~\bibnamefont {Pratzer}}, \bibinfo {author} {\bibfnamefont
  {M.}~\bibnamefont {Richter}}, \bibinfo {author} {\bibfnamefont
  {J.}~\bibnamefont {Kellner}}, \bibinfo {author} {\bibfnamefont
  {M.}~\bibnamefont {Eschbach}}, \bibinfo {author} {\bibfnamefont
  {B.}~\bibnamefont {Kaufmann}}, \bibinfo {author} {\bibfnamefont
  {L.}~\bibnamefont {Plucinski}}, \bibinfo {author} {\bibfnamefont {C.~M.}\
  \bibnamefont {Schneider}}, \bibinfo {author} {\bibfnamefont {M.}~\bibnamefont
  {Ruck}}, \bibinfo {author} {\bibfnamefont {J.}~\bibnamefont {van~den
  Brink}},\ and\ \bibinfo {author} {\bibfnamefont {M.}~\bibnamefont
  {Morgenstern}},\ }\bibfield  {title} {\bibinfo {title} {Subnanometre-wide
  electron channels protected by topology},\ }\href
  {https://doi.org/10.1038/nphys3264} {\bibfield  {journal} {\bibinfo
  {journal} {Nature Physics}\ }\textbf {\bibinfo {volume} {11}},\ \bibinfo
  {pages} {338–343} (\bibinfo {year} {2015})}\BibitemShut {NoStop}%
\bibitem [{\citenamefont {Reis}\ \emph {et~al.}(2017)\citenamefont {Reis},
  \citenamefont {Li}, \citenamefont {Dudy}, \citenamefont {Bauernfeind},
  \citenamefont {Glass}, \citenamefont {Hanke}, \citenamefont {Thomale},
  \citenamefont {Sch\"{a}fer},\ and\ \citenamefont {Claessen}}]{Reis2017}%
  \BibitemOpen
  \bibfield  {author} {\bibinfo {author} {\bibfnamefont {F.}~\bibnamefont
  {Reis}}, \bibinfo {author} {\bibfnamefont {G.}~\bibnamefont {Li}}, \bibinfo
  {author} {\bibfnamefont {L.}~\bibnamefont {Dudy}}, \bibinfo {author}
  {\bibfnamefont {M.}~\bibnamefont {Bauernfeind}}, \bibinfo {author}
  {\bibfnamefont {S.}~\bibnamefont {Glass}}, \bibinfo {author} {\bibfnamefont
  {W.}~\bibnamefont {Hanke}}, \bibinfo {author} {\bibfnamefont
  {R.}~\bibnamefont {Thomale}}, \bibinfo {author} {\bibfnamefont
  {J.}~\bibnamefont {Sch\"{a}fer}},\ and\ \bibinfo {author} {\bibfnamefont
  {R.}~\bibnamefont {Claessen}},\ }\bibfield  {title} {\bibinfo {title}
  {Bismuthene on a sic substrate: A candidate for a high-temperature quantum
  spin hall material},\ }\href {https://doi.org/10.1126/science.aai8142}
  {\bibfield  {journal} {\bibinfo  {journal} {Science}\ }\textbf {\bibinfo
  {volume} {357}},\ \bibinfo {pages} {287–290} (\bibinfo {year}
  {2017})}\BibitemShut {NoStop}%
\bibitem [{\citenamefont {Tang}\ \emph {et~al.}(2017)\citenamefont {Tang},
  \citenamefont {Zhang}, \citenamefont {Wong}, \citenamefont {Pedramrazi},
  \citenamefont {Tsai}, \citenamefont {Jia}, \citenamefont {Moritz},
  \citenamefont {Claassen}, \citenamefont {Ryu}, \citenamefont {Kahn},
  \citenamefont {Jiang}, \citenamefont {Yan}, \citenamefont {Hashimoto},
  \citenamefont {Lu}, \citenamefont {Moore}, \citenamefont {Hwang},
  \citenamefont {Hwang}, \citenamefont {Hussain}, \citenamefont {Chen},
  \citenamefont {Ugeda}, \citenamefont {Liu}, \citenamefont {Xie},
  \citenamefont {Devereaux}, \citenamefont {Crommie}, \citenamefont {Mo},\ and\
  \citenamefont {Shen}}]{Tang2017}%
  \BibitemOpen
  \bibfield  {author} {\bibinfo {author} {\bibfnamefont {S.}~\bibnamefont
  {Tang}}, \bibinfo {author} {\bibfnamefont {C.}~\bibnamefont {Zhang}},
  \bibinfo {author} {\bibfnamefont {D.}~\bibnamefont {Wong}}, \bibinfo {author}
  {\bibfnamefont {Z.}~\bibnamefont {Pedramrazi}}, \bibinfo {author}
  {\bibfnamefont {H.-Z.}\ \bibnamefont {Tsai}}, \bibinfo {author}
  {\bibfnamefont {C.}~\bibnamefont {Jia}}, \bibinfo {author} {\bibfnamefont
  {B.}~\bibnamefont {Moritz}}, \bibinfo {author} {\bibfnamefont
  {M.}~\bibnamefont {Claassen}}, \bibinfo {author} {\bibfnamefont
  {H.}~\bibnamefont {Ryu}}, \bibinfo {author} {\bibfnamefont {S.}~\bibnamefont
  {Kahn}}, \bibinfo {author} {\bibfnamefont {J.}~\bibnamefont {Jiang}},
  \bibinfo {author} {\bibfnamefont {H.}~\bibnamefont {Yan}}, \bibinfo {author}
  {\bibfnamefont {M.}~\bibnamefont {Hashimoto}}, \bibinfo {author}
  {\bibfnamefont {D.}~\bibnamefont {Lu}}, \bibinfo {author} {\bibfnamefont
  {R.~G.}\ \bibnamefont {Moore}}, \bibinfo {author} {\bibfnamefont {C.-C.}\
  \bibnamefont {Hwang}}, \bibinfo {author} {\bibfnamefont {C.}~\bibnamefont
  {Hwang}}, \bibinfo {author} {\bibfnamefont {Z.}~\bibnamefont {Hussain}},
  \bibinfo {author} {\bibfnamefont {Y.}~\bibnamefont {Chen}}, \bibinfo {author}
  {\bibfnamefont {M.~M.}\ \bibnamefont {Ugeda}}, \bibinfo {author}
  {\bibfnamefont {Z.}~\bibnamefont {Liu}}, \bibinfo {author} {\bibfnamefont
  {X.}~\bibnamefont {Xie}}, \bibinfo {author} {\bibfnamefont {T.~P.}\
  \bibnamefont {Devereaux}}, \bibinfo {author} {\bibfnamefont {M.~F.}\
  \bibnamefont {Crommie}}, \bibinfo {author} {\bibfnamefont {S.-K.}\
  \bibnamefont {Mo}},\ and\ \bibinfo {author} {\bibfnamefont {Z.-X.}\
  \bibnamefont {Shen}},\ }\bibfield  {title} {\bibinfo {title} {Quantum spin
  hall state in monolayer 1t’-wte2},\ }\href
  {https://doi.org/10.1038/nphys4174} {\bibfield  {journal} {\bibinfo
  {journal} {Nature Physics}\ }\textbf {\bibinfo {volume} {13}},\ \bibinfo
  {pages} {683–687} (\bibinfo {year} {2017})}\BibitemShut {NoStop}%
\bibitem [{\citenamefont {Song}\ \emph {et~al.}(2018)\citenamefont {Song},
  \citenamefont {Jia}, \citenamefont {Zhang}, \citenamefont {Zhu},
  \citenamefont {Shi}, \citenamefont {Wang}, \citenamefont {Zhu}, \citenamefont
  {Yuan}, \citenamefont {Zhang}, \citenamefont {Xing},\ and\ \citenamefont
  {Li}}]{Song2018}%
  \BibitemOpen
  \bibfield  {author} {\bibinfo {author} {\bibfnamefont {Y.-H.}\ \bibnamefont
  {Song}}, \bibinfo {author} {\bibfnamefont {Z.-Y.}\ \bibnamefont {Jia}},
  \bibinfo {author} {\bibfnamefont {D.}~\bibnamefont {Zhang}}, \bibinfo
  {author} {\bibfnamefont {X.-Y.}\ \bibnamefont {Zhu}}, \bibinfo {author}
  {\bibfnamefont {Z.-Q.}\ \bibnamefont {Shi}}, \bibinfo {author} {\bibfnamefont
  {H.}~\bibnamefont {Wang}}, \bibinfo {author} {\bibfnamefont {L.}~\bibnamefont
  {Zhu}}, \bibinfo {author} {\bibfnamefont {Q.-Q.}\ \bibnamefont {Yuan}},
  \bibinfo {author} {\bibfnamefont {H.}~\bibnamefont {Zhang}}, \bibinfo
  {author} {\bibfnamefont {D.-Y.}\ \bibnamefont {Xing}},\ and\ \bibinfo
  {author} {\bibfnamefont {S.-C.}\ \bibnamefont {Li}},\ }\bibfield  {title}
  {\bibinfo {title} {Observation of coulomb gap in the quantum spin hall
  candidate single-layer 1t’-wte2},\ }\bibfield  {journal} {\bibinfo
  {journal} {Nature Communications}\ }\textbf {\bibinfo {volume} {9}},\ \href
  {https://doi.org/10.1038/s41467-018-06635-x} {10.1038/s41467-018-06635-x}
  (\bibinfo {year} {2018})\BibitemShut {NoStop}%
\bibitem [{\citenamefont {Ugeda}\ \emph {et~al.}(2018)\citenamefont {Ugeda},
  \citenamefont {Pulkin}, \citenamefont {Tang}, \citenamefont {Ryu},
  \citenamefont {Wu}, \citenamefont {Zhang}, \citenamefont {Wong},
  \citenamefont {Pedramrazi}, \citenamefont {Mart{\'\i}n-Recio}, \citenamefont
  {Chen}, \citenamefont {Wang}, \citenamefont {Shen}, \citenamefont {Mo},
  \citenamefont {Yazyev},\ and\ \citenamefont {Crommie}}]{Ugeda2018}%
  \BibitemOpen
  \bibfield  {author} {\bibinfo {author} {\bibfnamefont {M.~M.}\ \bibnamefont
  {Ugeda}}, \bibinfo {author} {\bibfnamefont {A.}~\bibnamefont {Pulkin}},
  \bibinfo {author} {\bibfnamefont {S.}~\bibnamefont {Tang}}, \bibinfo {author}
  {\bibfnamefont {H.}~\bibnamefont {Ryu}}, \bibinfo {author} {\bibfnamefont
  {Q.}~\bibnamefont {Wu}}, \bibinfo {author} {\bibfnamefont {Y.}~\bibnamefont
  {Zhang}}, \bibinfo {author} {\bibfnamefont {D.}~\bibnamefont {Wong}},
  \bibinfo {author} {\bibfnamefont {Z.}~\bibnamefont {Pedramrazi}}, \bibinfo
  {author} {\bibfnamefont {A.}~\bibnamefont {Mart{\'\i}n-Recio}}, \bibinfo
  {author} {\bibfnamefont {Y.}~\bibnamefont {Chen}}, \bibinfo {author}
  {\bibfnamefont {F.}~\bibnamefont {Wang}}, \bibinfo {author} {\bibfnamefont
  {Z.-X.}\ \bibnamefont {Shen}}, \bibinfo {author} {\bibfnamefont {S.-K.}\
  \bibnamefont {Mo}}, \bibinfo {author} {\bibfnamefont {O.~V.}\ \bibnamefont
  {Yazyev}},\ and\ \bibinfo {author} {\bibfnamefont {M.~F.}\ \bibnamefont
  {Crommie}},\ }\bibfield  {title} {\bibinfo {title} {Observation of
  topologically protected states at crystalline phase boundaries in
  single-layer {WSe2}},\ }\href {https://doi.org/10.1038/s41467-018-05672-w}
  {\bibfield  {journal} {\bibinfo  {journal} {Nat. Commun.}\ }\textbf {\bibinfo
  {volume} {9}},\ \bibinfo {pages} {3401} (\bibinfo {year} {2018})}\BibitemShut
  {NoStop}%
\bibitem [{\citenamefont {Efros}\ and\ \citenamefont
  {Shklovskii}(1975)}]{Efros1975}%
  \BibitemOpen
  \bibfield  {author} {\bibinfo {author} {\bibfnamefont {A.~L.}\ \bibnamefont
  {Efros}}\ and\ \bibinfo {author} {\bibfnamefont {B.~I.}\ \bibnamefont
  {Shklovskii}},\ }\bibfield  {title} {\bibinfo {title} {Coulomb gap and low
  temperature conductivity of disordered systems},\ }\href
  {https://doi.org/10.1088/0022-3719/8/4/003} {\bibfield  {journal} {\bibinfo
  {journal} {Journal of Physics C: Solid State Physics}\ }\textbf {\bibinfo
  {volume} {8}},\ \bibinfo {pages} {L49} (\bibinfo {year} {1975})}\BibitemShut
  {NoStop}%
\bibitem [{\citenamefont {Voit}(1995)}]{Voit1995}%
  \BibitemOpen
  \bibfield  {author} {\bibinfo {author} {\bibfnamefont {J.}~\bibnamefont
  {Voit}},\ }\bibfield  {title} {\bibinfo {title} {One-dimensional fermi
  liquids},\ }\href {https://doi.org/10.1088/0034-4885/58/9/002} {\bibfield
  {journal} {\bibinfo  {journal} {Reports on Progress in Physics}\ }\textbf
  {\bibinfo {volume} {58}},\ \bibinfo {pages} {977} (\bibinfo {year}
  {1995})}\BibitemShut {NoStop}%
\bibitem [{\citenamefont {Becke}\ and\ \citenamefont
  {Johnson}(2006)}]{Becke2006}%
  \BibitemOpen
  \bibfield  {author} {\bibinfo {author} {\bibfnamefont {A.~D.}\ \bibnamefont
  {Becke}}\ and\ \bibinfo {author} {\bibfnamefont {E.~R.}\ \bibnamefont
  {Johnson}},\ }\bibfield  {title} {\bibinfo {title} {{A simple effective
  potential for exchange}},\ }\href {https://doi.org/10.1063/1.2213970}
  {\bibfield  {journal} {\bibinfo  {journal} {The Journal of Chemical Physics}\
  }\textbf {\bibinfo {volume} {124}},\ \bibinfo {pages} {221101} (\bibinfo
  {year} {2006})}\BibitemShut {NoStop}%
\bibitem [{\citenamefont {Tran}\ and\ \citenamefont {Blaha}(2009)}]{Tran2009}%
  \BibitemOpen
  \bibfield  {author} {\bibinfo {author} {\bibfnamefont {F.}~\bibnamefont
  {Tran}}\ and\ \bibinfo {author} {\bibfnamefont {P.}~\bibnamefont {Blaha}},\
  }\bibfield  {title} {\bibinfo {title} {Accurate band gaps of semiconductors
  and insulators with a semilocal exchange-correlation potential},\ }\href
  {https://doi.org/10.1103/PhysRevLett.102.226401} {\bibfield  {journal}
  {\bibinfo  {journal} {Phys. Rev. Lett.}\ }\textbf {\bibinfo {volume} {102}},\
  \bibinfo {pages} {226401} (\bibinfo {year} {2009})}\BibitemShut {NoStop}%
\bibitem [{\citenamefont {Reiss}\ \emph {et~al.}(2022)\citenamefont {Reiss},
  \citenamefont {Friedemann},\ and\ \citenamefont {Grosche}}]{Reiss2022}%
  \BibitemOpen
  \bibfield  {author} {\bibinfo {author} {\bibfnamefont {P.}~\bibnamefont
  {Reiss}}, \bibinfo {author} {\bibfnamefont {S.}~\bibnamefont {Friedemann}},\
  and\ \bibinfo {author} {\bibfnamefont {F.~M.}\ \bibnamefont {Grosche}},\
  }\bibfield  {title} {\bibinfo {title} {{Ab initio electronic structure of
  metallized ${\mathrm{NiS}}_{2}$ in the noncollinear magnetic phase}},\ }\href
  {https://doi.org/10.1103/PhysRevB.106.205131} {\bibfield  {journal} {\bibinfo
   {journal} {Phys. Rev. B}\ }\textbf {\bibinfo {volume} {106}},\ \bibinfo
  {pages} {205131} (\bibinfo {year} {2022})}\BibitemShut {NoStop}%
\bibitem [{\citenamefont {Xu}\ \emph {et~al.}(2020)\citenamefont {Xu},
  \citenamefont {Elcoro}, \citenamefont {Song}, \citenamefont {Wieder},
  \citenamefont {Vergniory}, \citenamefont {Regnault}, \citenamefont {Chen},
  \citenamefont {Felser},\ and\ \citenamefont {Bernevig}}]{Xu2020}%
  \BibitemOpen
  \bibfield  {author} {\bibinfo {author} {\bibfnamefont {Y.}~\bibnamefont
  {Xu}}, \bibinfo {author} {\bibfnamefont {L.}~\bibnamefont {Elcoro}}, \bibinfo
  {author} {\bibfnamefont {Z.-D.}\ \bibnamefont {Song}}, \bibinfo {author}
  {\bibfnamefont {B.~J.}\ \bibnamefont {Wieder}}, \bibinfo {author}
  {\bibfnamefont {M.~G.}\ \bibnamefont {Vergniory}}, \bibinfo {author}
  {\bibfnamefont {N.}~\bibnamefont {Regnault}}, \bibinfo {author}
  {\bibfnamefont {Y.}~\bibnamefont {Chen}}, \bibinfo {author} {\bibfnamefont
  {C.}~\bibnamefont {Felser}},\ and\ \bibinfo {author} {\bibfnamefont {B.~A.}\
  \bibnamefont {Bernevig}},\ }\bibfield  {title} {\bibinfo {title}
  {High-throughput calculations of magnetic topological materials},\ }\href
  {https://doi.org/10.1038/s41586-020-2837-0} {\bibfield  {journal} {\bibinfo
  {journal} {Nature}\ }\textbf {\bibinfo {volume} {586}},\ \bibinfo {pages}
  {702} (\bibinfo {year} {2020})}\BibitemShut {NoStop}%
\bibitem [{\citenamefont {Elcoro}\ \emph {et~al.}(2021)\citenamefont {Elcoro},
  \citenamefont {Wieder}, \citenamefont {Song}, \citenamefont {Xu},
  \citenamefont {Bradlyn},\ and\ \citenamefont {Bernevig}}]{Elcoro2021}%
  \BibitemOpen
  \bibfield  {author} {\bibinfo {author} {\bibfnamefont {L.}~\bibnamefont
  {Elcoro}}, \bibinfo {author} {\bibfnamefont {B.~J.}\ \bibnamefont {Wieder}},
  \bibinfo {author} {\bibfnamefont {Z.}~\bibnamefont {Song}}, \bibinfo {author}
  {\bibfnamefont {Y.}~\bibnamefont {Xu}}, \bibinfo {author} {\bibfnamefont
  {B.}~\bibnamefont {Bradlyn}},\ and\ \bibinfo {author} {\bibfnamefont {B.~A.}\
  \bibnamefont {Bernevig}},\ }\bibfield  {title} {\bibinfo {title} {Magnetic
  topological quantum chemistry},\ }\href
  {https://doi.org/10.1038/s41467-021-26241-8} {\bibfield  {journal} {\bibinfo
  {journal} {Nature Communications}\ }\textbf {\bibinfo {volume} {12}},\
  \bibinfo {pages} {5965} (\bibinfo {year} {2021})}\BibitemShut {NoStop}%
\bibitem [{\citenamefont {Song}\ \emph {et~al.}(2017)\citenamefont {Song},
  \citenamefont {Fang},\ and\ \citenamefont {Fang}}]{Song2017}%
  \BibitemOpen
  \bibfield  {author} {\bibinfo {author} {\bibfnamefont {Z.}~\bibnamefont
  {Song}}, \bibinfo {author} {\bibfnamefont {Z.}~\bibnamefont {Fang}},\ and\
  \bibinfo {author} {\bibfnamefont {C.}~\bibnamefont {Fang}},\ }\bibfield
  {title} {\bibinfo {title} {$(d\ensuremath{-}2)$-dimensional edge states of
  rotation symmetry protected topological states},\ }\href
  {https://doi.org/10.1103/PhysRevLett.119.246402} {\bibfield  {journal}
  {\bibinfo  {journal} {Phys. Rev. Lett.}\ }\textbf {\bibinfo {volume} {119}},\
  \bibinfo {pages} {246402} (\bibinfo {year} {2017})}\BibitemShut {NoStop}%
\bibitem [{\citenamefont {Rhim}\ \emph {et~al.}(2017)\citenamefont {Rhim},
  \citenamefont {Behrends},\ and\ \citenamefont {Bardarson}}]{Rhim2017}%
  \BibitemOpen
  \bibfield  {author} {\bibinfo {author} {\bibfnamefont {J.-W.}\ \bibnamefont
  {Rhim}}, \bibinfo {author} {\bibfnamefont {J.}~\bibnamefont {Behrends}},\
  and\ \bibinfo {author} {\bibfnamefont {J.~H.}\ \bibnamefont {Bardarson}},\
  }\bibfield  {title} {\bibinfo {title} {Bulk-boundary correspondence from the
  intercellular zak phase},\ }\href
  {https://doi.org/10.1103/PhysRevB.95.035421} {\bibfield  {journal} {\bibinfo
  {journal} {Phys. Rev. B}\ }\textbf {\bibinfo {volume} {95}},\ \bibinfo
  {pages} {035421} (\bibinfo {year} {2017})}\BibitemShut {NoStop}%
\bibitem [{\citenamefont {Benalcazar}\ \emph {et~al.}(2019)\citenamefont
  {Benalcazar}, \citenamefont {Li},\ and\ \citenamefont
  {Hughes}}]{Benalcazar2019}%
  \BibitemOpen
  \bibfield  {author} {\bibinfo {author} {\bibfnamefont {W.~A.}\ \bibnamefont
  {Benalcazar}}, \bibinfo {author} {\bibfnamefont {T.}~\bibnamefont {Li}},\
  and\ \bibinfo {author} {\bibfnamefont {T.~L.}\ \bibnamefont {Hughes}},\
  }\bibfield  {title} {\bibinfo {title} {Quantization of fractional corner
  charge in ${C}_{n}$-symmetric higher-order topological crystalline
  insulators},\ }\href {https://doi.org/10.1103/PhysRevB.99.245151} {\bibfield
  {journal} {\bibinfo  {journal} {Phys. Rev. B}\ }\textbf {\bibinfo {volume}
  {99}},\ \bibinfo {pages} {245151} (\bibinfo {year} {2019})}\BibitemShut
  {NoStop}%
\bibitem [{\citenamefont {Schindler}\ \emph {et~al.}(2019)\citenamefont
  {Schindler}, \citenamefont {Brzezi\ifmmode~\acute{n}\else \'{n}\fi{}ska},
  \citenamefont {Benalcazar}, \citenamefont {Iraola}, \citenamefont {Bouhon},
  \citenamefont {Tsirkin}, \citenamefont {Vergniory},\ and\ \citenamefont
  {Neupert}}]{Schindler2019}%
  \BibitemOpen
  \bibfield  {author} {\bibinfo {author} {\bibfnamefont {F.}~\bibnamefont
  {Schindler}}, \bibinfo {author} {\bibfnamefont {M.}~\bibnamefont
  {Brzezi\ifmmode~\acute{n}\else \'{n}\fi{}ska}}, \bibinfo {author}
  {\bibfnamefont {W.~A.}\ \bibnamefont {Benalcazar}}, \bibinfo {author}
  {\bibfnamefont {M.}~\bibnamefont {Iraola}}, \bibinfo {author} {\bibfnamefont
  {A.}~\bibnamefont {Bouhon}}, \bibinfo {author} {\bibfnamefont {S.~S.}\
  \bibnamefont {Tsirkin}}, \bibinfo {author} {\bibfnamefont {M.~G.}\
  \bibnamefont {Vergniory}},\ and\ \bibinfo {author} {\bibfnamefont
  {T.}~\bibnamefont {Neupert}},\ }\bibfield  {title} {\bibinfo {title}
  {Fractional corner charges in spin-orbit coupled crystals},\ }\href
  {https://doi.org/10.1103/PhysRevResearch.1.033074} {\bibfield  {journal}
  {\bibinfo  {journal} {Phys. Rev. Res.}\ }\textbf {\bibinfo {volume} {1}},\
  \bibinfo {pages} {033074} (\bibinfo {year} {2019})}\BibitemShut {NoStop}%
\bibitem [{\citenamefont {Xu}\ \emph {et~al.}(2021)\citenamefont {Xu},
  \citenamefont {Elcoro}, \citenamefont {Li}, \citenamefont {Song},
  \citenamefont {Regnault}, \citenamefont {Yang}, \citenamefont {Sun},
  \citenamefont {Parkin}, \citenamefont {Felser},\ and\ \citenamefont
  {Bernevig}}]{Xu2021}%
  \BibitemOpen
  \bibfield  {author} {\bibinfo {author} {\bibfnamefont {Y.}~\bibnamefont
  {Xu}}, \bibinfo {author} {\bibfnamefont {L.}~\bibnamefont {Elcoro}}, \bibinfo
  {author} {\bibfnamefont {G.}~\bibnamefont {Li}}, \bibinfo {author}
  {\bibfnamefont {Z.-D.}\ \bibnamefont {Song}}, \bibinfo {author}
  {\bibfnamefont {N.}~\bibnamefont {Regnault}}, \bibinfo {author}
  {\bibfnamefont {Q.}~\bibnamefont {Yang}}, \bibinfo {author} {\bibfnamefont
  {Y.}~\bibnamefont {Sun}}, \bibinfo {author} {\bibfnamefont {S.}~\bibnamefont
  {Parkin}}, \bibinfo {author} {\bibfnamefont {C.}~\bibnamefont {Felser}},\
  and\ \bibinfo {author} {\bibfnamefont {B.~A.}\ \bibnamefont {Bernevig}},\
  }\bibfield  {title} {\bibinfo {title} {Three-dimensional real space
  invariants, obstructed atomic insulators and a new principle for active
  catalytic sites},\ }\href {https://arxiv.org/abs/2111.02433} {\bibfield
  {journal} {\bibinfo  {journal} {arXiv e-prints arXiv:2111.02433}\ } (\bibinfo
  {year} {2021})},\ \Eprint {https://arxiv.org/abs/2111.02433}
  {arXiv:2111.02433 [cond-mat.mtrl-sci]} \BibitemShut {NoStop}%
\bibitem [{\citenamefont {Liu}\ \emph {et~al.}(2024)\citenamefont {Liu},
  \citenamefont {Zhu}, \citenamefont {Feng}, \citenamefont {Zeng},
  \citenamefont {Ma}, \citenamefont {Hao}, \citenamefont {Dai}, \citenamefont
  {Luo}, \citenamefont {Yamagami}, \citenamefont {Liu}, \citenamefont {Cui},
  \citenamefont {Sun}, \citenamefont {Liu}, \citenamefont {Liu}, \citenamefont
  {Ye}, \citenamefont {Shen}, \citenamefont {Li},\ and\ \citenamefont
  {Liu}}]{Lu2024}%
  \BibitemOpen
  \bibfield  {author} {\bibinfo {author} {\bibfnamefont {X.-R.}\ \bibnamefont
  {Liu}}, \bibinfo {author} {\bibfnamefont {M.-Y.}\ \bibnamefont {Zhu}},
  \bibinfo {author} {\bibfnamefont {Y.}~\bibnamefont {Feng}}, \bibinfo {author}
  {\bibfnamefont {M.}~\bibnamefont {Zeng}}, \bibinfo {author} {\bibfnamefont
  {X.-M.}\ \bibnamefont {Ma}}, \bibinfo {author} {\bibfnamefont {Y.-J.}\
  \bibnamefont {Hao}}, \bibinfo {author} {\bibfnamefont {Y.}~\bibnamefont
  {Dai}}, \bibinfo {author} {\bibfnamefont {R.-H.}\ \bibnamefont {Luo}},
  \bibinfo {author} {\bibfnamefont {K.}~\bibnamefont {Yamagami}}, \bibinfo
  {author} {\bibfnamefont {Y.}~\bibnamefont {Liu}}, \bibinfo {author}
  {\bibfnamefont {S.}~\bibnamefont {Cui}}, \bibinfo {author} {\bibfnamefont
  {Z.}~\bibnamefont {Sun}}, \bibinfo {author} {\bibfnamefont {J.-Y.}\
  \bibnamefont {Liu}}, \bibinfo {author} {\bibfnamefont {Z.}~\bibnamefont
  {Liu}}, \bibinfo {author} {\bibfnamefont {M.}~\bibnamefont {Ye}}, \bibinfo
  {author} {\bibfnamefont {D.}~\bibnamefont {Shen}}, \bibinfo {author}
  {\bibfnamefont {B.}~\bibnamefont {Li}},\ and\ \bibinfo {author}
  {\bibfnamefont {C.}~\bibnamefont {Liu}},\ }\bibfield  {title} {\bibinfo
  {title} {Observation of floating surface state in obstructed atomic insulator
  candidate nip$_2$},\ }\href {https://arxiv.org/abs/2406.05380} {\bibfield
  {journal} {\bibinfo  {journal} {arXiv e-prints arXiv:2406.05380}\ } (\bibinfo
  {year} {2024})},\ \Eprint {https://arxiv.org/abs/2406.05380}
  {arXiv:2406.05380 [cond-mat.mtrl-sci]} \BibitemShut {NoStop}%
\bibitem [{\citenamefont {Pizzi}\ \emph {et~al.}(2020)\citenamefont {Pizzi},
  \citenamefont {Vitale}, \citenamefont {Arita}, \citenamefont {Blügel},
  \citenamefont {Freimuth}, \citenamefont {G{\'{e}}ranton}, \citenamefont
  {Gibertini}, \citenamefont {Gresch}, \citenamefont {Johnson}, \citenamefont
  {Koretsune}, \citenamefont {Iba{\~{n}}ez-Azpiroz}, \citenamefont {Lee},
  \citenamefont {Lihm}, \citenamefont {Marchand}, \citenamefont {Marrazzo},
  \citenamefont {Mokrousov}, \citenamefont {Mustafa}, \citenamefont {Nohara},
  \citenamefont {Nomura}, \citenamefont {Paulatto}, \citenamefont
  {Ponc{\'{e}}}, \citenamefont {Ponweiser}, \citenamefont {Qiao}, \citenamefont
  {Thöle}, \citenamefont {Tsirkin}, \citenamefont {Wierzbowska}, \citenamefont
  {Marzari}, \citenamefont {Vanderbilt}, \citenamefont {Souza}, \citenamefont
  {Mostofi},\ and\ \citenamefont {Yates}}]{Pizzi2020}%
  \BibitemOpen
  \bibfield  {author} {\bibinfo {author} {\bibfnamefont {G.}~\bibnamefont
  {Pizzi}}, \bibinfo {author} {\bibfnamefont {V.}~\bibnamefont {Vitale}},
  \bibinfo {author} {\bibfnamefont {R.}~\bibnamefont {Arita}}, \bibinfo
  {author} {\bibfnamefont {S.}~\bibnamefont {Blügel}}, \bibinfo {author}
  {\bibfnamefont {F.}~\bibnamefont {Freimuth}}, \bibinfo {author}
  {\bibfnamefont {G.}~\bibnamefont {G{\'{e}}ranton}}, \bibinfo {author}
  {\bibfnamefont {M.}~\bibnamefont {Gibertini}}, \bibinfo {author}
  {\bibfnamefont {D.}~\bibnamefont {Gresch}}, \bibinfo {author} {\bibfnamefont
  {C.}~\bibnamefont {Johnson}}, \bibinfo {author} {\bibfnamefont
  {T.}~\bibnamefont {Koretsune}}, \bibinfo {author} {\bibfnamefont
  {J.}~\bibnamefont {Iba{\~{n}}ez-Azpiroz}}, \bibinfo {author} {\bibfnamefont
  {H.}~\bibnamefont {Lee}}, \bibinfo {author} {\bibfnamefont {J.-M.}\
  \bibnamefont {Lihm}}, \bibinfo {author} {\bibfnamefont {D.}~\bibnamefont
  {Marchand}}, \bibinfo {author} {\bibfnamefont {A.}~\bibnamefont {Marrazzo}},
  \bibinfo {author} {\bibfnamefont {Y.}~\bibnamefont {Mokrousov}}, \bibinfo
  {author} {\bibfnamefont {J.~I.}\ \bibnamefont {Mustafa}}, \bibinfo {author}
  {\bibfnamefont {Y.}~\bibnamefont {Nohara}}, \bibinfo {author} {\bibfnamefont
  {Y.}~\bibnamefont {Nomura}}, \bibinfo {author} {\bibfnamefont
  {L.}~\bibnamefont {Paulatto}}, \bibinfo {author} {\bibfnamefont
  {S.}~\bibnamefont {Ponc{\'{e}}}}, \bibinfo {author} {\bibfnamefont
  {T.}~\bibnamefont {Ponweiser}}, \bibinfo {author} {\bibfnamefont
  {J.}~\bibnamefont {Qiao}}, \bibinfo {author} {\bibfnamefont {F.}~\bibnamefont
  {Thöle}}, \bibinfo {author} {\bibfnamefont {S.~S.}\ \bibnamefont {Tsirkin}},
  \bibinfo {author} {\bibfnamefont {M.}~\bibnamefont {Wierzbowska}}, \bibinfo
  {author} {\bibfnamefont {N.}~\bibnamefont {Marzari}}, \bibinfo {author}
  {\bibfnamefont {D.}~\bibnamefont {Vanderbilt}}, \bibinfo {author}
  {\bibfnamefont {I.}~\bibnamefont {Souza}}, \bibinfo {author} {\bibfnamefont
  {A.~A.}\ \bibnamefont {Mostofi}},\ and\ \bibinfo {author} {\bibfnamefont
  {J.~R.}\ \bibnamefont {Yates}},\ }\bibfield  {title} {\bibinfo {title}
  {Wannier90 as a community code: new features and applications},\ }\href
  {https://doi.org/10.1088/1361-648x/ab51ff} {\bibfield  {journal} {\bibinfo
  {journal} {Journal of Physics: Condensed Matter}\ }\textbf {\bibinfo {volume}
  {32}},\ \bibinfo {pages} {165902} (\bibinfo {year} {2020})}\BibitemShut
  {NoStop}%
\bibitem [{\citenamefont {C{\u a}lug{\u a}ru}\ \emph
  {et~al.}(2025)\citenamefont {C{\u a}lug{\u a}ru}, \citenamefont {Jiang},
  \citenamefont {Guo}, \citenamefont {Sajan}, \citenamefont {Wang},
  \citenamefont {Hu}, \citenamefont {Yu}, \citenamefont {Bernevig},
  \citenamefont {de~Juan},\ and\ \citenamefont {Ugeda}}]{Calugaru2025}%
  \BibitemOpen
  \bibfield  {author} {\bibinfo {author} {\bibfnamefont {D.}~\bibnamefont {C{\u
  a}lug{\u a}ru}}, \bibinfo {author} {\bibfnamefont {Y.}~\bibnamefont {Jiang}},
  \bibinfo {author} {\bibfnamefont {H.}~\bibnamefont {Guo}}, \bibinfo {author}
  {\bibfnamefont {S.}~\bibnamefont {Sajan}}, \bibinfo {author} {\bibfnamefont
  {Y.}~\bibnamefont {Wang}}, \bibinfo {author} {\bibfnamefont {H.}~\bibnamefont
  {Hu}}, \bibinfo {author} {\bibfnamefont {J.}~\bibnamefont {Yu}}, \bibinfo
  {author} {\bibfnamefont {B.~A.}\ \bibnamefont {Bernevig}}, \bibinfo {author}
  {\bibfnamefont {F.}~\bibnamefont {de~Juan}},\ and\ \bibinfo {author}
  {\bibfnamefont {M.~M.}\ \bibnamefont {Ugeda}},\ }\bibfield  {title} {\bibinfo
  {title} {Probing the quantized berry phases in 1h-nbse$_2$ using scanning
  tunneling microscopy},\ }\href {https://arxiv.org/abs/2501.09063} {\bibfield
  {journal} {\bibinfo  {journal} {arXiv e-prints arXiv:2501.09063}\ } (\bibinfo
  {year} {2025})},\ \Eprint {https://arxiv.org/abs/2501.09063}
  {arXiv:2501.09063 [cond-mat.mes-hall]} \BibitemShut {NoStop}%
\bibitem [{\citenamefont {{Jiang}}\ \emph {et~al.}(2023)\citenamefont
  {{Jiang}}, \citenamefont {{Qi}}, \citenamefont {{Weng}},\ and\ \citenamefont
  {{Hu}}}]{Jiang2023}%
  \BibitemOpen
  \bibfield  {author} {\bibinfo {author} {\bibfnamefont {K.}~\bibnamefont
  {{Jiang}}}, \bibinfo {author} {\bibfnamefont {Z.}~\bibnamefont {{Qi}}},
  \bibinfo {author} {\bibfnamefont {H.}~\bibnamefont {{Weng}}},\ and\ \bibinfo
  {author} {\bibfnamefont {J.}~\bibnamefont {{Hu}}},\ }\bibfield  {title}
  {\bibinfo {title} {{Mottness in obstructed atomic insulators without Mott
  transition}},\ }\href {https://doi.org/10.1103/PhysRevB.108.195102}
  {\bibfield  {journal} {\bibinfo  {journal} {\prb}\ }\textbf {\bibinfo
  {volume} {108}},\ \bibinfo {eid} {195102} (\bibinfo {year} {2023})},\ \Eprint
  {https://arxiv.org/abs/2301.13048} {arXiv:2301.13048 [cond-mat.str-el]}
  \BibitemShut {NoStop}%
\bibitem [{\citenamefont {Horcas}\ \emph {et~al.}(2007)\citenamefont {Horcas},
  \citenamefont {Fernández}, \citenamefont {Gómez-Rodríguez}, \citenamefont
  {Colchero}, \citenamefont {Gómez-Herrero},\ and\ \citenamefont
  {Baro}}]{Horcas2007}%
  \BibitemOpen
  \bibfield  {author} {\bibinfo {author} {\bibfnamefont {I.}~\bibnamefont
  {Horcas}}, \bibinfo {author} {\bibfnamefont {R.}~\bibnamefont {Fernández}},
  \bibinfo {author} {\bibfnamefont {J.~M.}\ \bibnamefont {Gómez-Rodríguez}},
  \bibinfo {author} {\bibfnamefont {J.}~\bibnamefont {Colchero}}, \bibinfo
  {author} {\bibfnamefont {J.}~\bibnamefont {Gómez-Herrero}},\ and\ \bibinfo
  {author} {\bibfnamefont {A.~M.}\ \bibnamefont {Baro}},\ }\bibfield  {title}
  {\bibinfo {title} {{WSXM: A software for scanning probe microscopy and a tool
  for nanotechnology}},\ }\href {https://doi.org/10.1063/1.2432410} {\bibfield
  {journal} {\bibinfo  {journal} {Review of Scientific Instruments}\ }\textbf
  {\bibinfo {volume} {78}},\ \bibinfo {pages} {013705} (\bibinfo {year}
  {2007})}\BibitemShut {NoStop}%
\bibitem [{\citenamefont {Methfessel}\ and\ \citenamefont
  {Paxton}(1989)}]{Methfessel-Paxton}%
  \BibitemOpen
  \bibfield  {author} {\bibinfo {author} {\bibfnamefont {M.}~\bibnamefont
  {Methfessel}}\ and\ \bibinfo {author} {\bibfnamefont {A.~T.}\ \bibnamefont
  {Paxton}},\ }\bibfield  {title} {\bibinfo {title} {High-precision sampling
  for brillouin-zone integration in metals},\ }\href
  {https://doi.org/10.1103/PhysRevB.40.3616} {\bibfield  {journal} {\bibinfo
  {journal} {Phys. Rev. B}\ }\textbf {\bibinfo {volume} {40}},\ \bibinfo
  {pages} {3616} (\bibinfo {year} {1989})}\BibitemShut {NoStop}%
\bibitem [{\citenamefont {Becke}\ and\ \citenamefont
  {Roussel}(1989)}]{Becke1989}%
  \BibitemOpen
  \bibfield  {author} {\bibinfo {author} {\bibfnamefont {A.~D.}\ \bibnamefont
  {Becke}}\ and\ \bibinfo {author} {\bibfnamefont {M.~R.}\ \bibnamefont
  {Roussel}},\ }\bibfield  {title} {\bibinfo {title} {Exchange holes in
  inhomogeneous systems: A coordinate-space model},\ }\href
  {https://doi.org/10.1103/PhysRevA.39.3761} {\bibfield  {journal} {\bibinfo
  {journal} {Phys. Rev. A}\ }\textbf {\bibinfo {volume} {39}},\ \bibinfo
  {pages} {3761} (\bibinfo {year} {1989})}\BibitemShut {NoStop}%
\end{thebibliography}%
\end{document}